\documentclass[10pt,oneside,a4paper,onecolumn,hidelinks]{elsarticle}
\usepackage{times}
\usepackage{fullpage}

\usepackage{gensymb}
\usepackage{tikz}
\usepackage{alltt}
\usepackage{upquote}
\usepackage{float}
\usepackage{array}
\usepackage{pgfplots}
\usepackage{graphicx}
\usepackage[toc,page]{appendix}
\usepackage{amsmath,mathtools}
\usetikzlibrary{decorations.markings}
\usetikzlibrary{matrix,shapes,arrows,positioning,chains,shapes.geometric,calc,patterns,decorations.pathmorphing,decorations.markings,fit}
\usepackage{amsthm}
\usepackage{amsfonts}
\usepackage{enumitem}
\usepackage{subcaption}
\usepackage{caption}
\usepackage{xcolor}
\usepackage{scalerel}
\usepackage{hyperref}
\usepackage{setspace}
\usepackage{multicol}
\usepackage{mathtools}
\usetikzlibrary{patterns}
\theoremstyle{remark}
\usepackage{wrapfig}

\makeatletter

\def\pgfsys@declarepattern@alt#1#2#3#4#5#6#7{%
  \pgf@xa=#2\relax%
  \pgf@ya=#3\relax%
  \pgf@xb=#4\relax%
  \pgf@yb=#5\relax%
  \pgf@xc=#6\relax%
  \pgf@yc=#7\relax%
  \pgf@sys@bp@correct\pgf@xa%
  \pgf@sys@bp@correct\pgf@ya%
  \pgf@sys@bp@correct\pgf@xb%
  \pgf@sys@bp@correct\pgf@yb%
  \pgf@sys@bp@correct\pgf@xc%
  \pgf@sys@bp@correct\pgf@yc%
  \pgfsys@@declarepattern@alt{#1}}

\def\pgfsys@@declarepattern@alt#1#2#3#4#5#6#7#8#9{%
   \pgfutil@tempdima=#6\relax%
   \pgfutil@tempdimb=#7\relax%
   \pgf@sys@bp@correct\pgf@xa%
   \pgf@sys@bp@correct\pgf@ya%
   \immediate\pdfobj stream
   attr
   {
     /Type /Pattern
     /PatternType 1
     /PaintType \ifnum#9=0 2 \else 1 \fi
     /TilingType 1
     /BBox [\pgf@sys@tonumber\pgf@xa\space\pgf@sys@tonumber\pgf@ya\space\pgf@sys@tonumber\pgf@xb\space\pgf@sys@tonumber\pgf@yb]
     /XStep \pgf@sys@tonumber\pgf@xc\space
     /YStep \pgf@sys@tonumber\pgf@yc\space
     /Matrix [#2\space#3\space#4\space#5\space\pgf@sys@tonumber\pgfutil@tempdima\space\pgf@sys@tonumber\pgfutil@tempdimb]
     /Resources << >> 
   }
   {#8}%
   \pgfutil@addpdfresource@patterns{/pgfpat#1\space \the\pdflastobj\space 0 R}%
 }

\pgfkeys{/pgf/patterns/.cd,
  name/.code=\edef\pgf@pat@name{#1},
  type/.is choice,
  type/uncolored/.code=\def\pgf@pat@type{0},
  type/colored/.code=\def\pgf@pat@type{1},
  parameters/.store in=\pgf@pat@parameters,
  bottom left/.store in=\pgf@pat@bottomleft,
  top right/.store in=\pgf@pat@topright,
  tile size/.store in=\pgf@pat@tilesize,
  transformation/.store in=\pgf@pat@transformation,
  code/.store in=\pgf@pat@code,
  name=,
  type=uncolored,
  parameters=,
  bottom left=,
  top right=,
  transformation=,
  code=,
  points/.style={/pgf/patterns/points/.cd, #1},
  transformations/.style={/pgf/patterns/transformations/.cd,#1},
  /pgf/patterns/points/.cd,
    x/.store in=\pgf@pat@x,
    y/.store in=\pgf@pat@y,
    width/.store in=\pgf@pat@x,
    height/.store in=\pgf@path@y,
  /pgf/patterns/transformations/.cd,
    rotate/.code=\pgftransformrotate{#1},
    xscale/.code=\pgftransformxscale{#1},
    yscale/.code=\pgftransformyscale{#1},
}

\def\pgf@pat@processpoint#1{%
  \def\pgf@marshal{\pgfutil@in@=}%
  \expandafter\pgf@marshal\expandafter{#1}%
  \ifpgfutil@in@%
    \pgfkeys{/pgf/patterns/points/.expanded=#1}%
    \pgf@process{\pgfpoint{\pgf@pat@x}{\pgf@pat@y}}%
  \else%
    \pgf@process{#1}%
  \fi%
}

\def\pgf@pat@processtransformations#1{%
  \def\pgf@marshal{\pgfutil@in@=}%
  \expandafter\pgf@marshal\expandafter{#1}%
  \ifpgfutil@in@%
    \pgfkeys{/pgf/patterns/transformations/.expanded=#1}%
  \else%
    #1%
  \fi%
}

\def\pgfdeclarepattern#1{%
  \begingroup%
    \def\pgf@pat@opts{#1}%
    \pgfkeys{/pgf/patterns/.cd,#1}%
    \pgfutil@ifundefined{pgf@pattern@name\pgf@pat@name}{%
      \ifx\pgf@pat@parameters\pgfutil@empty%
        \expandafter\global\expandafter\let\csname pgf@pattern@name@\pgf@pat@name @parameters\endcsname=\pgfutil@empty%
        \pgf@declarepattern%
      \else%
        \expandafter\global\expandafter\let\csname pgf@pattern@name@\pgf@pat@name @parameters\endcsname=\pgf@pat@parameters
        \expandafter\global\expandafter\let\csname pgf@pattern@name@\pgf@pat@name\endcsname=\pgf@pat@opts%
      \fi%
    }{%
       \pgferror{Pattern `\pgf@pat@type' already defined}%
    }%
  \endgroup%
}

\def\pgf@declarepattern{%
   \pgfsysprotocol@getcurrentprotocol\pgf@pattern@temp%
   {%
     \pgfinterruptpath%
       \pgfpicturetrue%
       \pgf@relevantforpicturesizefalse%
       \pgftransformreset%
       \pgfsysprotocol@setcurrentprotocol\pgfutil@empty%
       \pgfsysprotocol@bufferedtrue%
       \pgfsys@beginscope%
       \pgfsetarrows{-}%
       \pgf@pat@code%
       \pgfsys@endscope%
       \pgfsysprotocol@getcurrentprotocol\pgf@pattern@code%
       \global\let\pgf@pattern@code=\pgf@pattern@code%
     \endpgfinterruptpath%
     \pgf@pat@processpoint{\pgf@pat@bottomleft}%
     \pgf@xa=\pgf@x%
     \pgf@ya=\pgf@y%
     \pgf@pat@processpoint{\pgf@pat@topright}%
     \pgf@xb=\pgf@x%
     \pgf@yb=\pgf@y%
     \pgf@pat@processpoint{\pgf@pat@tilesize}%
     \pgf@xc=\pgf@x%
     \pgf@yc=\pgf@y%
     \begingroup%
       \pgftransformreset%
       \pgf@pat@processtransformations\pgf@pat@transformation%
       \pgfgettransformentries\aa\ab\ba\bb\shiftx\shifty%
       \global\edef\pgf@pattern@matrix{{\aa}{\ab}{\ba}{\bb}{\shiftx}{\shifty}}%
     \endgroup%
     \pgfutil@tempcnta=\pgf@pattern@number%
     \advance\pgfutil@tempcnta by1\relax%
     \xdef\pgf@pattern@number{\the\pgfutil@tempcnta}%
     \expandafter\xdef\csname pgf@pattern@name@\pgf@pat@name\endcsname{\the\pgfutil@tempcnta}%
     \expandafter\xdef\csname pgf@pattern@type@\pgf@pat@name\endcsname{\pgf@pat@type}%
     \xdef\pgf@marshal{\noexpand\pgfsys@declarepattern@alt%
       {\csname pgf@pattern@name@\pgf@pat@name\endcsname}
       {\the\pgf@xa}{\the\pgf@ya}{\the\pgf@xb}{\the\pgf@yb}{\the\pgf@xc}{\the\pgf@yc}\pgf@pattern@matrix{\pgf@pattern@code}{\pgf@pat@type}}%
   }%
   \pgf@marshal%
   \pgfsysprotocol@setcurrentprotocol\pgf@pattern@temp%
 }

\def\pgfsetfillpattern#1#2{%
  \pgfutil@ifundefined{pgf@pattern@name@#1}%
  {%
    \pgferror{Undefined pattern `#1'}%
  }%
  {%
     \pgfutil@ifundefined{pgf@pattern@name@#1@parameters}{%
        \pgf@set@fillpattern{#1}{#2}%
     }{%
     \expandafter\ifx\csname pgf@pattern@name@#1@parameters\endcsname\pgfutil@empty%
       \pgf@set@fillpattern{#1}{#2}%
     \else
       \edef\pgf@pat@currentparameters{\csname pgf@pattern@name@#1@parameters\endcsname}%
       \edef\pgf@pat@mutablename{#1@\pgf@pat@currentparameters}%
       \pgfutil@ifundefined{pgf@pattern@name@\pgf@pat@mutablename}%
       {%
         \expandafter\expandafter\expandafter\pgfdeclarepattern\expandafter\expandafter\expandafter{\csname pgf@pattern@name@#1\endcsname,
           name=\pgf@pat@mutablename,parameters=}%
       }%
       {}%
       \expandafter\pgf@set@fillpattern\expandafter{\pgf@pat@mutablename}{#2}%
     \fi%
    }%
  }%
}

 \def\pgf@set@fillpattern#1#2{%
    \ifcase\csname pgf@pattern@type@#1\endcsname\relax%
       \pgfutil@colorlet{pgf@tempcolor}{#2}%
       \pgfutil@ifundefined{applycolormixins}{}{\applycolormixins{pgf@tempcolor}}%
       \pgfutil@extractcolorspec{pgf@tempcolor}{\pgf@tempcolor}%
       \expandafter\pgfutil@convertcolorspec\pgf@tempcolor{rgb}{\pgf@rgbcolor}%
       \expandafter\pgf@set@fill@patternuncolored\pgf@rgbcolor\relax{#1}%
    \or
     \pgfsys@setpatterncolored{\csname pgf@pattern@name@#1\endcsname}%
    \else
    \fi
 }

\def\tikzdeclarepattern#1{%
   \begingroup%
     \pgfkeys{/pgf/patterns/code/.code={\def\pgf@pat@code{%
       \let\tikz@transform=\relax\tikz@installcommands##1}}}
     \pgfdeclarepattern{#1,type=colored}%
   \endgroup%
 }
\makeatother

\pgfdeclarepattern{name=hatch,
  type=uncolored,
  parameters={\hatchsize, \hatchangle, \hatchlinewidth},
  bottom left={x=-.1pt, y=-.1pt}, 
  top right={x=\hatchsize+.1pt, y=\hatchsize+.1pt},
  tile size={width=\hatchsize, height=\hatchsize},
  transformation={rotate=\hatchangle},
  code={
    \pgfsetlinewidth{\hatchlinewidth}
    \pgfpathmoveto{\pgfpoint{-.1pt}{-.1pt}}
    \pgfpathlineto{\pgfpoint{\hatchsize+.1pt}{\hatchsize+.1pt}}
    \pgfpathmoveto{\pgfpoint{-.1pt}{\hatchsize+.1pt}}
    \pgfpathlineto{\pgfpoint{\hatchsize+.1pt}{-.1pt}}
    \pgfusepath{stroke}
  }}

\tikzset{%
  hatch size/.store in=\hatchsize,
  hatch angle/.store in=\hatchangle,
  hatch line width/.store in=\hatchlinewidth,
  hatch size=5pt,
  hatch angle=0pt,
  hatch line width=.5pt,
}

\def\vec#1{\mbox{\boldmath $#1$}}

\newcommand{\redcolor}[1]{\textcolor{red}{#1}}

\newcolumntype{M}[1]{>{\centering\arraybackslash}m{#1}}
\newcolumntype{N}{@{}m{0pt}@{}}
\newtheorem{remark}{Remark}

\tikzset{>=latex}
\graphicspath{ {Pics/} }
\def\@author#1{\g@addto@macro\elsauthors{\normalsize%
    \def\baselinestretch{1}%
    \upshape\authorsep#1\unskip\textsuperscript{%
      \ifx\@fnmark\@empty\else\unskip\sep\@fnmark\let\sep=,\fi
      \ifx\@corref\@empty\else\unskip\sep\@corref\let\sep=,\fi
      }%
    \def\authorsep{\unskip,\space}%
    \global\let\@fnmark\@empty
    \global\let\@corref\@empty  
    \global\let\sep\@empty}%
    \@eadauthor={#1}
}

\usepackage[english]{babel}
\begin{document}
\begin{frontmatter}

\title{A variational flexible multibody formulation for partitioned fluid-structure interaction: Application to bat-inspired drones and unmanned air-vehicles}
\author[nus]{Vaibhav Joshi\corref{cor1}}
\ead{vaibhav.joshi@ubc.ca}
\cortext[cor1]{Corresponding author}

\author[nus]{Rajeev K. Jaiman}
\ead{rjaiman@mech.ubc.ca}
\address[nus]{Department of Mechanical Engineering, The University of British Columbia, Vancouver, Canada}

\author[nus]{Carl Ollivier-Gooch}
\ead{cfog@mech.ubc.ca}

\begin{abstract}
We present a three-dimensional (3D) partitioned aeroelastic formulation for a flexible multibody system interacting with incompressible turbulent fluid flow. While the incompressible Navier-Stokes system is discretized using a stabilized Petrov-Galerkin procedure, the multibody structural system consists of a generic interaction of multiple components such as rigid body, beams and flexible thin shells along with various types of joints and connections among them. A co-rotational framework is utilized for the category of small strain problems where the displacement of the body is decomposed into a rigid body rotation and a small strain component. This assumption simplifies the structural equations and allows for the incorporation of multiple bodies (rigid as well as flexible) in the system. The displacement and rotation constraints at the joints are imposed by a Lagrange multiplier method.  The equilibrium conditions at the fluid-structure interface are satisfied by the transfer of tractions and structural displacements via the radial basis function approach, a scattered data interpolation technique, which is globally conservative. For the coupled stability in low structure-to-fluid mass ratio regimes, a nonlinear iterative force correction scheme is employed in the partitioned staggered predictor-corrector scheme. The convergence and generality of the radial basis function mapping are analyzed by carrying out systematic error analysis of the transfer of fluid traction across the non-matching fluid-structure interface where a third-order of convergence is observed. The proposed aeroelastic framework is then validated by considering a flow across a flexible pitching plate configuration with serration at the trailing edge. Finally, we demonstrate the flow across a flexible flapping wing of a bat modeling the bone fingers as beams and the flexible membrane as thin shells in the multibody system along with the joints.
\end{abstract}

\begin{keyword} 
Flexible multibody analysis, bat-inspired flapping, aeroelasticity, radial basis function, partitioned iterative, non-matching meshes
\end{keyword}

\end{frontmatter}

\section{Introduction}
In recent years, there has been an increasing interest in biologically inspired micro-air vehicles (MAVs) and drones with applications in search and rescue, surveillance and transportation of payload in remote locations. These vehicles normally operate at a low speed in a Reynolds number regime of $10^4$-$10^5$ or lower. Aerodynamic performance of rigid fixed wings based on quasi-steady flow decreases dramatically in this flow regime.  Moreover, with their small size, it is challenging to have flight stability and proper control of these vehicles along with maneuverability and agility, especially in the presence of wind gusts. Therefore, it is essential to explore alternate flight mechanisms which can improve the flight performance of these vehicles while imparting enhanced agility and control.

One such mechanism which is observed in natural flyers such as insects, birds and bats is the flapping of flexible wings which has the potential to be an effective biologically inspired mechanism \cite{Ho}. Bio-inspired wings can offer superior aerodynamic performance with improved lift-to-drag ratio, delayed stall and
enhanced flight stability while maintaining light-weight structures, compared to conventional rigid fixed wings. Large amplitude flapping motion of the wings induces periodically varying acceleration leading to large inertial forces and unsteady effects, which lie beyond the scope of quasi-steady aerodynamics. At such low Reynolds numbers, small changes can trigger transition to turbulence and affect the flow separation leading to fluctuation in performance of the wings \cite{Tobalske263}. Mechanisms involving an unsteady leading edge vortex and its reattachment are worth investigating, because such a flow can generate much higher lift than the steady case \cite{vanBerg1997}. Unsteady aerodynamics combined with flexibility and varying wing stiffness generate highly nonlinear fluid-structure interactions; proper treatment of these interactions is essential in modeling flapping flight \cite{SmithAIAA}. It is crucial to study how these natural flyers remain aloft and have better flight control to improve the design of artificial flying vehicles. Such investigation requires a high-fidelity fluid-structure interaction modeling to understand the role of articulated wing  kinematics on the aerodynamics performance  and flight stability.

The flexible wings of natural flyers like birds and bats are prime examples of a multibody aeroelastic system. Such systems are found in applications ranging from biological, such as the musculoskeletal system of animals (bones connected by ligaments), to industrial, like underwater robotics, marine/offshore, rotor dynamics, dual aerial-aquatic vehicles and various other automotive/aerospace applications. Generally, the system consists of multiple components of bodies (rigid or flexible) which are connected by joints and their three-dimensional motion is constrained based upon the type of joint. Of particular interest to the present study is the anatomical wing structure of a bird or a bat which consists of bones (humerus, radius, metacarpals and phalanges) connected to each other by joints. Flexible structures such as feathers (for birds) or membranes (for bats) are attached to these bones. These joints along with the interconnecting muscles create more than 40 degrees of freedom in the wing kinematics of a bat or a bird which can be either active or passive \cite{RISKIN2008604}.

Among animals, bats have the most complicated flight mechanism due to their ability to independently control the movement of the joints to change the wing span as well as the flexibility of the membrane connecting the bone fingers and joints. These wings have evolved over millions of years to sustain flight by the flapping mechanism. They utilize wing flexibility, wing span morphing and complex wing kinematics for maneuverability and agility during flight \cite{Azuma, Tian_2006, Swartz2008}. The flight mechanism of a bat can be categorized into these three effects, viz., adaptive flexibility, articulated kinematics and wing morphing. The anisotropic behaviour of the structural properties along the bone fingers and the membranes gives the benefit of wing flexibility. Some of the experimental works to quantify the wing kinematics of bat and the analysis of the wake patterns generated during flight can be found in \cite{Wolf2142, Muijres2014, Norberg207}. One of the findings was the delayed shedding of the leading-edge vortex due to adaptive flexibility of the wing membrane. Works in \cite{Chen2014, Alireza2017} discussed the replication of the wing kinematics in a robotic wing via articulation of the joints. The effects of changing wing span of the wing on the aerodynamic performance were studied in \cite{Wang2015}. Apart from agility and hovering capabilities, the wings are flexible and soft and have a lower frequency of flapping compared to the rotating blades of a quadcopter drone which are considered a safety hazard \cite{Alireza2017}. Therefore, an in-depth understanding of the flapping mechanism of a bat can be useful to enhance the designs of bio-inspired drones and MAVs.

While a vast amount of literature exists on the experimental studies of bat flight, numerical computations are limited owing to the high deformations of the wing and anisotropic structural properties along the wing span as well as the chord. A flexible multibody aeroelastic framework is required for such problems \cite{ALTENBUCHNER201823, ALTENBUCHNER201873}.  
There have been some numerical studies pertaining to the multibody system of veins and membranes of an insect flapping wing \cite{Chimakurthi2009, Cho2016, Masarati2011, Cho2019}. In these studies, nonlinear beam and shell elements were employed for the structural modeling of the insect wings. 
However, such studies are not directly applicable for the wing modeling of a bat due to the various number of joints across the bone fingers and their interaction with the flexible membrane. Furthermore, the wing kinematics of a bat is a challenge to replicate because of the active articulation of the joints. 

The flexible flapping wing kinematics of a bat wing can be idealized as a small strain problem with large rotations. For such a category of problems, a co-rotational structural framework can be employed. The idea is to decompose the structural displacement as a combination of rigid body rotation and small strains of each multibody component in the system with a rotated frame of reference attached to each multibody \cite{DEVEUBEKE1976895, Cardona_2, Bauchau}.  The recent study in \cite{Cho2019} applied such co-rotational framework for the structural modeling of the insect wing and demonstrated their coupling with the fluid solver.
A co-rotational framework with constraints at the joints applied to an offshore multibody system in  \cite{GURUGUBELLI2018160} dealt with the nonlinear beam elements and rigid body formulations. It was extended to pure shell elements in \cite{LI201996} where a full-scale numerical simulation of a wing was demonstrated. Carrying forward, the current work focuses on the realization of the fully flexible multibody formulation consisting of beams, shells and joints for the three-dimensional computational modeling of bat-inspired wings.

Dealing with such three-dimensional fluid-structure interaction problems is a challenge in terms of the Eulerian-Lagrangian conflict. In the field of computational continuum mechanics, structures are typically simulated using Lagrangian methods with moving material nodes and fluids using an Eulerian spatial grid. This conflict is partially rooted in the fact that for a structure, the stress depends on the total deformation which is computed from the relative positions of the material points, whereas for a fluid, the stress is a function of the deformation rate which is obtained from the numerical derivatives of the velocity field on a fixed mesh. Furthermore, fluid flows involve phenomena which require Eulerian description, like mixing, vortex stretching, turbulence, inflow and outflow boundaries, whereas the structural deformation is inherently Lagrangian, characterized by a relatively smaller total strains and boundary conditions that move with the deforming structure. Therefore, fluid-structure interaction is a prime example whereby the above inconsistency is problematic to solve the coupled system of equations described by the fluid and structural domains. By far, the arbitrary Lagrangian-Eulerian (ALE) moving mesh technique \cite{ALE_2} with the body-fitted interface is the most accurate procedure for fluid-structure interaction problems. The boundary layer can be modeled accurately along with the accurate and conservative satisfaction of the kinematic and dynamic equilibrium conditions at the interface. Other techniques such as immersed boundary method \cite{peskin_2002}, fictitious domain methods \cite{YU20051} and techniques based on Eulerian description of both the structure and the fluid are other alternatives which can handle large deformations of the structure as a result of a fixed grid, but lack the ability to capture boundary layer and high Reynolds number phenomena.

In a typical Eulerian-Lagrangian FSI simulation,  surface meshes at the fluid-structure interface are generally non-matching i.e. connectivity arrangements are different and their geometric coordinates may not be coincident due to  discretization requirements \cite{JAIMAN2006372}.  It is essential that the fluid tractions and structural displacements are transferred across the fluid-structure interface in a locally accurate and conservative manner. For generic non-matching meshes in the structural and fluid domains at the interface, local and global conservative methods such as quadrature projection \cite{Cebral_1997} and common-refinement technique \cite{JAIMAN2006372, LI2018163} as well as globally conservative methods such as radial basis function mapping \cite{BECKERT2001125, Rendall2008} can be utilized. However, for the flexible multibody system considered in the present study, the local conservative methods of projection and common-refinement can become  somewhat complicated to implement considering multiple structural components along with the joints and connections in between them. 
In particular, it is challenging to construct a common-refinement surface since the geometrical realizations of the fluid mesh and the flexible multibody structural meshes 
are defined by distinct surfaces, lines and points with arbitrary mesh intersections.
Therefore, the globally conservative radial basis function interpolation with local support can be a reliable choice for such flexible multibody FSI applications involving non-matching grids.
The advantages of point-to-point mapping using radial basis function (RBF) with compact support is three-fold. First, it avoids the requirements of continuous projection for the transfer of tractions as well as displacement across the fluid-structure interface with non-matching meshes. Second, it is an efficient way of transfer of data for the current application of flapping wing of a bat where high amplitude large deformation of the wings is expected. Thus, the deteriorating quality of the moving mesh can be avoided as time progresses in numerical computation. Third, the RBF technique does not require information of the connectivity of the data points as it is a scattered point interpolation.


In the current work, we extend and apply the recently developed three-dimensional flexible multibody aeroelastic framework for bat-inspired flying vehicles. The structural components of the multibody system are modeled with the help of a co-rotational framework in the Lagrangian description for application to small strain problems with large rigid body rotations. The connections between different components are considered as constraints and incorporated by a Lagrange multiplier technique. The flexible multibody structural framework with constraints at joints is then coupled with the incompressible flow equation in a partitioned staggered manner. We independently construct the meshes for the fluid and the solid subdomains.
The flow equations, written in an arbitrary Lagrangian-Eulerian reference frame, are variationally discretized using the Petrov-Galerkin stabilization technique with Newton-Raphson linearization. The coupling at the fluid-structure interface is carried out by the globally conservative radial basis function technique along with the nonlinear iterative force correction (NIFC) employed for stability at low structure-to-fluid mass ratio regimes \cite{NIFC_1, NIFC_2}.  In a nutshell, the key contributions of this work are as follows:
\begin{itemize}
	\item Development of 3D parallel flexible multibody fluid-structure interaction solver using the stabilized Petrov-Galerkin finite element framework and a generalized partitioned staggered coupling;
	\item Systematic analysis of the interface coupling using the compactly supported radial basis function for non-matching fluid and solid meshes;
	\item Accurate and robust integration of the RBF-based interface scheme with the NIFC procedure for low-mass ratio thin elastic structure interacting with turbulent flow;
	\item Illustration of our 3D RBF-NIFC formulation for a full-scale bat-inspired flying vehicle with flexible multibody constraints and joints.
\end{itemize}

The organization of the paper is as follows. We begin with the description of the flexible multibody aeroelastic equations in Section \ref{FMAE}. The co-rotational flexible multibody framework along with the constraints at the joints connecting the multibodies is discussed. The flow equations and the equilibrium conditions to be satisfied at the fluid-structure interface are reviewed. Semi-discrete variational discretization of the equations and the linearized matrix form are the topic of discussion in Section \ref{SDVF}. The coupling procedure at the fluid-structure interface via the radial basis function mapping is the topic of discussion of Section \ref{FSI_coupling}. In Section \ref{RBF_Error_analysis}, convergence study of the RBF mapping for static load transfer from the fluid to the structural interface is systematically carried out. Mesh convergence and validation tests pertaining to the flow across a flexible pitching plate are performed in Section \ref{Num_tests}. Three-dimensional flapping dynamics of a bat model consisting of multiple bodies is then demonstrated in Section \ref{demons}. Finally, we conclude with some key findings in Section \ref{conclusion}.

\section{Flexible multibody aeroelastic equations}
\label{FMAE}
Let us begin with a discussion about the formulation of the recently developed flexible multibody aeroelastic framework. It consists of the structural equations written in the co-rotational form for small strain problems along with the constraint equations at the connections of multiple bodies. The external fluid loads on the structure are modeled via the incompressible Navier-Stokes equations. Kinematic and dynamic equilibrium conditions are satisfied at the fluid-structure interface for the aeroelastic framework.

\subsection{Co-rotational multibody structural framework for small strain problems}
\label{co_rot_frame}


Consider a $d$-dimensional structural domain of a multibody component $i$ in a multibody system, $\Omega^\mathrm{s}_i(0) \subset \mathbb{R}^d$ with material coordinates $\boldsymbol{X}^\mathrm{s}$ at time $t=0$ consisting of a piecewise smooth boundary $\Gamma^\mathrm{s}$. Let $\boldsymbol{\varphi}^\mathrm{s}(\boldsymbol{X}^\mathrm{s},t):\Omega^\mathrm{s}_i(0) \to \Omega^\mathrm{s}_i(t)$ be a one-to-one mapping between the reference coordinates $\boldsymbol{X}^\mathrm{s}$ at $t=0$ and the respective position in the deformed configuration $\Omega^\mathrm{s}_i(t)$. If $\boldsymbol{\eta}^\mathrm{s}(\boldsymbol{X}^\mathrm{s},t)$ represents the structural displacement due to the deformation, $\boldsymbol{\varphi}^\mathrm{s}(\boldsymbol{X}^\mathrm{s},t) = \boldsymbol{X}^\mathrm{s} + \boldsymbol{\eta}^\mathrm{s}$. 

The continuum structural equation of the flexible structural component $i$ in the multibody system is given as,
\begin{align}
\rho^\mathrm{s}\frac{\partial^2 \boldsymbol{\varphi}^\mathrm{s}}{\partial t^2} &= \nabla\cdot\boldsymbol{\sigma}^\mathrm{s}(\boldsymbol{E}(\boldsymbol{\eta}^\mathrm{s})) + \boldsymbol{b}^\mathrm{s},\ &&\mathrm{on}\ \Omega^\mathrm{s}_i(0),  \label{Eqn_S}\\
	\boldsymbol{u}^\mathrm{s} &= \boldsymbol{u}^\mathrm{s}_D,\ &&\forall \boldsymbol{X}^\mathrm{s} \in \Gamma^\mathrm{s}_D,\\
	\boldsymbol{\sigma}^\mathrm{s}\cdot\mathbf{n}^\mathrm{s} &= \boldsymbol{h}^\mathrm{s},\ &&\forall \boldsymbol{X}^\mathrm{s} \in \Gamma^\mathrm{s}_N,\\
	\boldsymbol{\varphi}^\mathrm{s} &= \boldsymbol{\varphi}^\mathrm{s}_0,\ &&\mathrm{on}\ \Omega^\mathrm{s}_i(0),\\
	\boldsymbol{u}^\mathrm{s} &= \boldsymbol{u}^\mathrm{s}_0,\ &&\mathrm{on}\ \Omega^\mathrm{s}_i(0),
\end{align}
where $\rho^\mathrm{s}$ is the structural density, $\boldsymbol{b}^\mathrm{s}$ is the body force acting on $\Omega^\mathrm{s}_i$ and $\boldsymbol{\sigma}^\mathrm{s}$ is the first Piola-Kirchhoff stress tensor which is a function of the Cauchy-Green Lagrangian strain tensor $\boldsymbol{E}(\boldsymbol{\eta}^\mathrm{s}) = (1/2)[(\boldsymbol{I} + \nabla\boldsymbol{\eta}^\mathrm{s})^T(\boldsymbol{I}+\nabla\boldsymbol{\eta}^\mathrm{s}) - \boldsymbol{I}]$, $\boldsymbol{I}$ being the identity tensor. The Dirichlet condition on the structural velocity $\boldsymbol{u}^\mathrm{s} = \partial \boldsymbol{\varphi}^\mathrm{s}/\partial t = \partial\boldsymbol{\eta}^\mathrm{s}/\partial t$ on the Dirichlet boundary $\Gamma^\mathrm{s}_D$ is given by $\boldsymbol{u}^\mathrm{s}_D$. The Neumann condition on the stress tensor is denoted by $\boldsymbol{h}^\mathrm{s}$ on $\Gamma^\mathrm{s}_N$ with $\mathbf{n}^\mathrm{s}$ being the outward normal to $\Gamma^\mathrm{s}_N$. The initial conditions on the position vector and the structural velocity are given by $\boldsymbol{\varphi}^\mathrm{s}_0$ and $\boldsymbol{u}^\mathrm{s}_0$ respectively. 

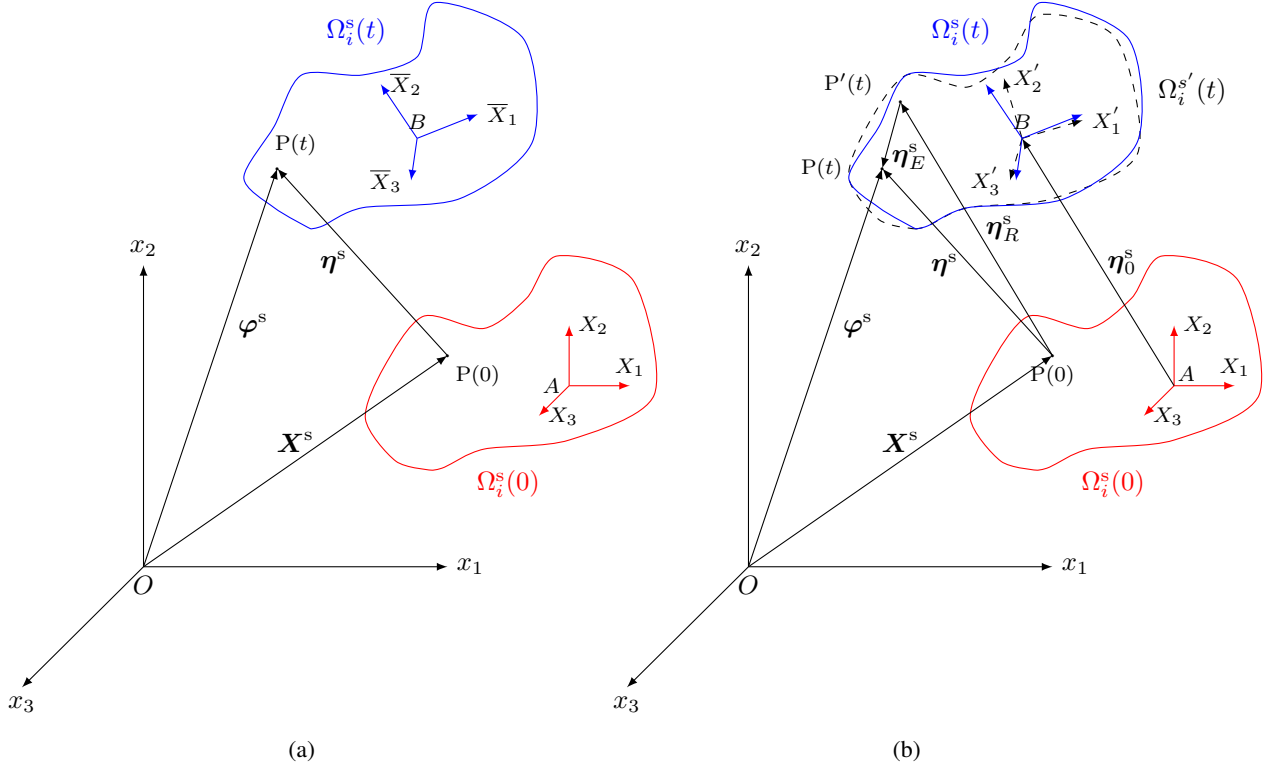
\begin{figure}[!htbp]
	\centering
	\hspace{-0.5cm}
	\begin{subfigure}[b]{0.5\textwidth}
		\def\th{4}
		\def\tx{2}
		\def\tt{0.2}
		\begin{tikzpicture}[decoration={markings,mark=at position 0.5 with {\arrow[scale=2]{>}}},scale=0.8]
		\coordinate (o) at (-2+1,0);
		\coordinate (i2) at (-2+1,5);
		\coordinate (i1) at (5-2+1,0);
		\coordinate (i3) at (-2-2+1,-2);
		
		\coordinate (a) at (6,3);
		\coordinate (ai1) at (7,3);
		\coordinate (ai2) at (6,4);
		\coordinate (ai3) at (5.5,2.5);
		
		\coordinate (b) at (6-\tx-0.5,3+\th+0.1);
		\coordinate (bi1) at (7-\tx-0.5,3.5+\th-0.1);
		\coordinate (bi2) at (5.7-\tx-0.5,4+\th+0.1);
		\coordinate (bi3) at (5.8-\tx-0.5,2.5+\th-0.1);
		
		\coordinate (bi11) at (7-\tx-0.5,3.5+\th);
		\coordinate (bi21) at (5.7-\tx-0.8,4+\th);
		\coordinate (bi31) at (5.8-\tx-0.4,2.5+\th-0.1);
		
		\draw[->] (o) -- (i1);
		\draw[->] (o) -- (i2);
		\draw[->] (o) -- (i3);
		\draw[red] plot [smooth] coordinates  {(7.5/2,3.2/2) (6.3/2,3.5/2) (5.3/2,4.9/2) (5.9/2,6.6/2) (7.2/2,8.3/2) (9.2/2,7.9/2) (10.6/2,8.9/2) (11.6/2,10.3/2) (14.1/2,9.5/2) (14.8/2,5.7/2) (12/2,4.2/2) (9.2/2,3.9/2) (7.9/2,3.3/2) (7.5/2,3.2/2)};
		
		\draw[blue] plot [smooth] coordinates  {(7.5/2-\tx,3.2/2+\th) (6.3/2-\tx,3.5/2+\th+\tt) (5.3/2-\tx,4.9/2+\th) (5.9/2-\tx+\tt,6.6/2+\th) (7.2/2-\tx,8.3/2+\th) (9.2/2-\tx,7.9/2+\th+\tt) (10.6/2-\tx+\tt,8.9/2+\th) (11.6/2-\tx,10.3/2+\th+\tt) (14.1/2-\tx+\tt,9.5/2+\th) (14.8/2-\tx,5.7/2+\th+\tt) (12/2-\tx+\tt,4.2/2+\th) (9.2/2-\tx,3.9/2+\th) (7.9/2-\tx,3.3/2+\th) (7.5/2-\tx,3.2/2+\th)};
		
		\coordinate (P) at (4,3.5);
		\coordinate (Pd) at (1.5,7.7);
		\coordinate (Pt) at (1.2,6.6);
		\draw[->] (o) -- (P);
		\draw (P) node {.};
		\draw (Pt) node {.};
		\draw (4.5,3.5)[anchor=north] node {\footnotesize{$\mathrm{P}(0)$}};
		\draw (2,7)[anchor=east] node {\footnotesize{$\mathrm{P}(t)$}};

		\draw[red,->] (a) -- (ai1);
		\draw[red,->] (a) -- (ai2);
		\draw[red,->] (a) -- (ai3);
		
		\draw[blue,->] plot [smooth] coordinates {(b) (bi11)};
		\draw[blue,->] plot [smooth] coordinates {(b) (bi21)};
		\draw[blue,->] plot [smooth] coordinates {(b) (bi31)};
		
		\draw (o)[anchor=north] node {$O$};
		\draw (i1)[anchor=west] node {$x_1$};
		\draw (i2)[anchor=south] node {$x_2$};
		\draw (i3)[anchor=north] node {$x_3$};
		
		\draw (a)[anchor=east] node {\footnotesize{$A$}};
		\draw (ai1)[anchor=south] node {\footnotesize{$X_1$}};
		\draw (ai2)[anchor=west] node {\footnotesize{$X_2$}};
		\draw (ai3)[anchor=west] node {\footnotesize{$X_3$}};
		
		\draw (b)[anchor=south] node {\footnotesize{$B$}};
		\draw (bi11)[anchor=west] node {\footnotesize{$\overline{X}_{1}$}};
		\draw (bi21)[anchor=west] node {\footnotesize{$\overline{X}_{2}$}};
		\draw (bi31)[anchor=east] node {\footnotesize{$\overline{X}_{3}$}};
		
		\draw (1.5,1.75) node[anchor=south] {$\boldsymbol{X}^\mathrm{s}$};
		
		\draw[->] (P) -- (Pt);
		\draw (2.5,5.05) node[anchor=east] {$\boldsymbol{\eta}^\mathrm{s}$};
		\draw[->] (o) -- (Pt);
		\draw (1.2,4) node[anchor=east] {$\boldsymbol{\varphi}^\mathrm{s}$};
		
		\draw[red] (5,1) node[anchor=south] {$\Omega^\mathrm{s}_i(0)$};
		\draw[blue] (2.5,8.5) node[anchor=south] {$\Omega^\mathrm{s}_i(t)$};
		\end{tikzpicture}
		\caption{}
	\end{subfigure}%
	\begin{subfigure}[b]{0.5\textwidth}
		\def\th{4}
		\def\tx{2}
		\def\tt{0.2}
		\begin{tikzpicture}[decoration={markings,mark=at position 0.5 with {\arrow[scale=2]{>}}},scale=0.8]
		\coordinate (o) at (-2+1,0);
		\coordinate (i2) at (-2+1,5);
		\coordinate (i1) at (5-2+1,0);
		\coordinate (i3) at (-2-2+1,-2);
		
		\coordinate (a) at (6,3);
		\coordinate (ai1) at (7,3);
		\coordinate (ai2) at (6,4);
		\coordinate (ai3) at (5.5,2.5);
		
		\coordinate (b) at (6-\tx-0.5,3+\th+0.1);
		\coordinate (bi1) at (7-\tx-0.5,3.5+\th-0.1);
		\coordinate (bi2) at (5.7-\tx-0.5,4+\th+0.1);
		\coordinate (bi3) at (5.8-\tx-0.5,2.5+\th-0.1);
		
		\coordinate (bi11) at (7-\tx-0.5,3.5+\th);
		\coordinate (bi21) at (5.7-\tx-0.8,4+\th);
		\coordinate (bi31) at (5.8-\tx-0.4,2.5+\th-0.1);
		
		\draw[->] (o) -- (i1);
		\draw[->] (o) -- (i2);
		\draw[->] (o) -- (i3);
		\draw[red] plot [smooth] coordinates  {(7.5/2,3.2/2) (6.3/2,3.5/2) (5.3/2,4.9/2) (5.9/2,6.6/2) (7.2/2,8.3/2) (9.2/2,7.9/2) (10.6/2,8.9/2) (11.6/2,10.3/2) (14.1/2,9.5/2) (14.8/2,5.7/2) (12/2,4.2/2) (9.2/2,3.9/2) (7.9/2,3.3/2) (7.5/2,3.2/2)};
		
		\draw[dashed] plot [smooth] coordinates  {(7.5/2-\tx,3.2/2+\th) (6.3/2-\tx,3.5/2+\th) (5.3/2-\tx,4.9/2+\th) (5.9/2-\tx,6.6/2+\th) (7.2/2-\tx,8.3/2+\th) (9.2/2-\tx,7.9/2+\th) (10.6/2-\tx,8.9/2+\th) (11.6/2-\tx,10.3/2+\th) (14.1/2-\tx,9.5/2+\th) (14.8/2-\tx,5.7/2+\th) (12/2-\tx,4.2/2+\th) (9.2/2-\tx,3.9/2+\th) (7.9/2-\tx,3.3/2+\th) (7.5/2-\tx,3.2/2+\th)};
		
		\draw[blue] plot [smooth] coordinates  {(7.5/2-\tx,3.2/2+\th) (6.3/2-\tx,3.5/2+\th+\tt) (5.3/2-\tx,4.9/2+\th) (5.9/2-\tx+\tt,6.6/2+\th) (7.2/2-\tx,8.3/2+\th) (9.2/2-\tx,7.9/2+\th+\tt) (10.6/2-\tx+\tt,8.9/2+\th) (11.6/2-\tx,10.3/2+\th+\tt) (14.1/2-\tx+\tt,9.5/2+\th) (14.8/2-\tx,5.7/2+\th+\tt) (12/2-\tx+\tt,4.2/2+\th) (9.2/2-\tx,3.9/2+\th) (7.9/2-\tx,3.3/2+\th) (7.5/2-\tx,3.2/2+\th)};
		
		\coordinate (P) at (4,3.5);
		\coordinate (Pd) at (1.5,7.7);
		\coordinate (Pt) at (1.2,6.6);
		\draw[->] (o) -- (P);
		\draw (P) node {.};
		\draw (Pd) node {.};
		\draw (Pt) node {.};
		\draw (P)[anchor=north] node {\footnotesize{$\mathrm{P}(0)$}};
		\draw (1.2,8)[anchor=east] node {\footnotesize{$\mathrm{P}'(t)$}};
		\draw (0.7,6.6)[anchor=east] node {\footnotesize{$\mathrm{P}(t)$}};

		\draw[red,->] (a) -- (ai1);
		\draw[red,->] (a) -- (ai2);
		\draw[red,->] (a) -- (ai3);
		
		\draw[dashed,->] (b) -- (bi1);
		\draw[dashed,->] (b) -- (bi2);
		\draw[dashed,->] (b) -- (bi3);
		
		\draw[blue,->] plot [smooth] coordinates {(b) (bi11)};
		\draw[blue,->] plot [smooth] coordinates {(b) (bi21)};
		\draw[blue,->] plot [smooth] coordinates {(b) (bi31)};

		\draw (o)[anchor=north] node {$O$};
		\draw (i1)[anchor=west] node {$x_1$};
		\draw (i2)[anchor=south] node {$x_2$};
		\draw (i3)[anchor=north] node {$x_3$};
		
		\draw (6.2,3.2) node {\footnotesize{$A$}};
		\draw (ai1)[anchor=south] node {\footnotesize{$X_1$}};
		\draw (ai2)[anchor=west] node {\footnotesize{$X_2$}};
		\draw (ai3)[anchor=west] node {\footnotesize{$X_3$}};
		
		\draw (b)[anchor=south] node {\footnotesize{$B$}};
		\draw (bi1)[anchor=west] node {\footnotesize{$X^{'}_{1}$}};
		\draw (bi2)[anchor=west] node {\footnotesize{$X^{'}_{2}$}};
		\draw (bi3)[anchor=east] node {\footnotesize{$X^{'}_{3}$}};
		
		\draw (1.5,1.75) node[anchor=south] {$\boldsymbol{X}^\mathrm{s}$};
		\draw[->] (P) -- (Pd);
		\draw (2.75,5.6) node[anchor=west] {$\boldsymbol{\eta}^\mathrm{s}_{R}$};
		\draw[->] (a) -- (b);
		\draw (4.75,5.05) node[anchor=west] {$\boldsymbol{\eta}^\mathrm{s}_0$};
		
		\draw[->] (Pd) -- (Pt);
		\draw (1.2,6.8) node[anchor=west] {$\boldsymbol{\eta}^\mathrm{s}_{E}$};
		\draw[->] (P) -- (Pt);
		\draw (2.6,5) node[anchor=east] {$\boldsymbol{\eta}^\mathrm{s}$};
		
		\draw[->] (o) -- (Pt);
		\draw (1.2,4) node[anchor=east] {$\boldsymbol{\varphi}^\mathrm{s}$};
		
		\draw[red] (5,1) node[anchor=south] {$\Omega^\mathrm{s}_i(0)$};
		\draw[blue] (2.5,8.5) node[anchor=south] {$\Omega^\mathrm{s}_i(t)$};
		\draw (6.3,7.5) node[anchor=south] {$\Omega^{s'}_i(t)$};
		\end{tikzpicture}
		\caption{}
	\end{subfigure}%
	\caption{The reference and deformed configurations ($\Omega^\mathrm{s}_i(0)$ and $\Omega^\mathrm{s}_i(t)$ respectively) of a component $i$ of the multibody system in the inertial coordinate system $Ox_1 x_2 x_3$.}
	\label{CRV1}
\end{figure}

To simplify the above equation for small strain problems, consider a component $i$ of the multibody system along with the fixed inertial frame $Ox_1x_2x_3$. At time $t=0$, the body occupies a reference configuration $\Omega^\mathrm{s}_i(0)$ with a material coordinate system $AX_1X_2X_3$ as shown in Fig. \ref{CRV1}(a). The coordinate axes form an orthonormal bases. This configuration can also be referred as the undeformed configuration of the body. Under the action of internal as well as external forces, the body then transforms to the deformed configuration $\Omega^\mathrm{s}_i(t)$ at time $t$. The deformed basis vectors forming the coordinate system $B\overline{X}_1\overline{X}_2\overline{X}_3$ in $\Omega^\mathrm{s}_i(t)$ are neither orthogonal nor unit vectors as the material lines are deformed along with the deformed structure. Let P be any point which undergoes a displacement of $\boldsymbol{\eta}^\mathrm{s}$ from the reference (P$(0)$) to the deformed configuration (P$(t)$). The position of point P in the deformed state can be written as $\boldsymbol{\varphi}^\mathrm{s} = \boldsymbol{X}^\mathrm{s} + \boldsymbol{\eta}^\mathrm{s}$. Therefore, the deformation gradient tensor for the mapping from $AX_1X_2X_3$ to $B\overline{X}_1\overline{X}_2\overline{X}_3$ can be written as $\boldsymbol{F} = \boldsymbol{I} + \nabla\boldsymbol{\eta}^\mathrm{s}$, where $\nabla(\cdot)$ is the gradient with respect to the material coordinates $\boldsymbol{X}^\mathrm{s}$.

Note that the basis vectors in $B\overline{X}_1\overline{X}_2\overline{X}_3$ are not orthonormal. Let us define a new orthogonal coordinate system in the deformed configuration $\Omega^{s'}_i(t)$ as $BX^{'}_1X^{'}_2X^{'}_3$ (Fig. \ref{CRV1}(b)). The deformation gradient tensor can now be expressed as $\boldsymbol{F} = \boldsymbol{R}\hat{\boldsymbol{F}}$, where $\hat{\boldsymbol{F}}$ is the deformation mapping from $BX^{'}_1X^{'}_2X^{'}_3$ to $B\overline{X}_1\overline{X}_2\overline{X}_3$ and $\boldsymbol{R}$ is the rotation matrix which maps the orthonormal coordinates $AX_1X_2X_3$ to $BX^{'}_1X^{'}_2X^{'}_3$. This decomposition expresses the deformation tensor as a product of a rotation tensor, signifying the rigid body rotation and a deformation gradient tensor $\hat{\boldsymbol{F}}$, which depicts the deformation of the body after rigid body motion.

In a flexible multibody system, individual flexible bodies usually undergo small deformations although there are large relative motions at the joints. Consequently, the displacement of any point on the flexible body can be decomposed into a rigid body displacement and an elastic displacement \cite{DEVEUBEKE1976895}. Therefore, the total displacement can be decomposed into the rigid body motion $\boldsymbol{\eta}^\mathrm{s}_{R}$ and the elastic deformation relative to the configuration $\Omega^{s'}_i(t)$, $\boldsymbol{\eta}^\mathrm{s}_{E}$ as $\boldsymbol{\eta}^\mathrm{s} = \boldsymbol{\eta}^\mathrm{s}_{R} + \boldsymbol{\eta}^\mathrm{s}_E$. The rigid body displacement can be expressed as $\boldsymbol{\eta}^\mathrm{s}_R = \boldsymbol{\eta}^\mathrm{s}_0 + \boldsymbol{R}\boldsymbol{X}^\mathrm{s} - \boldsymbol{X}^\mathrm{s}$. Thus, the gradient of the displacement with respect to material coordinates $\boldsymbol{X}^\mathrm{s}$ is
\begin{align}
	\nabla\boldsymbol{\eta}^\mathrm{s} &= \nabla\boldsymbol{\eta}^\mathrm{s}_R + \nabla\boldsymbol{\eta}^\mathrm{s}_E\\
							&= \boldsymbol{R} - \boldsymbol{I} + \nabla\boldsymbol{\eta}^\mathrm{s}_E\\
	\boldsymbol{I} + \nabla\boldsymbol{\eta}^\mathrm{s} &= \boldsymbol{R} + \nabla\boldsymbol{\eta}^\mathrm{s}_E
\end{align}
The Cauchy-Green Lagrangian strain tensor can thus be written as,
\begin{align}
\boldsymbol{E} &= \frac{1}{2}[ (\boldsymbol{R} + \nabla\boldsymbol{\eta}^\mathrm{s}_\mathrm{E})^T (\boldsymbol{R} + \nabla\boldsymbol{\eta}^\mathrm{s}_\mathrm{E}) - \boldsymbol{I}],\\
	&= \frac{1}{2}[ \boldsymbol{R}^T\boldsymbol{R} + \boldsymbol{R}^T\nabla\boldsymbol{\eta}^\mathrm{s}_E + (\nabla\boldsymbol{\eta}^\mathrm{s}_E)^T \boldsymbol{R} + (\nabla\boldsymbol{\eta}^\mathrm{s}_E)^T\nabla\boldsymbol{\eta}^\mathrm{s}_E - \boldsymbol{I} ],\\
	&= \frac{1}{2}[ \boldsymbol{R}^T\nabla\boldsymbol{\eta}^\mathrm{s}_E + (\nabla\boldsymbol{\eta}^\mathrm{s}_E)^T \boldsymbol{R} + (\nabla\boldsymbol{\eta}^\mathrm{s}_E)^T\nabla\boldsymbol{\eta}^\mathrm{s}_E ],
\end{align}
which can be simplified after neglecting higher order terms in $\nabla\boldsymbol{\eta}^\mathrm{s}_\mathrm{E}$ for small strain problems as
\begin{align}
\boldsymbol{E} \approx \frac{1}{2}[\boldsymbol{R}^T \nabla\boldsymbol{\eta}^\mathrm{s}_\mathrm{E} + (\nabla\boldsymbol{\eta}^\mathrm{s}_\mathrm{E})^T\boldsymbol{R}].
\end{align}
The small elastic deformations $\boldsymbol{\eta}^\mathrm{s}_{E}$ can be evaluated in the deformed coordinate system based on the modeling characteristics of the multibody component such as cable, beam, membrane, shell, etc.

\begin{remark}
The rotation tensor $\boldsymbol{R}$ is parameterized using the conformal rotation vector.
Let the vector parameterization of rotation of angle $\phi$ about the normal $\boldsymbol{n}$ be defined as $\boldsymbol{p} = p(\phi) \hat{\boldsymbol{n}}$, where $\hat{\boldsymbol{n}}$ is the unit vector along $\boldsymbol{n}$. The rotation tensor is thus given by Rodrigues' rotation formula, $\boldsymbol{R} = \boldsymbol{I} + \xi_1(\phi)\tilde{\boldsymbol{p}} + \xi_2(\phi)\tilde{\boldsymbol{p}}\tilde{\boldsymbol{p}}$, 
where $\tilde{\boldsymbol{a}}$ is a skew-symmetric tensor formed of the components of a vector $\boldsymbol{a} = [a_1\ \ a_2\ \ a_3]^T$ as,
\begin{align} \label{SSTensor}
	\tilde{\boldsymbol{a}} = \begin{bmatrix}
										0 & -a_{3} & a_{2} \\
										a_{3} & 0 & -a_{1}\\
										-a_{2} & a_{1} & 0\\
										\end{bmatrix},
\end{align}
and $\xi_1(\phi) = \mathrm{sin} \phi/p(\phi)$ and $\xi_2(\phi) = (1-\mathrm{cos}\phi)/(p(\phi))^2$ are even functions of the rotation angle. Based on the choice of the generating function $p(\phi)$, different variants of the parameterization can be obtained such as Cartesian rotation vector, linear, Cayley-Gibbs-Rodrigues and so forth. In this work, the rotation matrix $\boldsymbol{R}$ is parameterized by Wiener-Milenkovi$\mathrm{\acute{c}}$ parameters, which is a technique based on conformal transformation on Euler parameters, also known as conformal rotation vector (CRV) \cite{Wiener1962, Milenkovic}. It is obtained when the generating function is chosen as $p(\phi) = 4 \mathrm{tan}(\phi/4)$. Thus, the rotation tensor can be recast as \cite{Cardona_2},
\begin{align}
	\boldsymbol{R} = \frac{1}{(4-c_0)^2}[(c_0^2+8c_0-16)\boldsymbol{I} + 2\boldsymbol{c}\boldsymbol{c}^T + 2c_0\tilde{\boldsymbol{c}}],
\end{align}
where $\boldsymbol{c} = 4\hat{\boldsymbol{n}}\mathrm{tan}(\phi/4)$ and $c_0 = (1/8)(16-||\boldsymbol{c}||^2)$.
This choice avoids singularities for particular values of rotation and a minimal set of parameters (three in this case) also avoids redundancies in the description of the rotation tensor \cite{Bauchau, Cardona}.
\end{remark}

This completes the co-rotational framework description for the flexible multibody system. Next, we formulate the constraint equations to be satisfied at the connections of multiple bodies.

\subsection{Constraints for joints connecting multiple bodies}
\label{constraint_formulation}
\begin{figure}[!htbp]
	\centering
		\def\th{1}
		\def\tx{4.8}
		\def\tt{0.2}
		\def\thh{5}
		\def\txx{3.7}
		\def\ttt{0.2}
		\def\thhh{5}
		\def\txxx{-1.5}
		\begin{tikzpicture}[decoration={markings,mark=at position 0.5 with {\arrow[scale=2]{>}}},scale=0.8]
		\coordinate (o) at (-5,0);
		\coordinate (i2) at (-5,5);
		\coordinate (i1) at (0,0);
		\coordinate (i3) at (-7,-2);
		
		\draw[blue, ultra thick] (2.725,6.6/2+0.3) circle (2ex);
		\draw[blue, ultra thick] (2.725+1.21,6.6/2+4.7) circle (2ex);		
		
		\coordinate (a) at (5.9/2,6.6/2);
		\coordinate (ai2) at (5.9/2+0.8,6.6/2-0.2);
		\coordinate (ai3) at (5.9/2+0.7,6.6/2+0.6);
		\coordinate (ai1) at (5.9/2-0.1,6.6/2-0.7);
		
		\coordinate (b) at (6-\tx-0.5+1.8,3+\th-0.2);		
		\coordinate (bi11) at (6-\tx-0.5+1.8-0.3,3+\th-0.2+0.8);
		\coordinate (bi21) at (5.7-\tx-0.8+1.8-0.2,4+\th-0.3-0.5);
		\coordinate (bi31) at (5.8-\tx-0.4+1.8-0.6,2.5+\th+0.2);
		
		\coordinate (c) at (5.9/2+1.25,6.6/2+4.5);
		\coordinate (ci2) at (5.9/2+0.8+1.25,6.6/2-0.2+4.5);
		\coordinate (ci3) at (5.9/2+0.7+1.25,6.6/2+0.6+4.5);
		\coordinate (ci1) at (5.9/2-0.1+1.25,6.6/2-0.7+4.5);
		
		\coordinate (d) at (6-\tx-0.5+1.8+1.25,3+\th-0.2+4.5);		
		\coordinate (di11) at (6-\tx-0.5+1.8-0.3+1.25,3+\th-0.2+0.8+4.5);
		\coordinate (di21) at (5.7-\tx-0.8+1.8-0.2+1.25,4+\th-0.3-0.5+4.5);
		\coordinate (di31) at (5.8-\tx-0.4+1.8-0.6+1.25,2.5+\th+0.2+4.5);
		
		\draw[->] (o) -- (i1);
		\draw[->] (o) -- (i2);
		\draw[->] (o) -- (i3);
		\draw[red] plot [smooth] coordinates  {(7.5/2,3.2/2) (6.3/2,3.5/2) (5.3/2,4.9/2) (5.9/2,6.6/2) (7.2/2,8.3/2) (9.2/2,7.9/2) (10.6/2,8.9/2) (11.6/2,10.3/2) (14.1/2,9.5/2) (14.8/2,5.7/2) (12/2,4.2/2) (9.2/2,3.9/2) (7.9/2,3.3/2) (7.5/2,3.2/2)};
		
		\draw[red] plot [smooth] coordinates  {(7.5/2-\tx,3.2/2+\th) (6.3/2-\tx,3.5/2+\th+\tt) (5.3/2-\tx,4.9/2+\th) (5.9/2-\tx+\tt,6.6/2+\th) (7.2/2-\tx,8.3/2+\th) (9.2/2-\tx,7.9/2+\th+\tt) (10.6/2-\tx+\tt,8.9/2+\th) (11.6/2-\tx,10.3/2+\th+\tt) (14.1/2-\tx+\tt,9.5/2+\th) (14.8/2-\tx,5.7/2+\th+\tt) (12/2-\tx+\tt,4.2/2+\th) (9.2/2-\tx,3.9/2+\th) (7.9/2-\tx,3.3/2+\th) (7.5/2-\tx,3.2/2+\th)};
		
		\draw[blue] plot [smooth] coordinates  {(7.5/2-\txxx,3.2/2+\thhh) (6.3/2-\txxx,3.5/2+\thhh) (5.3/2-\txxx,4.9/2+\thhh) (5.9/2-\txxx,6.6/2+\thhh) (7.2/2-\txxx,8.3/2+\thhh) (9.2/2-\txxx,7.9/2+\thhh) (10.6/2-\txxx,8.9/2+\thhh) (11.6/2-\txxx,10.3/2+\thhh) (14.1/2-\txxx,9.5/2+\thhh) (14.8/2-\txxx,5.7/2+\thhh) (12/2-\txxx,4.2/2+\thhh) (9.2/2-\txxx,3.9/2+\thhh) (7.9/2-\txxx,3.3/2+\thhh) (7.5/2-\txxx,3.2/2+\thhh)};
		
		\draw[blue] plot [smooth] coordinates  {(7.5/2-\txx,3.2/2+\thh) (6.3/2-\txx,3.5/2+\thh+\ttt) (5.3/2-\txx,4.9/2+\thh) (5.9/2-\txx+\ttt,6.6/2+\thh) (7.2/2-\txx,8.3/2+\thh) (9.2/2-\txx,7.9/2+\thh+\ttt) (10.6/2-\txx+\ttt,8.9/2+\thh) (11.6/2-\txx,10.3/2+\thh+\ttt) (14.1/2-\txx+\ttt,9.5/2+\thh) (14.8/2-\txx,5.7/2+\thh+\ttt) (12/2-\txx+\ttt,4.2/2+\thh) (9.2/2-\txx,3.9/2+\thh) (7.9/2-\txx,3.3/2+\thh) (7.5/2-\txx,3.2/2+\thh)};
		\draw[->] (o) -- (a);
		\draw[->] (o) -- (b);
		\draw[->] (a) -- (c);
		\draw[->] (b) -- (d);

		\draw[red,->] (a) -- (ai1);
		\draw[red,->] (a) -- (ai2);
		\draw[red,->] (a) -- (ai3);
		
		\draw[red,->] plot [smooth] coordinates {(b) (bi11)};
		\draw[red,->] plot [smooth] coordinates {(b) (bi21)};
		\draw[red,->] plot [smooth] coordinates {(b) (bi31)};
		
		\draw[blue,->] (c) -- (ci1);
		\draw[blue,->] (c) -- (ci2);
		\draw[blue,->] (c) -- (ci3);
		
		\draw[blue,->] (d) -- (di11);
		\draw[blue,->] (d) -- (di21);
		\draw[blue,->] (d) -- (di31);
		
		\draw (o)[anchor=north] node {$O$};
		\draw (i1)[anchor=west] node {$x_1$};
		\draw (i2)[anchor=south] node {$x_2$};
		\draw (i3)[anchor=north] node {$x_3$};
		
		\draw (5.9/2+0.15,6.6/2-0.2) node {\footnotesize{$A$}};
		\draw (5.9/2+0.1,6.6/2-0.7-0.2) node {\footnotesize{$X_1$}};
		\draw (5.9/2+0.8+0.2,6.6/2-0.2) node {\footnotesize{$X_2$}};
		\draw (5.9/2+0.7+0.2,6.6/2+0.6) node {\footnotesize{$X_3$}};
		
		\draw (6-\tx-0.5+1.8,3+\th-0.2+0.3) node {\footnotesize{$B$}};
		\draw (6-\tx-0.5+1.8-0.3-0.1,3+\th-0.2+0.8+0.3) node {\footnotesize{$Y_1$}};
		\draw (5.7-\tx-0.8+1.8-0.2-0.2,4+\th-0.3-0.5+0.2)node {\footnotesize{$Y_2$}};
		\draw (5.8-\tx-0.4+1.8-0.6-0.3,2.5+\th+0.2) node {\footnotesize{$Y_3$}};
		
		\draw (5.9/2+0.15+1.25,6.6/2-0.2+4.4) node {\footnotesize{$C$}};
		\draw (5.9/2+0.1+1.25,6.6/2-0.7-0.2+4.5) node {\footnotesize{$X'_1$}};
		\draw (5.9/2+0.8+0.2+1.25,6.6/2-0.2+4.5) node {\footnotesize{$X'_2$}};
		\draw (5.9/2+0.7+0.2+1.25,6.6/2+0.6+4.5) node {\footnotesize{$X'_3$}};
		
		\draw (6-\tx-0.5+1.8+1.5,3+\th-0.2+0.3+4.5) node {\footnotesize{$D$}};
		\draw (6-\tx-0.5+1.8-0.3-0.1+1.25,3+\th-0.2+0.8+0.3+4.5) node {\footnotesize{$Y'_1$}};
		\draw (5.7-\tx-0.8+1.8-0.2-0.2+1.25,4+\th-0.3-0.5+0.2+4.5)node {\footnotesize{$Y'_2$}};
		\draw (5.8-\tx-0.4+1.8-0.6-0.3+1.25,2.5+\th+0.2+4.5) node {\footnotesize{$Y'_3$}};
		
		\draw (0.8,1.5) node[anchor=south] {$\boldsymbol{X}^\mathrm{s}(A)$};
		\draw (-2.5,1.5) node[anchor=south] {$\boldsymbol{X}^\mathrm{s}(B)$};
		
		\draw[red] (5,3) node[anchor=south] {$\Omega^\mathrm{s}_1(0)$};
		\draw[red] (-0.5,4) node[anchor=south] {$\Omega^\mathrm{s}_2(0)$};
		\draw[blue] (5+1.25,3+4.5) node[anchor=south] {$\Omega^\mathrm{s}_1(t)$};
		\draw[blue] (-0.5+1.25,4+4) node[anchor=south] {$\Omega^\mathrm{s}_2(t)$};
		
		\draw (4,5.7) node {$\boldsymbol{\eta}_A$};
		\draw (2.8,6.2) node {$\boldsymbol{\eta}_B$};
		\end{tikzpicture}
	\caption{The reference and deformed configurations of a constraint joint connecting two bodies $\Omega^\mathrm{s}_1$ and $\Omega^\mathrm{s}_2$ of a multibody system in the inertial coordinate system $Ox_1 x_2 x_3$.}
	\label{Constraint_schematic}
\end{figure}
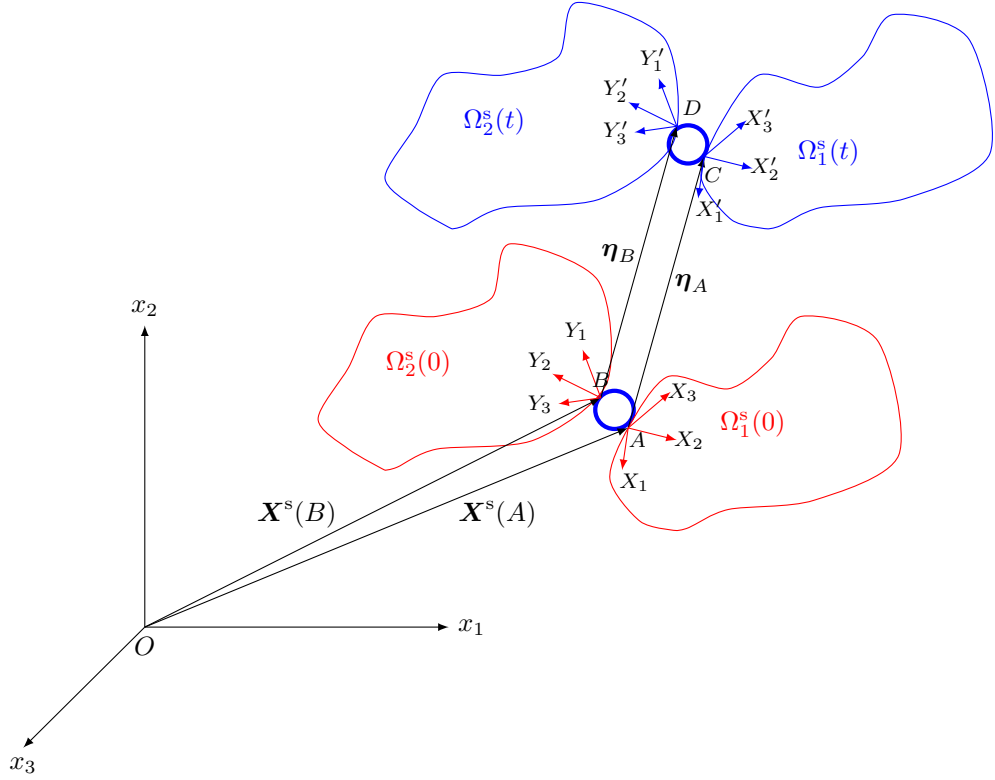

Suppose a constraint joint is connected to bodies $\Omega^\mathrm{s}_1$ and $\Omega^\mathrm{s}_2$ at points $A$ and $B$ respectively. Let $AX_1X_2X_3$ and $BY_1Y_2Y_3$ be the coordinate systems at the reference configurations of $\Omega^\mathrm{s}_1(0)$ and $\Omega^\mathrm{s}_2(0)$ respectively with $CX_1'X_2'X_3'$ and $DY_1'Y_2'Y_3'$ being the corresponding co-rotated coordinate systems in the deformed configuration as shown in Fig. \ref{Constraint_schematic}. The unit vectors of the deformed coordinate systems can be written as $\boldsymbol{e}^A_k = \boldsymbol{R}^A\boldsymbol{R}^A_0\boldsymbol{i}_k$ and $\boldsymbol{e}^B_l = \boldsymbol{R}^B\boldsymbol{R}^B_0\boldsymbol{i}_l$, where $\boldsymbol{R}^A$ and $\boldsymbol{R}^A_0$ denote the rotation matrix mapping $AX_1X_2X_3$ to $CX_1'X_2'X_3'$ and the inertial coordinate system $Ox_1x_2x_3$ to $AX_1X_2X_3$ respectively for the point $A$, $\boldsymbol{i}_k$ being the unit vectors in the inertial coordinate system. 

Let us define the scalar product between the unit vectors as $s_{kl} = (\boldsymbol{e}^A_{k})^T\boldsymbol{e}^B_{l}$ and the vector product as $\boldsymbol{v}_{kl} = \tilde{\boldsymbol{e}}^A_{k} \boldsymbol{e}^B_{l}$. At the reference configuration, the position vector of the two locations at the joint $A$ and $B$ are the same, i.e., $\boldsymbol{X}^\mathrm{s}(A) = \boldsymbol{X}^\mathrm{s}(B)$ and $\boldsymbol{R}^A_0 = \boldsymbol{R}^B_0$. Let the prescribed relative displacement between the two bodies at the joint in the deformed configuration be denoted by $\boldsymbol{\eta}_m$ along the unit vector $\boldsymbol{e}^A_{m}$ and $\phi_m$ be the prescribed relative rotation about the same unit vector. The relative displacement at the joint of point $B$ with respect to point $A$ is given by $\boldsymbol{\eta}_{B/A} = \boldsymbol{\eta}_B - \boldsymbol{\eta}_A$. Thus, the relative displacement and rotational constraints can be written as
\begin{align}
	(\boldsymbol{e}^{A}_m)^T\boldsymbol{\eta}_{B/A} - \boldsymbol{\eta}_m &= 0, \label{rel_disp}\\
	s_{kk} \mathrm{sin}(\phi_m) + s_{kl}\mathrm{cos}(\phi_m) &= 0. \label{rel_rot}
\end{align}
Therefore, when there is no relative displacement at the joint, $\boldsymbol{\eta}_m = 0$, i.e., $(\boldsymbol{e}^A_k)^T\boldsymbol{\eta}_{B/A} = 0$ and in the case of no relative rotation, $\phi_m = 0$ implying that $s_{kl} = 0$. More details can be found in \cite{Bauchau}.

\subsection{Fluid loading on the structure}
\label{eqns_fluid}
The flexible multibody structure along with constraints at joints presented in the previous section is subjected to fluid loading in the aeroelastic framework. This loading is obtained by solving the incompressible Navier-Stokes equations in the arbitrary Lagrangian-Eulerian (ALE) coordinate system. Consider a $d$-dimensional spatial fluid domain $\Omega^\mathrm{f}(t) \subset \mathbb{R}^d$ comprising of a piecewise smooth boundary $\Gamma^\mathrm{f}(t)$. The governing equation for an incompressible, viscous fluid flow are given as
\begin{align} \label{RANS}
	\rho^\mathrm{f}\frac{\partial \bar{\boldsymbol{u}}^\mathrm{f}}{\partial t}\bigg|_{\boldsymbol{\chi}} + \rho^\mathrm{f}(\bar{\boldsymbol{u}}^\mathrm{f} - {{\boldsymbol{u}^\mathrm{m}}})\cdot\nabla\bar{\boldsymbol{u}}^\mathrm{f} &= \nabla\cdot \bar{\boldsymbol{\sigma}}^\mathrm{f} + \nabla\cdot{\boldsymbol{\sigma}}^\mathrm{des} + \boldsymbol{b}^\mathrm{f}\  &&\mathrm{on}\  \Omega^\mathrm{f}(t),\\
	\nabla\cdot\bar{\boldsymbol{u}}^\mathrm{f} &= 0,\ &&\mathrm{on}\ \Omega^\mathrm{f}(t),\\
	\bar{\boldsymbol{u}}^\mathrm{f} &= \boldsymbol{u}^\mathrm{f}_D,\ &&\forall \boldsymbol{x}^\mathrm{f} \in \Gamma^\mathrm{f}_D(t),\\
	\bar{\boldsymbol{\sigma}}^\mathrm{f}\cdot\mathbf{n}^\mathrm{f} &= \boldsymbol{h}^\mathrm{f},\ &&\forall \boldsymbol{x}^\mathrm{f} \in \Gamma^\mathrm{f}_N(t),\\
	\bar{\boldsymbol{u}}^\mathrm{f} &= \boldsymbol{u}^\mathrm{f}_0,\ &&\mathrm{on}\ \Omega^\mathrm{f}(0), 
\end{align}
where $\bar{\boldsymbol{u}}^\mathrm{f} = \bar{\boldsymbol{u}}^\mathrm{f}(\boldsymbol{x}^\mathrm{f},t)$ and $\boldsymbol{u}^\mathrm{m}=\boldsymbol{u}^\mathrm{m}(\boldsymbol{x}^\mathrm{f},t)$ denote the fluid and mesh velocities defined for each spatial point $\boldsymbol{x}^\mathrm{f} \in \Omega^\mathrm{f}(t)$ respectively, $\rho^\mathrm{f}$ is the fluid density, $\boldsymbol{b}^\mathrm{f}$ is the body force applied on the fluid and $\bar{\boldsymbol{\sigma}}^\mathrm{f}$ is the Cauchy stress tensor for a Newtonian fluid, given as $\bar{\boldsymbol{\sigma}}^\mathrm{f} = -\bar{p}\boldsymbol{I} + \mu^\mathrm{f}( \nabla\bar{\boldsymbol{u}}^\mathrm{f} + (\nabla\bar{\boldsymbol{u}}^\mathrm{f})^T)$, 
$\bar{p}$ being the time averaged fluid pressure, $\mu^\mathrm{f}$ is the dynamic viscosity of the fluid, and $\boldsymbol{\sigma}^\mathrm{des}$ is the turbulent stress term. The first term in Eq.~(\ref{RANS}) represents the partial derivative of $\bar{\boldsymbol{u}}^\mathrm{f}$ with respect to time while the ALE reference coordinate $\boldsymbol{\chi}$ is kept fixed. The Dirichlet condition on the fluid velocity is denoted by $\boldsymbol{u}^\mathrm{f}_D$ on $\Gamma^\mathrm{f}_D(t)$. The Neumann condition on the boundary $\Gamma^\mathrm{f}_N$ is represented by $\boldsymbol{h}^\mathrm{f}$ and the initial condition on the velocity field is $\boldsymbol{u}^\mathrm{f}_0$.

\subsection{The fluid-structure interface}
The coupling between the fluid and the multibody system involves the satisfaction of the continuity of velocity as well as equilibrium of forces along the fluid-structure interface for each multibody component. Let $\Gamma^\mathrm{fs}_i(0) = \partial \Omega^\mathrm{f}(0) \cap \partial \Omega^\mathrm{s}_i(0)$ be the fluid-structure interface at $t=0$ for the multibody $i$ and $\Gamma^\mathrm{fs}_i(t) = \vec{\varphi}^\mathrm{s}(\Gamma^\mathrm{fs}_i,t)$  be the evolved interface at time $t$. The kinematic and dynamic equilibrium conditions can be written as,
\begin{align}
\bar{\boldsymbol{u}}^\mathrm{f}(\vec{\varphi}^\mathrm{s}(\boldsymbol{X}^\mathrm{s},t),t) &= \boldsymbol{u}^\mathrm{s}(\boldsymbol{X}^\mathrm{s},t),\ &&\forall \boldsymbol{X}^\mathrm{s} \in \Gamma^\mathrm{fs}_i(0),\\
\int_{\vec{\varphi}^\mathrm{s}(\gamma,t)} \bar{\boldsymbol{\sigma}}^\mathrm{f}(\boldsymbol{x}^\mathrm{f},t)\cdot \mathbf{n} \mathrm{d\Gamma} + \int_\gamma  \boldsymbol{t}^\mathrm{s} \mathrm{d}\Gamma &= 0,\ &&\forall \gamma \in \Gamma^\mathrm{fs}_i(0),
\end{align}
where $\vec{\varphi}^\mathrm{s}$ is the position vector mapping the initial position $\boldsymbol{X}^\mathrm{s}$ of the flexible multibody component $i$ to its position at time $t$ and $\boldsymbol{t}^\mathrm{s}$ is the traction on the structure along the interface $\gamma$. Here, $\mathbf{n}$ is the outer normal to the fluid-structure interface, $\gamma$ is any part of the interface $\Gamma^\mathrm{fs}_i(0)$ in the reference configuration and $\boldsymbol{\varphi}^\mathrm{s}(\gamma,t)$ is the corresponding fluid part at time $t$.
The above interface conditions are enforced such that the fluid velocity matches exactly equal to the velocity of the deformable solid body. The structural motion is determined by the fluid forces which includes the integration of pressure and shear stress effects on the body surface. 

\section{Semi-discrete variational discretization}
\label{SDVF}
For the sake of completeness, we briefly present the variational formulation for the structure and fluid fields in the aeroelastic framework in this section. This formulation then leads to the linearized system of equations for the unknown degrees of freedom for each equation.

\subsection{Structural system}
\label{weak_structure}
The weak form of the structural equation to find the displacement in a solution space $\mathcal{S}^\mathrm{h}_{\boldsymbol{\eta}^\mathrm{s}}$ can be written by projecting Eq. (\ref{Eqn_S}) on a set of finite weighting function space $\mathcal{V}^\mathrm{h}_{\boldsymbol{\psi}^\mathrm{s}}$ and integrating it over the whole domain $\Omega^\mathrm{s}_i$ and time $t$. The variational statement thus reads: find $\boldsymbol{\eta}^\mathrm{s}_\mathrm{h} \in \mathcal{S}^\mathrm{h}_{\boldsymbol{\eta}^\mathrm{s}}$ such that $\forall \boldsymbol{\psi}^\mathrm{s}_\mathrm{h} \in \mathcal{V}^\mathrm{h}_{\boldsymbol{\psi}^\mathrm{s}}$,
\begin{align}
	\int^{t^\mathrm{n+1}}_{t^\mathrm{n}}\bigg( \int_{\Omega^\mathrm{s}_i} \rho^\mathrm{s}\frac{\partial^2\boldsymbol{\eta}^\mathrm{s}_\mathrm{h}}{\partial t^2}\cdot \boldsymbol{\psi}^\mathrm{s}_\mathrm{h} \mathrm{d}\Omega + \int_{\Omega^\mathrm{s}_i} \boldsymbol{\sigma}^\mathrm{s}: \nabla\boldsymbol{\psi}^\mathrm{s}_\mathrm{h} \mathrm{d}\Omega \bigg) \mathrm{d}t = \int^{t^\mathrm{n+1}}_{t^\mathrm{n}}\bigg( \int_{\Omega^\mathrm{s}_i} \boldsymbol{b}^\mathrm{s}\cdot\boldsymbol{\psi}^\mathrm{s}_\mathrm{h} \mathrm{d}\Omega + \int_{\Gamma^\mathrm{s}_i} \boldsymbol{h}^\mathrm{s}\cdot\boldsymbol{\psi}^\mathrm{s}_\mathrm{h} \mathrm{d}\Gamma \bigg) \mathrm{d}t,
\end{align}
where $\boldsymbol{h}^\mathrm{s}$ is the Neumann boundary condition on the boundary $\Gamma^\mathrm{s}_N$.

The kinematic joints between different multibody components are formulated as a constraint equation, $\boldsymbol{c}_{J}(\boldsymbol{\eta}^\mathrm{s}) = 0$ which is discussed in detail in the next section. A Lagrange multipler technique is employed to impose the constraint on the governing equation of the multibody system. The variational statement thus is transformed as follows: find $\boldsymbol{\eta}^\mathrm{s}_\mathrm{h} \in \mathcal{S}^\mathrm{h}_{\boldsymbol{\eta}^\mathrm{s}}$ such that $\forall \boldsymbol{\psi}^\mathrm{s}_\mathrm{h} \in \mathcal{V}^\mathrm{h}_{\boldsymbol{\psi}^\mathrm{s}}$,
\begin{align} \label{VMS}
	&\int^{t^\mathrm{n+1}}_{t^\mathrm{n}}\bigg( \int_{\Omega^\mathrm{s}_i} \rho^\mathrm{s}\frac{\partial^2\boldsymbol{\eta}^\mathrm{s}_\mathrm{h}}{\partial t^2}\cdot \boldsymbol{\psi}^\mathrm{s}_\mathrm{h} \mathrm{d}\Omega + \int_{\Omega^\mathrm{s}_i} \boldsymbol{\sigma}^\mathrm{s}: \nabla\boldsymbol{\psi}^\mathrm{s}_\mathrm{h} \mathrm{d}\Omega + \int_{\Gamma^\mathrm{s}_i} \boldsymbol{c}'_J(\boldsymbol{\eta}^\mathrm{s}_\mathrm{h})^T \lambda_\mathrm{h}\cdot\boldsymbol{\psi}^\mathrm{s}_\mathrm{h}\mathrm{d}\Gamma \bigg) \mathrm{d}t \nonumber \\
 &= \int^{t^\mathrm{n+1}}_{t^\mathrm{n}}\bigg( \int_{\Omega^\mathrm{s}_i} \boldsymbol{b}^\mathrm{s}\cdot\boldsymbol{\psi}^\mathrm{s}_\mathrm{h} \mathrm{d}\Omega + \int_{\Gamma^\mathrm{s}_i} \boldsymbol{h}^\mathrm{s}\cdot\boldsymbol{\psi}^\mathrm{s}_\mathrm{h} \mathrm{d}\Gamma \bigg) \mathrm{d}t,\\
 \boldsymbol{c}_{J}(\boldsymbol{\eta}^\mathrm{s}_\mathrm{h}) &= 0,
\end{align}
 where $\lambda_\mathrm{h}$ is the Lagrange multiplier corresponding to the constraints and $\boldsymbol{c}'_{J}$ represents the Jacobian of $\boldsymbol{c}_{J}$.
 
The structural equations presented above are discretized in time with the help of an unconditionally stable energy decaying scheme based on linear time discontinuous Galerkin approximation. Further details can be found in \cite{Bauchau, BAUCHAU20033253, BAUCHAU199961, BAUCHAU199637}. The above variational form can be written in the simplified matrix form as
\begin{align} \label{MFS}
	\int^{t^\mathrm{n+1}}_{t^\mathrm{n}} \bigg(\boldsymbol{M}(\Delta\ddot{\boldsymbol{\eta}}^\mathrm{s}(t)) + \boldsymbol{K}(\Delta\boldsymbol{\eta}^\mathrm{s}(t)) + \boldsymbol{C}(\Delta\boldsymbol{\eta}^\mathrm{s}(t)) \bigg) \mathrm{d}t = \int^{t^\mathrm{n+1}}_{t^\mathrm{n}} \boldsymbol{F}^\mathrm{s}(t) \mathrm{d} t,
\end{align}
where $\boldsymbol{M}$, $\boldsymbol{K}$ and $\boldsymbol{C}$ are the mass, stiffness and constraint matrices of the multibody system respectively after the Newton-Raphson linearization, $\Delta $ represents the increment of the unknowns and $\boldsymbol{F}^\mathrm{s}$ consists of the body forces and the external fluid forces acting on the system. The construction of these matrices will differ depending on the type of the multibody component like beam, cable, shell, etc. Next, we discuss the details about the variational form of the constraint equations at the joints.

\subsection{Constraints for joints}
Continuing the discussion of a joint between two bodies at points $A$ and $B$ in Section \ref{constraint_formulation}, we write the variational and matrix form of the constraint equation for the same joint. The virtual rotation vector for the relative motion at the joints can be written as
\begin{align}
	\delta\boldsymbol{\psi}^A &= \mathrm{axial}(\delta\boldsymbol{R}^A(\boldsymbol{R}^A)^T),\\
	\delta\boldsymbol{\psi}^B &= \mathrm{axial}(\delta\boldsymbol{R}^B(\boldsymbol{R}^B)^T),
\end{align}
where $\delta\boldsymbol{R}\boldsymbol{R}^T$ is the skew-symmetric tensor related to the variation and $\mathrm{axial(\cdot)}$ is the vector associated with it (similar to Eq. (\ref{SSTensor})). The variation in the unit vectors in the deformed coordinate system is given as $\delta\boldsymbol{e}^A_{k} = (\tilde{\boldsymbol{e}}^A_{k})^T\delta \boldsymbol{\psi}^A$ and $\delta\boldsymbol{e}^B_{l} = (\tilde{\boldsymbol{e}}^B_{l})^T\delta \boldsymbol{\psi}^B$. Furthermore, the variation in the scalar and vector products defined in Section \ref{constraint_formulation} are
\begin{align}
	\delta s_{kl} &= (\delta\boldsymbol{\psi}^A - \delta\boldsymbol{\psi}^B)^T\boldsymbol{v}_{kl},\label{var_scal}\\
	\delta \boldsymbol{v}_{kl} &= (\delta\boldsymbol{\psi}^A)^T\boldsymbol{D}^{AB}_{kl} - (\delta\boldsymbol{\psi}^B)^T\boldsymbol{D}^{BA}_{lk}, \label{var_vec}
\end{align}
where $\boldsymbol{D}^{AB}_{kl} = \tilde{\boldsymbol{e}}^A_k\tilde{\boldsymbol{e}}^B_l$ and $\boldsymbol{D}^{BA}_{lk} = \tilde{\boldsymbol{e}}^B_l\tilde{\boldsymbol{e}}^A_k$. Based on the constraint at the joint $C_{AB} = 0$ (which could be any one of the constraints defined in Eqs. (\ref{rel_disp}) and (\ref{rel_rot})), the potential function for the constraint can be written as $V = \lambda C_{AB}$ and its variation is $\delta V = \delta\lambda C_{AB} + \lambda \delta C_{AB}$, where $\lambda$ is the Lagrange multiplier for the constraint. The virtual work done is $\delta W = \lambda \delta C_{AB} = (\delta \boldsymbol{q})^T\boldsymbol{F}_{AB}$, where $\delta\boldsymbol{q}$ and $\boldsymbol{F}_{AB}$ are the variation of generalized coordinates and the constraint forces respectively. Based on the type of constraint and the variations in Eqs. (\ref{var_scal}) and (\ref{var_vec}) and $\delta C_{AB} = \boldsymbol{B}\delta \boldsymbol{q}$, one can write the constraint forces as $\boldsymbol{F}_{AB} = \lambda \boldsymbol{B}^T$. The matrix $\boldsymbol{B}^T$ is the Jacobian of the constraint denoted by $\boldsymbol{c}'(\boldsymbol{\eta}^\mathrm{s}_\mathrm{h})^T$ in Eq. (\ref{VMS}). 

With the help of Newton-Raphson technique for nonlinear constraint relations, the increment of constraint forces is expressed as
\begin{align}
	\Delta \boldsymbol{F}_{AB} = \frac{\partial \boldsymbol{F}_{AB}}{\partial \boldsymbol{q}} \Delta\boldsymbol{q} = \boldsymbol{K}_{AB}\Delta\boldsymbol{q},
\end{align}
where $\boldsymbol{K}_{AB}$ is the stiffness matrix for the constraint.

\textbf{Case 1: Constraint for relative displacement}\\
In the case of relative displacement at the joint, the constraint is given by $C_{AB} = (\boldsymbol{e}^{A}_m)^T\boldsymbol{\eta}_{B/A} - \boldsymbol{\eta}_m = 0$. The incremental vector of unknowns and the stiffness matrix is given by
\begin{align}
	\Delta\boldsymbol{q} = \begin{bmatrix} \Delta\boldsymbol{\eta}_A \\ 
										\Delta \boldsymbol{\psi}_A\\
										\Delta \boldsymbol{\eta}_B\\
									 	\Delta \boldsymbol{\eta}_m\end{bmatrix},\qquad				 	
	\boldsymbol{K}_{AB} = \lambda\begin{bmatrix}
								\boldsymbol{0} & \tilde{\boldsymbol{e}}^A_m & \boldsymbol{0} & \boldsymbol{0}\\
								-\tilde{\boldsymbol{e}}^A_m & \tilde{\boldsymbol{\eta}}_{A/B}\tilde{\boldsymbol{e}}^A_m & \tilde{\boldsymbol{e}}^A_m & \boldsymbol{0}\\
								\boldsymbol{0} & -\tilde{\boldsymbol{e}}^A_m & \boldsymbol{0} & \boldsymbol{0}\\
								\boldsymbol{0} & \boldsymbol{0} & \boldsymbol{0} & \boldsymbol{0}
								\end{bmatrix}								 	
\end{align}

\textbf{Case 2: Constraint for relative rotation}\\
For relative rotation, the constraint equation is given as $C_{AB} = s_{kk} \mathrm{sin}(\phi_m) + s_{kl}\mathrm{cos}(\phi_m) = 0$. The increment vector of the unknowns and the stiffness matrix are given as the following
\begin{align}
	\Delta\boldsymbol{q} = \begin{bmatrix} \Delta \boldsymbol{\psi}_A\\
										\Delta \boldsymbol{\eta}_B\\
									 	\Delta \boldsymbol{\phi}_m\end{bmatrix},\qquad				 	
	\boldsymbol{K}_{AB} = \lambda\begin{bmatrix}
								\boldsymbol{E}^T & \boldsymbol{E} & \boldsymbol{z}\\
								-\boldsymbol{E}^T & \boldsymbol{E} & -\boldsymbol{z}\\
								\boldsymbol{z}^T & -\boldsymbol{z}^T & -C_{AB}
								\end{bmatrix}								 	
\end{align}
where $\boldsymbol{z} = \boldsymbol{v}_{kk}\mathrm{cos}(\phi_m) - \boldsymbol{v}_{kl}\mathrm{sin}(\phi_m)$ and $\boldsymbol{E} = \boldsymbol{D}^{AB}_{kk}\mathrm{sin}(\phi_m) + \boldsymbol{D}^{AB}_{kl}\mathrm{cos}(\phi_m)$.

Therefore, based on the type of joint, the constraint can be a combination of the two cases described above and the stiffness matrices are formed for each constraint with the help of the Lagrange multiplier technique. All the constraints at every joint are then assembled to form the global constraint matrix $\boldsymbol{C}$ in Eq. (\ref{MFS}).

\subsection{Flow system}
\label{weak_flow} 
We employ the generalized-$\alpha$ method for the time integration between $t \in [t^\mathrm{n}, t^\mathrm{n+1}]$, which can be unconditionally stable and second-order accurate for linear problems \cite{Gen_alpha}. The spatial discretization is carried out by a Petrov-Galerkin technique that circumvents the Babu$\mathrm{\check{s}}$ka-Brezzi condition that is required to be satisfied by any standard mixed Galerkin method \cite{Johnson}. Details of the semi-discrete and the variational form can be found in \cite{JOSHI201784}. 

Similarly, the semi-discrete flow equations are projected onto the weighting function space and integrated over the whole spatial domain. The variational statement of the semi-discrete form of the flow equations can be written as: find $[\boldsymbol{u}^\mathrm{f,n+\alpha^f}_\mathrm{h}, p^\mathrm{n+1}_\mathrm{h}] \in \mathcal{S}^\mathrm{h}_{\boldsymbol{u}^\mathrm{f}} \times \mathcal{S}^\mathrm{h}_p$ such that $\forall [\boldsymbol{\psi}^\mathrm{f}_\mathrm{h}, q_\mathrm{h}] \in \mathcal{V}^\mathrm{h}_{\boldsymbol{\psi}^\mathrm{f}} \times \mathcal{V}^\mathrm{h}_q$,
\begin{align}
\label{PG_FEM_Fluid}
&\int_{\Omega^\mathrm{f}(t^\mathrm{n+1})} \bigg( \rho^\mathrm{f} \partial_t \boldsymbol{u}^\mathrm{f,n+\alpha^f_m}_\mathrm{h}\big|_{\boldsymbol{\chi}} + \rho^\mathrm{f}(\boldsymbol{u}_\mathrm{h}^\mathrm{f,n+\alpha^f} - \boldsymbol{u}^\mathrm{m}_\mathrm{h})\cdot\nabla\boldsymbol{u}^\mathrm{f,n+\alpha^f}_\mathrm{h} \bigg) \cdot \boldsymbol{\psi}^\mathrm{f}_\mathrm{h} \mathrm{d}\Omega  + \int_{\Omega^\mathrm{f}(t^\mathrm{n+1})} \boldsymbol{\sigma}^\mathrm{f,n+\alpha^f}_\mathrm{h}: \nabla\boldsymbol{\psi}^\mathrm{f}_\mathrm{h}  \mathrm{d}\Omega \nonumber \\
&+ \int_{\Omega^\mathrm{f}(t^\mathrm{n+1})} \boldsymbol{\sigma}^\mathrm{des,n+\alpha^f}_\mathrm{h}: \nabla\boldsymbol{\psi}^\mathrm{f}_\mathrm{h}  \mathrm{d}\Omega + \displaystyle\sum_{e=1}^\mathrm{n_{el}} \int_{\Omega^e} \frac{\tau_\mathrm{m}}{\rho^\mathrm{f}}\big( \rho^\mathrm{f}(\boldsymbol{u}^\mathrm{f,n+\alpha^f}_\mathrm{h} - \boldsymbol{u}^\mathrm{m}_\mathrm{h})\cdot \nabla\boldsymbol{\psi}^\mathrm{f}_\mathrm{h} + \nabla q_\mathrm{h}  \big)\cdot \boldsymbol{R}_\mathrm{m} \mathrm{d}\Omega^e \nonumber \\
&+ \int_{\Omega^\mathrm{f}(t^\mathrm{n+1})} q_\mathrm{h}(\nabla\cdot\boldsymbol{u}^\mathrm{f,n+\alpha^f}_\mathrm{h}) \mathrm{d}\Omega + \displaystyle\sum_{e=1}^\mathrm{n_{el}} \int_{\Omega^e} \nabla\cdot\boldsymbol{\psi}^\mathrm{f}_\mathrm{h} \tau_\mathrm{c} \rho^\mathrm{f} \boldsymbol{R}_\mathrm{c} d\Omega^e  \nonumber \\
&=\int_{\Omega^\mathrm{f}(t^\mathrm{n+1})} \boldsymbol{\psi}^\mathrm{f}_\mathrm{h}\cdot \boldsymbol{f}^\mathrm{f,n+\alpha^f}_\mathrm{h} \mathrm{d} \Omega + \int_{\Gamma^\mathrm{f}_N} \boldsymbol{\psi}^\mathrm{f}_\mathrm{h}\cdot \boldsymbol{h}^\mathrm{f,n+\alpha^f} \mathrm{d}\Gamma,
\end{align} 
where the second line represents the stabilization term for the momentum equation and the second term in the third line depicts the same for the continuity equation. $\boldsymbol{R}_\mathrm{m}$ and $\boldsymbol{R}_\mathrm{c}$ are the residual of the momentum and continuity equations respectively. The stabilization parameters $\tau_\mathrm{m}$ and $\tau_\mathrm{c}$ are the least-squares metrics added to the element-level integrals \cite{Hughes_X, Brooks, Tezduyar_1, France_II} defined as
\begin{align}
\tau_\mathrm{m} = \bigg[ \bigg( \frac{2}{\Delta t}\bigg)^2 + \big( \boldsymbol{u}^\mathrm{f,n+\alpha^f}_\mathrm{h} - \boldsymbol{u}^\mathrm{m}_\mathrm{h} \big)\cdot \boldsymbol{G} \big( \boldsymbol{u}^\mathrm{f,n+\alpha^f}_\mathrm{h} - \boldsymbol{u}^\mathrm{m}_\mathrm{h} \big) + C_I \bigg( \frac{\mu^\mathrm{f}}{\rho^\mathrm{f}}  \bigg)^2 \boldsymbol{G}:\boldsymbol{G}  \bigg]^{-1/2},\qquad \tau_\mathrm{c} = \frac{1}{\mathrm{tr}(\boldsymbol{G})\tau_\mathrm{m}},
\end{align}
where $C_I$ is a constant derived from inverse estimates \cite{Hughes_inv_est}, $\mathrm{tr}()$ denotes the trace and $\boldsymbol{G}$ is the contravariant metric tensor given by $\boldsymbol{G} = (\partial \boldsymbol{\xi}^T/\partial \boldsymbol{x})(\partial \boldsymbol{\xi}/\partial \boldsymbol{x})$,
where $\boldsymbol{x}$ and $\boldsymbol{\xi}$ are the physical and parametric coordinates respectively.

The nonlinear variational form of the flow equations is then linearized by Newton-Raphson iterative procedure where the increments in the velocity and pressure variables are computed. It leads to the following matrix form:
\begin{align} \label{matrix_NS}
	\begin{bmatrix}
		\boldsymbol{K}_{\Omega^\mathrm{f}} & -\boldsymbol{G}_{\Omega^\mathrm{f}} \\
		\boldsymbol{G}^T_{\Omega^\mathrm{f}} & \boldsymbol{C}_{\Omega^\mathrm{f}} 
	\end{bmatrix}
	\begin{pmatrix}
		\Delta \bar{\boldsymbol{u}}^\mathrm{f} \\
		\Delta p
	\end{pmatrix}
	= -\begin{pmatrix}
		\boldsymbol{\mathcal{R}}_\mathrm{m} \\
		\boldsymbol{\mathcal{R}}_\mathrm{c}
	  \end{pmatrix},
\end{align}
where $\boldsymbol{K}_{\Omega^\mathrm{f}}$ is the stiffness matrix for the momentum equations consisting of the transient, convection, diffusion and Petrov-Galerkin stabilization terms for the momentum equation, $\boldsymbol{G}_{\Omega^\mathrm{f}}$ is the gradient operator, $\boldsymbol{G}^T_{\Omega^\mathrm{f}}$ is the divergence operator for the continuity equation, $\boldsymbol{C}_{\Omega^\mathrm{f}}$ is the stabilization term dealing the pressure-pressure coupling and $\boldsymbol{\mathcal{R}}_\mathrm{m}$ and $\boldsymbol{\mathcal{R}}_\mathrm{c}$ are the weighted residuals of the stabilized momentum and continuity equations respectively.

This completes the variational formulation of the different field equations for the current flexible multibody aeroelastic framework. We next focus our attention to the treatment of the fluid-structure interface and the radial basis function mapping for the large deformation problems. We also discuss how the coupling between the fluid and structural fields is being carried out in a partitioned staggered iterative manner.



\section{Flexible multibody fluid-structure coupling}  
\label{FSI_coupling}
In this section, we present the coupling procedure between the fluid and structural domains at the fluid-structure interface.
We achieve the FSI coupling via a scattered point data interpolation technique using radial basis functions, which maps the structural displacements from the two-dimensional multibody structural component to the three-dimensional fluid mesh as well as the transfer of fluid forces from the fluid to the structural domain. 
  
\subsection{Treatment of the fluid-structure interface via radial basis function mapping}
\label{rbf}
For the interaction between the fluid and the multiple components of the flexible multibody system, it is imperative that the transfer of data between the fluid and the structure at the fluid-structure interface is carried out in a locally accurate and conservative manner. This becomes even more challenging when the discretized mesh at the interface is non-matching. As mentioned in the introduction, methods such as quadrature projection and common-refinement techniques give the advantage of local as well as global conservation property. However, their algorithms can become complex in terms of accurate projection of the quadrature points for large deformation problems considered in the current study. For such cases, global conservation methods such as point-to-point mapping (e.g., radial basis function with compact support) can be a good alternative, which does not require the connectivity data between the scattered points.

\begin{remark} For satisfying the global conservation property, the virtual work done by the forces in the structural domain $\delta W^\mathrm{s}$ should equal the work done by the fluid loads $\delta W^\mathrm{f}$ at the fluid-structure interface. The interpolation by the radial basis function mapping is a node-to-node mapping constructed in such a way that the global conservation property is maintained. Suppose the interpolation of fluid displacement at the fluid-structure interface is given by $\boldsymbol{\eta}^\mathrm{f} = \boldsymbol{H}\boldsymbol{\eta}^\mathrm{s}$, where $\boldsymbol{H}$ is some interpolation matrix and the force transfer to the structural nodes at the interface is written as $\boldsymbol{f}^\mathrm{s} = \boldsymbol{H}^T\boldsymbol{f}^\mathrm{f}$. Thus, the virtual work done at the fluid-structure interface is given by
\begin{align}
	\delta W^\mathrm{f} &= \delta\boldsymbol{\eta}^\mathrm{f}\cdot\boldsymbol{f}^\mathrm{f} = (\delta\boldsymbol{\eta}^\mathrm{f})^T\boldsymbol{f}^\mathrm{f},\\
		&= (\delta\boldsymbol{\eta}^\mathrm{s})^T \boldsymbol{H}^T\boldsymbol{f}^\mathrm{f},\\
		& = (\delta\boldsymbol{\eta}^\mathrm{s})^T \boldsymbol{f}^\mathrm{s},\\
		& = \delta\boldsymbol{\eta}^\mathrm{s}\cdot \boldsymbol{f}^\mathrm{s} = \delta W^\mathrm{s},
\end{align}
where $\boldsymbol{f}^\mathrm{s}$ and $\boldsymbol{f}^\mathrm{f}$ are the contribution of the forces from the structure and fluid domains respectively. Therefore, if an approximation $\boldsymbol{H}$ exists, it can be utilized for both the transformation of displacements and forces between the fluid and the structural domains \cite{BECKERT200013, Hounjet1994}. In the present formulation, we employ radial basis function to construct this interpolation matrix $\boldsymbol{H}$. We next review the concept of radial basis function interpolation.
\end{remark}

\subsubsection{Review of radial basis functions}

A radially basis function (RBF) is radially symmetric and forms the basis of the interpolation matrix. A $d$-variate interpolation function $g(\boldsymbol{x})$ from a given scattered data $\{g_1, g_2, ..., g_N \}$ at the arbitrary distinct source locations $X^\mathrm{c} = \{\boldsymbol{x}^\mathrm{c}_1, \boldsymbol{x}^\mathrm{c}_2, ..., \boldsymbol{x}^\mathrm{c}_N\} \subseteq \mathbb{R}^d$ is given as
\begin{align}
	g(\boldsymbol{x}) = \displaystyle\sum^{N}_{j=1} \alpha_j\phi(||\boldsymbol{x}-\boldsymbol{x}^\mathrm{c}_j||), \label{RBF}
\end{align} 
where $||\cdot||$ is the Euclidean norm and $\alpha_j$ are the weights for the basis functions. If the weights are known, the interpolated values at the requested target data points $X^\mathrm{t} = \{\boldsymbol{x}^\mathrm{t}_1, \boldsymbol{x}^\mathrm{t}_2, ..., \boldsymbol{x}^\mathrm{t}_M\} $ are thus given by
\begin{align}
	g(\boldsymbol{x}^\mathrm{t}_i) &= \displaystyle\sum^{N}_{j=1} \alpha_j\phi(||\boldsymbol{x}^\mathrm{t}_i-\boldsymbol{x}^\mathrm{c}_j||),\ 1 \leq i \leq M,
\end{align}
where the matrix $\phi(||\boldsymbol{x}^\mathrm{t}_i-\boldsymbol{x}^\mathrm{c}_j||)$ is also known as kernel matrix for the interpolation.

The coefficients $\alpha_j$ are determined by the condition, $g(\boldsymbol{x}^\mathrm{c}_j) = g_j,\ 1\leq j \leq N$. Therefore, the interpolation recovers to the exact values at the scattered control points, i.e., for $\boldsymbol{x}^\mathrm{t} = \boldsymbol{x}^\mathrm{c}$,
\begin{align}
	g_j = g(\boldsymbol{x}^\mathrm{c}_j) &= \displaystyle\sum^{N}_{j=1} \alpha_j\phi(||\boldsymbol{x}^\mathrm{c}_i-\boldsymbol{x}^\mathrm{c}_j||),\ 1 \leq i \leq N,
\end{align}
which is a system of linear equations with $\alpha_j$ as unknowns and let $(\boldsymbol{A}_\mathrm{cc})_{ij} = \phi(||\boldsymbol{x}^\mathrm{c}_i-\boldsymbol{x}^\mathrm{c}_j||)$. To have a guaranteed solvability for these coefficients, the symmetric matrix $\boldsymbol{A}_\mathrm{cc}$ should be positive definite which depends on the positive-definiteness of the radial basis function $\phi(||\cdot||)$.

\begin{remark} The positive definite property can be imparted to a radial basis function by adding polynomials to the interpolant as
\begin{align}
	g(\boldsymbol{x}^\mathrm{t}) &= \displaystyle\sum^{N}_{j=1} \alpha_j\phi(||\boldsymbol{x}^\mathrm{t}-\boldsymbol{x}^\mathrm{c}_j||) + q(\boldsymbol{x}^\mathrm{t}),
\end{align}
where $q(\boldsymbol{x}^\mathrm{t})$ is a polynomial. Now, with additional degrees of freedom of the coefficients of the polynomial, the condition $\displaystyle\sum_{j=1}^N\alpha_jp(\boldsymbol{x}^\mathrm{t}_j) = 0$, has to be satisfied for unique solvability of the coefficients, for any polynomial $p(\boldsymbol{x})$ with degree less than or equal to the degree of $q(\boldsymbol{x})$. A minimal degree of the polynomial $q(\boldsymbol{x}^\mathrm{t})$ depends on the radial basis function.
\end{remark}

The radial basis function can be selected among many alternatives, like Gaussian, multiquadric and polyharmonic, to name a few, with certain conditions for the positive definite property. Apart from the positive definiteness property, it is beneficial for the radial basis function to have localization or compact support. Such localization ensures that the system matrix is sparse, which is helpful in its inversion. A prime example of such functions having compact support is the class of Wendland's function, which are always positive definite up to a maximal space dimension and have smoothness $C^{2k}$. They are of the form
\begin{align}
	\phi(r) = \begin{cases}
											p(r), &0 \leq r \leq 1,\\
											0, & r > 1,
	\end{cases}
\end{align}
where $p(r)$ is a univariate polynomial. Considering a truncated power function $\phi_l(r) = (1-r)^l$ which satisfies the positive definiteness property for $l \geq \lfloor d/2 \rfloor +1$, $d$ being the spacial dimensions and $\lfloor x \rfloor$ represents an integer $n$ such that $n \leq x < n+1$. Using this definition, a new class of functions is constructed $\phi_{l,k}(r)$ which are positive definite in the dimension $d$ and are $C^{2k}$ with degree $\lfloor d/2 \rfloor + 3k+1$ \cite{Wendland}. If one defines an operator $I(f)(r) = \int_{r}^{\infty} f(t) t dt$, these new class of functions can be constructed via integration as $\phi_{l,k}(r) = I^k\phi_{l}(r)$, which can be represented by
\begin{align}
	\phi_{l,k}(r) = \displaystyle\sum_{n=0}^k \beta_{n,k} r^n \phi_{l+2k-n}(r),
\end{align} 
where $\beta_{0,0} = 1$ and 
\begin{align}
	\beta_{j,k+1} = \displaystyle\sum_{n=j-1}^{k} \beta_{n,k} \frac{[n+1]_{n-j+1}}{(l+2k-n+1)_{n-j+2}},
\end{align}
where $[q]_{-1} = 1/(q+1)$, $[q]_0 = 1$, $[q]_l = q(q-1)...(q-l+1)$ and $(q)_0 = 1$, $(q)_l = q(q+1)...(q+l-1)$. It has also been shown in \cite{Wendland} that a function $\Phi_{l,k}(x) = \phi_{l,k}(||x||)$ has strictly positive Fourier transform and produces a positive definite radial basis function, except when $d=1$ and $k=0$. Furthermore, it has been proven that the function is $2k$ times continuously differentiable and their polynomial degree is minimal for a given smoothness, and they are related to certain Sobolev spaces \cite{Wu1995MultivariateCS, Wendland, Wendland2}. We utilize Wendland's $C^2$ function for the current three-dimensional study ($d=3$) with $k=1$ and choosing $l = \lfloor d/2 \rfloor +k +1$ as $\phi_{3,1}(r)$ written as
\begin{align}
	\phi(||\boldsymbol{x}||) = \begin{cases}    (1-||\boldsymbol{x}||)^4(1+4||\boldsymbol{x}||),	&0\leq ||\boldsymbol{x}|| \leq 1,\\
	0, &||\boldsymbol{x}|| > 1.
	\end{cases}
\end{align}

\subsubsection{Application to aeroelastic framework}
For the three-dimensional flexible multibody aeroelastic problems and the use of Wendland's $C^2$ function, we can employ a linear polynomial $q(\boldsymbol{x}) = \lambda_0 + \lambda_1 x + \lambda_2 y + \lambda_3 z$, which is exactly reproduced and recovers any rigid body translation and rotation in the mapping. Note that the radial basis function can be scaled by a compact support radius $r$ which gives the advantage of covering enough number of interpolation points depending on the application. This scaling, however, does not have any effect on the positive definiteness and the compact support properties of the function. The scaled Wendland's $C^2$ function is given by
\begin{align}
	\phi(||\boldsymbol{x}||/r) = \begin{cases}    (1-||\boldsymbol{x}||/r)^4(1+4||\boldsymbol{x}||/r),	&0\leq ||\boldsymbol{x}||/r \leq 1,\\
	0, &||\boldsymbol{x}||/r > 1.
	\end{cases}
\end{align}

Next, we discuss the interpolation procedure to obtain the displacements and fluid tractions across the fluid-structure interface. Suppose the structural displacement field is known at the fluid-structure interface as $\boldsymbol{\eta}^\mathrm{s}_{I}$ and we want to interpolate the fluid displacement at the interface $\boldsymbol{\eta}^\mathrm{f}_{I}$. Based on the RBF interpolation, the coefficients $\beta_j$ can be found by the fact that $\boldsymbol{\eta}^\mathrm{s}_{I} = \boldsymbol{C}_\mathrm{ss} \boldsymbol{\beta}$, where $\boldsymbol{C}_\mathrm{ss}$ is given as
\begin{align}
	\boldsymbol{C}_\mathrm{ss} &= \begin{bmatrix}  
					0 & 0 & 0 & 0 & 1 & 1 & ... & 1\\ 
					0 & 0 & 0 & 0 & x^\mathrm{c}_{1} & x^\mathrm{c}_{2} & ... & x^\mathrm{c}_{N}\\
					0 & 0 & 0 & 0 & y^\mathrm{c}_{1} & y^\mathrm{c}_{2} & ... & y^\mathrm{c}_{N}\\
					0 & 0 & 0 & 0 & z^\mathrm{c}_{1} & z^\mathrm{c}_{2} & ... & z^\mathrm{c}_{N}\\
					1 & x^\mathrm{c}_{1} & y^\mathrm{c}_{1} & z^\mathrm{c}_{1} & \phi^\mathrm{c,c}_{1,1} & \phi^\mathrm{c,c}_{1,2} & ... & \phi^\mathrm{c,c}_{1,N}\\
					1 & x^\mathrm{c}_{2} & y^\mathrm{c}_{2} & z^\mathrm{c}_{2} & \phi^\mathrm{c,c}_{2,1} & \phi^\mathrm{c,c}_{2,2} & ... & \phi^\mathrm{c,c}_{2,N}\\
					. & . & . & . & . & . &.\ \  & .\\
					. & . & . & . & . & . & \ .\  & .\\
					. & . & . & . & . & . & \ \ \ . & .\\
					1 & x^\mathrm{c}_{N} & y^\mathrm{c}_{N} & z^\mathrm{c}_{N} & \phi^\mathrm{c,c}_{N,1} & \phi^\mathrm{c,c}_{N,2} & ... & \phi^\mathrm{c,c}_{N,N}\\
				     \end{bmatrix}, \label{Css}
\end{align}
where $\phi^\mathrm{c,c}_{i,j} = \phi(||\boldsymbol{x}^\mathrm{c}_i - \boldsymbol{x}^\mathrm{c}_j||)$, where $\boldsymbol{x}^\mathrm{c}_i$ are the coordinates of the points on the structural interface at $\Gamma^\mathrm{fs}_i$. The displacement field of the fluid nodes at the interface can be interpolated by $\boldsymbol{\eta}^\mathrm{f}_{I} = \boldsymbol{A}_\mathrm{fs}\boldsymbol{C}_\mathrm{ss}^{-1}\boldsymbol{\eta}^\mathrm{s}_{I}$, where $\boldsymbol{A}_\mathrm{fs}$ is given by
\begin{align}
\boldsymbol{A}_\mathrm{fs} &= \begin{bmatrix}  
					1 & x^\mathrm{t}_{1} & y^\mathrm{t}_{1} & z^\mathrm{t}_{1} & \phi^\mathrm{t,c}_{1,1} & \phi^\mathrm{t,c}_{1,2} & ... & \phi^\mathrm{t,c}_{1,N}\\
					1 & x^\mathrm{t}_{2} & y^\mathrm{t}_{2} & z^\mathrm{t}_{2} & \phi^\mathrm{t,c}_{2,1} & \phi^\mathrm{t,c}_{2,2} & ... & \phi^\mathrm{t,c}_{2,N}\\
					. & . & . & . & . & . &.\ \  & .\\
					. & . & . & . & . & . & \ .\  & .\\
					. & . & . & . & . & . & \ \ \ . & .\\
					1 & x^\mathrm{t}_{M} & y^\mathrm{t}_{M} & z^\mathrm{t}_{M} & \phi^\mathrm{t,c}_{M,1} & \phi^\mathrm{t,c}_{M,2} & ... & \phi^\mathrm{t,c}_{M,N}\\
				     \end{bmatrix} \label{Afs},
\end{align}
where $\phi^\mathrm{t,c}_{i,j}= \phi(||\boldsymbol{x}^\mathrm{t}_i - \boldsymbol{x}^\mathrm{c}_j||)$, $\boldsymbol{x}^\mathrm{t}_{i}$ being the coordinates of the fluid points on the interface. Therefore, the mapping of the fluid displacements to the structural displacements can be described as $\boldsymbol{\eta}^\mathrm{f}_{I} = \boldsymbol{A}_\mathrm{fs}\boldsymbol{C_\mathrm{ss}}^{-1}\boldsymbol{\eta}^\mathrm{s} = \boldsymbol{H}\boldsymbol{\eta}^\mathrm{s}_{I}$. 

The same technique can be applied for the transfer of tractions from fluid points to the structural points at the interface by considering the control points to be that of the fluid side of the interface and target points as that of the structural side. Similar to the finite element method, the local support of RBF makes the system matrices to be sparse along the interface. Moreover, 
the Wendland's function provides positive definite property for the matrix computation.

Apart from the transfer of data along the fluid-structure interface, the fluid mesh nodes inside the fluid domain can also be displaced based on the radial basis function mapping where the matrix $\boldsymbol{A}_\mathrm{fs}$ is constructed based on the volumetric nodes of the fluid domain rather than just the fluid-structure interface carried out previously, i.e., the target nodes now contain the volumetric data of the fluid domain.

\subsection{Coupling algorithm}
\label{algo}

The different field equations of the fluid and the structure are coupled in a partitioned staggered manner. The structural solver provides a predictor displacement which is then passed to the turbulent flow solver which updates the fluid forces as a corrector step on the fluid-structure interface.
Marching in time from $t^\mathrm{n}$ to $t^\mathrm{n+1}$, the algorithm consists of nonlinear iterations to correct the fluid forces. In a predictor-corrector iteration $\mathrm{k}$, with the input of the forces at time step $t^\mathrm{n}$ from the flow solver, we first solve the structural equation (Eq. (\ref{MFS})) to predict the structural displacement. These computed structural displacements are then transferred to the flow solver via the radial basis function mapping described in Section \ref{rbf} in the second step by satisfying the geometric mesh compatibility and the velocity continuity (kinematic equilibrium) at the fluid-structure interface $\Gamma_i^\mathrm{fs}$ for the multibody component $i$. As the third step, the turbulent flow equations (Eq. (\ref{matrix_NS})) are solved in the ALE moving mesh framework to get the velocity and pressure fields which are then used to construct the fluid forces on the structural component. These forces are then corrected using the nonlinear iterative force correction (NIFC) filter by successive approximation and the corrected forces are then mapped on the structural domain using the transpose of the constructed interpolation matrix $\boldsymbol{H}$ via RBF. Further details about the coupling can be found in \cite{LI201996}.

After the Newton-Raphson linearization, the coupled fluid-structure system with the equilibrium conditions at the fluid-structure interface can be written in the abstract matrix form as
\begin{align} \label{matrix_coupled_FSI}
	\begin{bmatrix} \boldsymbol{A}^\mathrm{ss} & \boldsymbol{0} & \boldsymbol{0} & \boldsymbol{A}^\mathrm{Is} \\
	\boldsymbol{A}^\mathrm{sI} & \boldsymbol{I} & \boldsymbol{0} & \boldsymbol{0} \\
	\boldsymbol{0} & \boldsymbol{A}^\mathrm{If} & \boldsymbol{A}^\mathrm{ff} & \boldsymbol{0} \\
	\boldsymbol{0} & \boldsymbol{0} & \boldsymbol{A}^\mathrm{fI} & \boldsymbol{I} 
	\end{bmatrix} 
	\begin{pmatrix}
	\Delta\boldsymbol{\eta}^\mathrm{s}\\
	\Delta\boldsymbol{\eta}^\mathrm{I}\\
	\Delta \boldsymbol{q}^\mathrm{f}\\
	\Delta \boldsymbol{f}^\mathrm{I}
	\end{pmatrix}
	= \begin{pmatrix}
	\boldsymbol{\mathcal{R}}^\mathrm{s}\\
	\boldsymbol{\mathcal{R}}^\mathrm{I}_\mathrm{D}\\
	\boldsymbol{\mathcal{R}}^\mathrm{f}\\
	\boldsymbol{\mathcal{R}}^\mathrm{I}_\mathrm{N}
	\end{pmatrix},
\end{align} 
where $\boldsymbol{A}^\mathrm{ss}$ is the block matrix consisting of mass, stiffness and constraint matrices of the structural equation for the non-interface structural degrees-of-freedom (DOFs), $\boldsymbol{A}^\mathrm{Is}$ represents the transfer of the fluid forces from the fluid-structure interface $\Gamma^\mathrm{fs}_i$ to the structure, $\boldsymbol{A}^\mathrm{sI}$ is the mapping of the structural displacement on the fluid interface nodes, $\boldsymbol{A}^\mathrm{ff}$ is the block matrix for the internal DOFs of the fluid domain (given by Eq. (\ref{matrix_NS})), $\boldsymbol{A}^\mathrm{fI}$ denotes the mapping of the fluid forces from the fluid domain to the fluid-structure interface and $\boldsymbol{A}^\mathrm{If}$ is the RBF mapping of the displacement to the fluid spatial points. The increments on the structural displacement for interior DOFs are given by $\Delta\boldsymbol{\eta}^\mathrm{s}$, and $\Delta \boldsymbol{\eta}^\mathrm{I}$ and $\Delta\boldsymbol{f}^\mathrm{I}$ denote the increments in the displacement and forces at $\Gamma^\mathrm{fs}_i$ respectively. The increment of the fluid variables (pressure and velocity) are given by $\Delta\boldsymbol{q}^\mathrm{f} = (\Delta\boldsymbol{u}^\mathrm{f}, \Delta p)$. On the right-hand side, $\boldsymbol{\mathcal{R}}^\mathrm{f}= (\boldsymbol{\mathcal{R}}_\mathrm{m}, \boldsymbol{\mathcal{R}}_\mathrm{c})$ and $\boldsymbol{\mathcal{R}}^\mathrm{s}$ denote the weighted residuals of the variational discretization of the fluid and structural equations respectively. $\boldsymbol{\mathcal{R}}^\mathrm{I}_\mathrm{D}$ and $\boldsymbol{\mathcal{R}}^\mathrm{I}_\mathrm{N}$ are the residuals in the kinematic and dynamic equilibrium conditions at the fluid-structure interface.

The nonlinear iterative force correction technique is essential for low structure-to-fluid mass ratio regimes.
It constructs the cross-coupling effect along the fluid-structure interface without forming the off-diagonal term $\boldsymbol{A}^\mathrm{Is}$ in Eq. (\ref{matrix_coupled_FSI}) with the help of static condensation in the above system. The increment on the fluid forces $\Delta \boldsymbol{f}^\mathrm{I}$ is based on an input-output relationship between the displacement and the force transfer at each sub-iteration and is evaluated by the generalization of Aitken's $\Delta^2$ extrapolation scheme via dynamic weighting parameter to transform a divergent fixed point iteration to a stable and convergent update of the forces associated with the interface degrees of freedom \cite{NIFC_1, IJNME_Joshi}. While the brute-force sub-iterations lead to severe numerical instabilities for low structure-to-fluid mass ratio, the NIFC-based correction provides a stability to the overall partitioned fluid-structure coupling \cite{NIFC_2,NIFC_1} without the need of direct evaluation of off-diagonal Jacobian.

\subsection{Implementation details}
\label{implement}

In the study, Newton-Raphson linearization of the Navier-Stokes equations results in a linear system of equations with the incremental variables for velocity and pressure as the unknowns. They are solved by the  Generalized Minimal RESidual (GMRES) algorithm proposed in \cite{saad1986}. The algorithm consists of Krylov subspace iterations via modified Gram-Schmidt orthogonalization. 
The global left-hand side matrix is not needed in an explicit form, but rather we perform the required matrix-vector product of each block matrix for the GMRES algorithm. The partitioned format of the employed scheme gives the advantage of flexibility in terms of numerical implementation and avoids ill-conditioning of the left-hand side matrix, which is considered a challenge in monolithic formulations \cite{Hron_Turek, Degroote} due to widely differing temporal and spatial scales of fluid and structural fields.

The solver depends on hybrid parallelism for parallel computing. It employs a standard master-slave strategy for distributed memory clusters via message passing interface (MPI) based on a domain decomposition strategy \cite{mpi}. The master process extracts the mesh and generates the partition of the mesh into subgrids via an automatic graph partitioner \cite{metis}. Each master process performs the computation for the root subgrid and the remaining subgrids behave as the slaves. While the local element matrices and the local right-hand vectors are evaluated by the slave processes, the resulting system is solved in parallel across different compute nodes. The hybrid or mixed approach provides the benefit of thread-level parallelism of multicore architecture and allows each MPI task to access the full memory of a compute node \cite{smith_mpi}.


\section{Convergence of RBF mapping: Static load data transfer}
\label{RBF_Error_analysis}
\begin{figure}[!htbp]
	\centering
	\begin{subfigure}[b]{0.33\textwidth}
		\includegraphics[trim={2cm 1cm 2cm 2cm},clip,width=6cm]{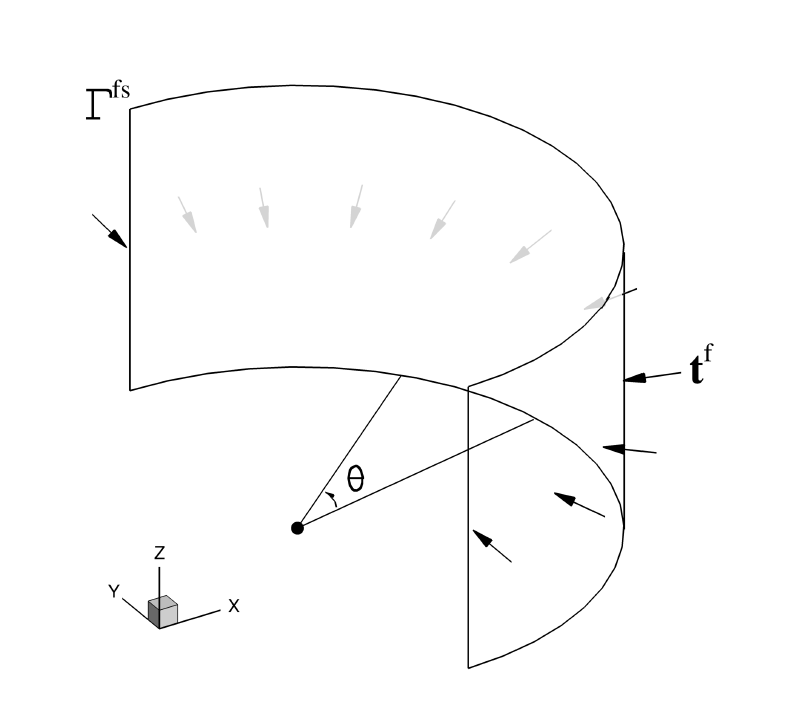}
		\caption{}
	\end{subfigure}%
	\begin{subfigure}[b]{0.33\textwidth}
		\hspace{0.5cm}
		\includegraphics[trim={2cm 1cm 2cm 2cm},clip,width=6cm]{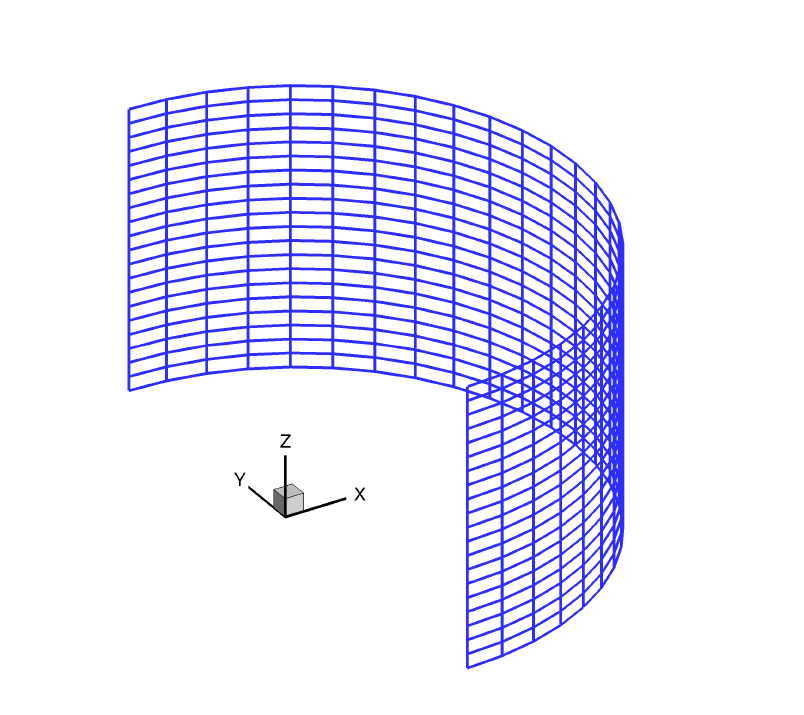}
		\caption{}
	\end{subfigure}%
	\begin{subfigure}[b]{0.33\textwidth}
		\hspace{0.5cm}
		\includegraphics[trim={2cm 1cm 2cm 2cm},clip,width=6cm]{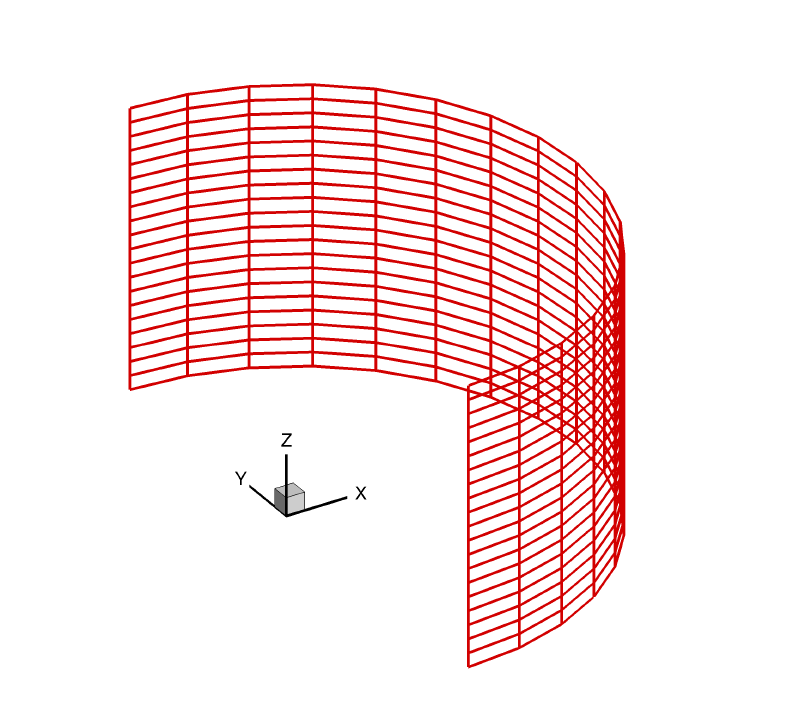}
		\caption{}
	\end{subfigure}%
	\caption{Convergence test for radial basis function mapping: (a) Schematic of the traction transfer from the fluid to the structural meshes along $\Gamma^\mathrm{fs}$, (b) fluid and (c) structural meshes for the case $A^\mathrm{f}/A^\mathrm{s} = 0.67$ and $n^\mathrm{s}_c = 16$.}
	\label{schematic_RBF_conv1}
\end{figure}
Prior to the verification and validation tests of the developed flexible multibody aeroelastic framework, we first carry out a systematic convergence analysis of the radial basis function mapping approach described in the previous section to transfer data across the fluid-structure interface. To accomplish this, we consider a semi-circular surface as the fluid-structure interface with varying discretization on the fluid and the structural sides of the interface as shown in Fig. \ref{schematic_RBF_conv1}. In the schematic, the outer domain consists of fluid ($\Omega^\mathrm{f}$) and the inner domain is the structure ($\Omega^\mathrm{s}$). A fluid traction acts on the fluid side of the fluid-structure interface $\Gamma^\mathrm{fs}$ as
\begin{align} \label{exact_trac}
	\mathbf{t}^\mathrm{f} = \mathbf{t}^\mathrm{f}(\theta,z) = -\bigg( \frac{1}{2}\rho^\mathrm{f}U_{\infty}^2(1-4\mathrm{sin}^2\theta) + \rho^\mathrm{f}gz \bigg)\begin{pmatrix}
	0.5\ \mathrm{cos}\ \theta\\ 0.5\ \mathrm{sin}\ \theta\\ 0
	\end{pmatrix},
\end{align}
where $\rho^\mathrm{f} = 1000$ kg/m$^3$, $U_{\infty} = 1.0$ m/s, $g = 9.81$ m/s$^2$, $z$ is the Z-coordinate and $\theta \in [-\pi/2,\pi/2]$ is the angle shown in Fig. \ref{schematic_RBF_conv1}(a). This prescribed load represents the static pressure along the Z-direction as a result of potential flow around a cylinder. We only consider the downstream half of the cylinder for the current study. The support radius $r$ for the radial basis function is selected as 2 for the study.
\begin{figure}[!htbp]
	\centering
	\begin{subfigure}[b]{\textwidth}
		\centering
		\includegraphics[trim={2cm 1cm 2cm 1.5cm},clip,width=5cm]{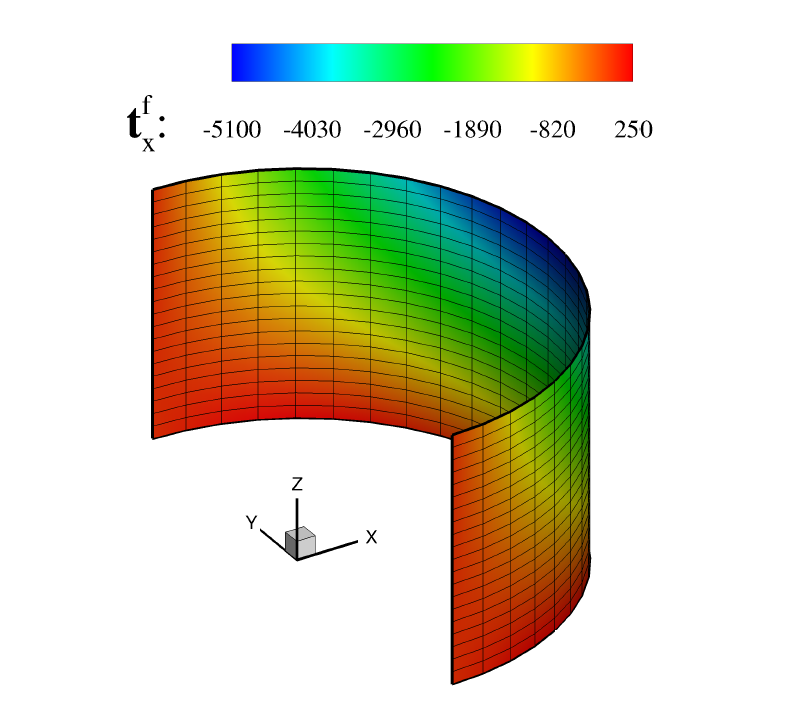}
		\includegraphics[trim={2cm 1cm 2cm 1.5cm},clip,width=5cm]{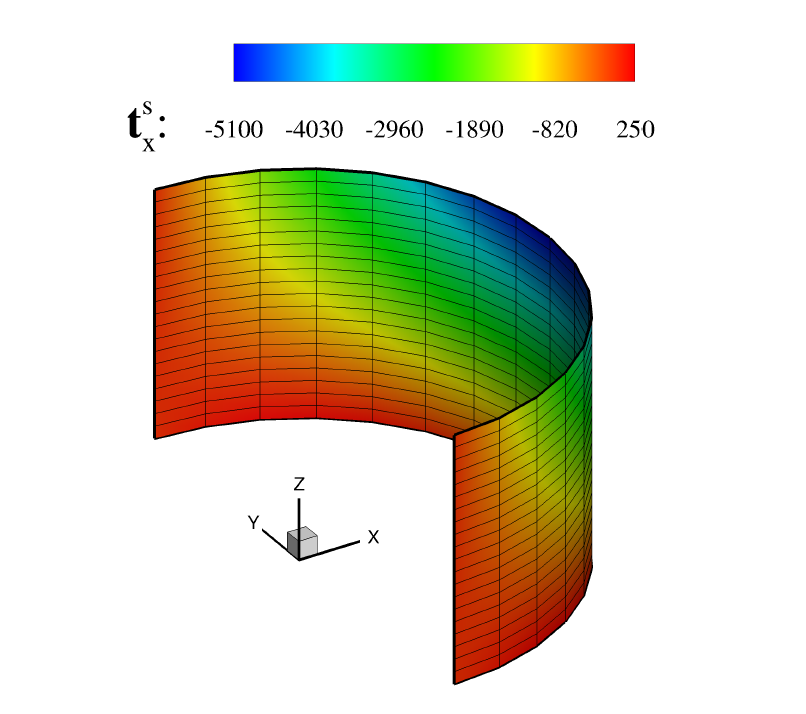}
		\includegraphics[trim={2cm 1cm 2cm 1.5cm},clip,width=5cm]{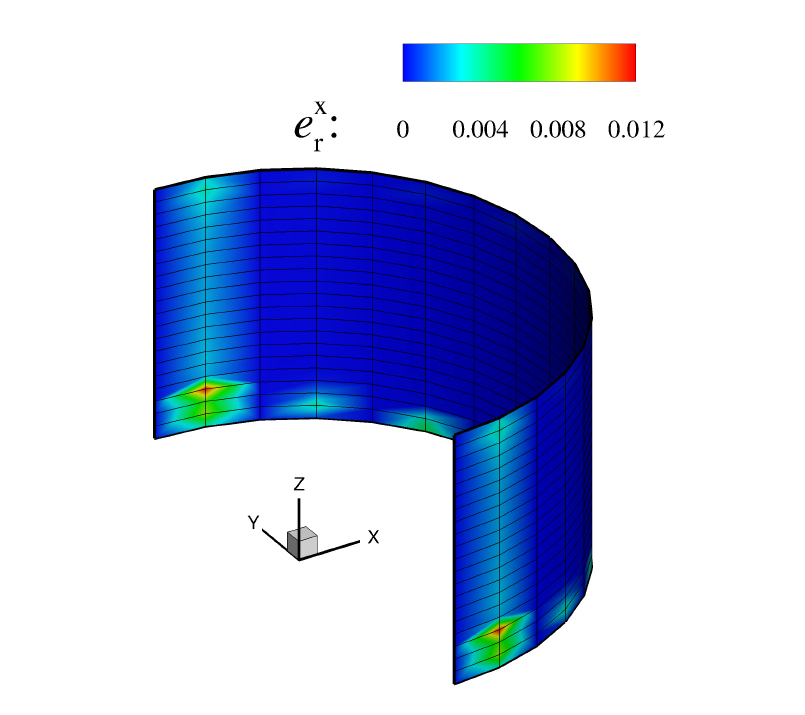}
		\caption{}
	\end{subfigure}%
	
	\begin{subfigure}[b]{\textwidth}
		\centering
		\includegraphics[trim={2cm 1cm 2cm 1.5cm},clip,width=5cm]{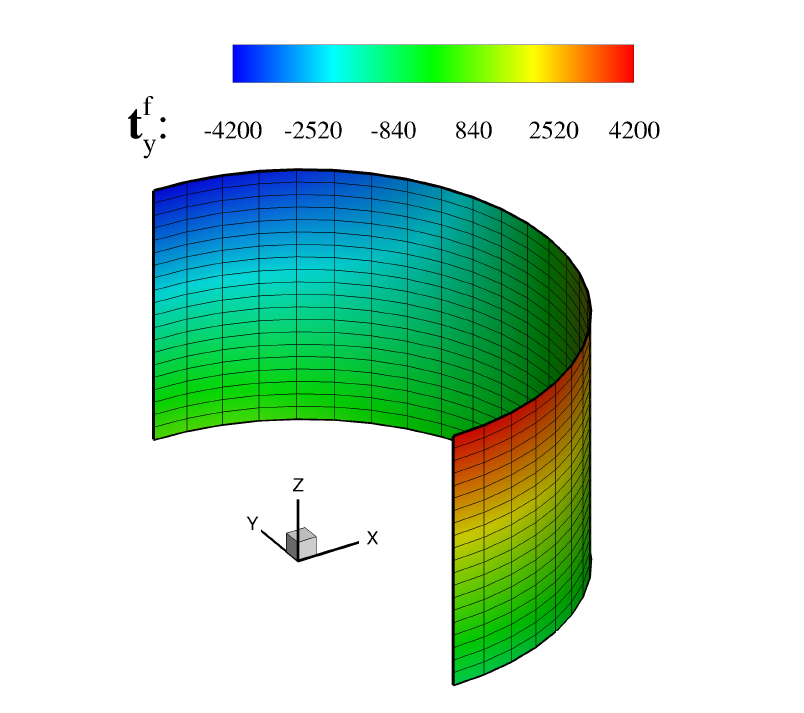}
		\includegraphics[trim={2cm 1cm 2cm 1.5cm},clip,width=5cm]{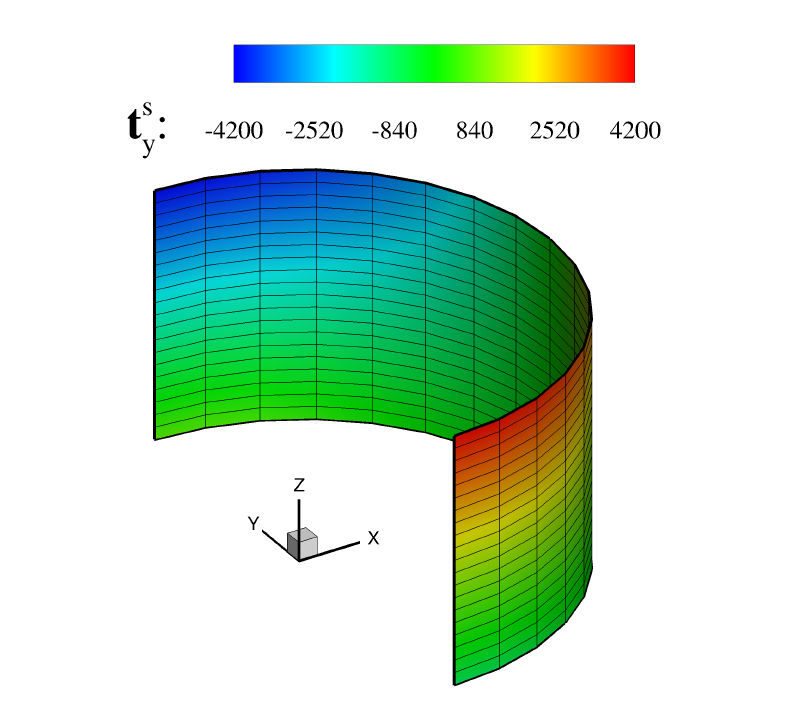}
		\includegraphics[trim={2cm 1cm 2cm 1.5cm},clip,width=5cm]{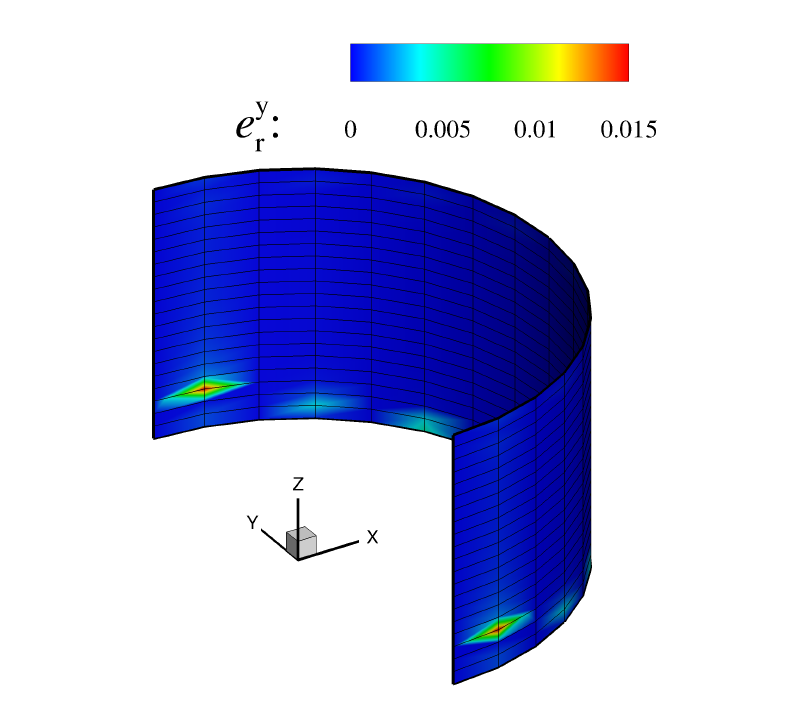}
		\caption{}
	\end{subfigure}%
	\caption{Comparison of interpolation via radial basis function mapping for $A^\mathrm{f}/A^\mathrm{s} = 0.67$ and $n^\mathrm{s}_c = 16$ for the fluid (left) and structural (middle) meshes for traction in (a) X direction, and (b) Y direction. The relative error of the traction values are also shown for the structural mesh (right).}
	\label{RBF_conv2}
\end{figure}
\begin{figure}[!htbp]
	\centering
	\begin{subfigure}[b]{\textwidth}
		\centering
		\includegraphics[trim={2cm 1cm 2cm 1.5cm},clip,width=5cm]{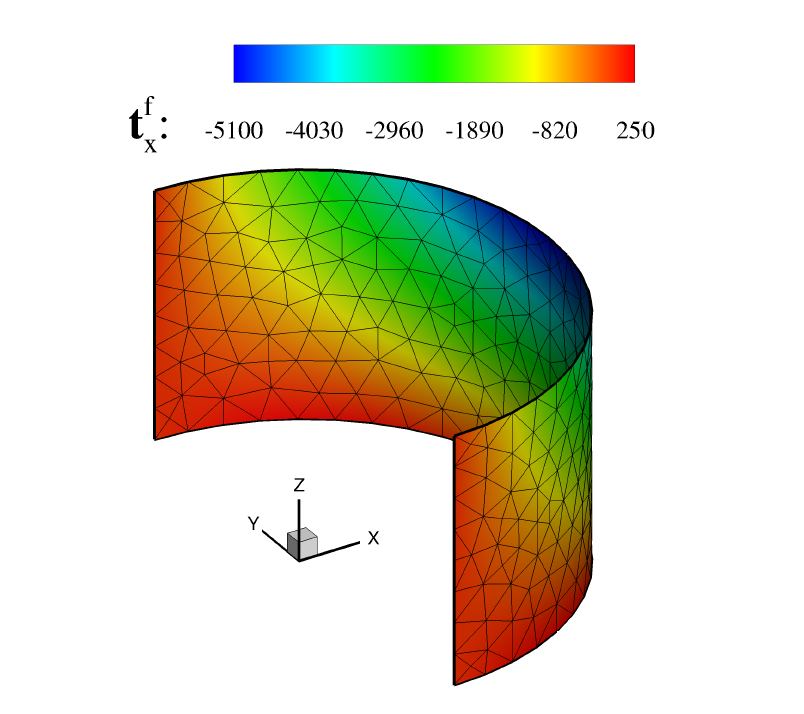}
		\includegraphics[trim={2cm 1cm 2cm 1.5cm},clip,width=5cm]{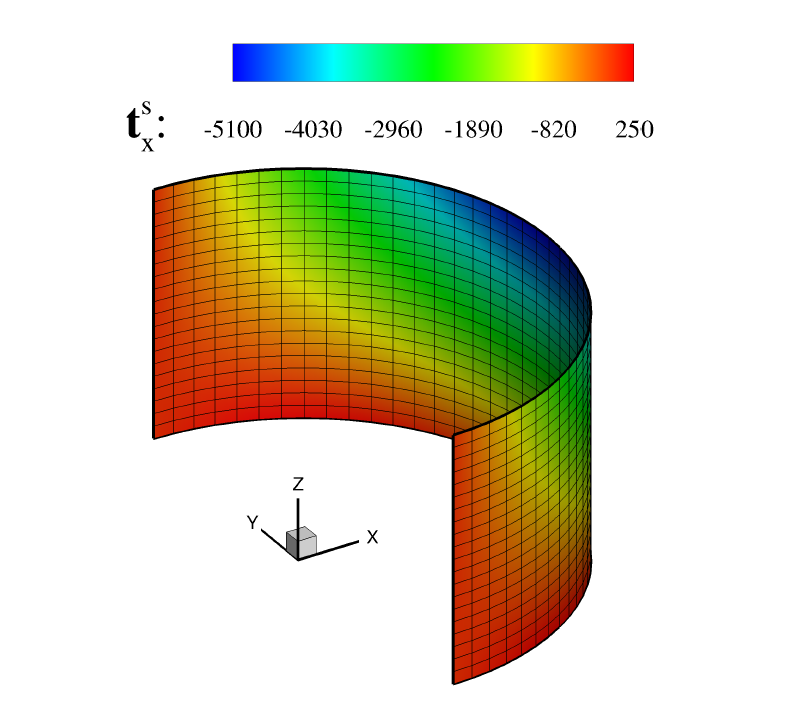}
		\includegraphics[trim={2cm 1cm 2cm 1.5cm},clip,width=5cm]{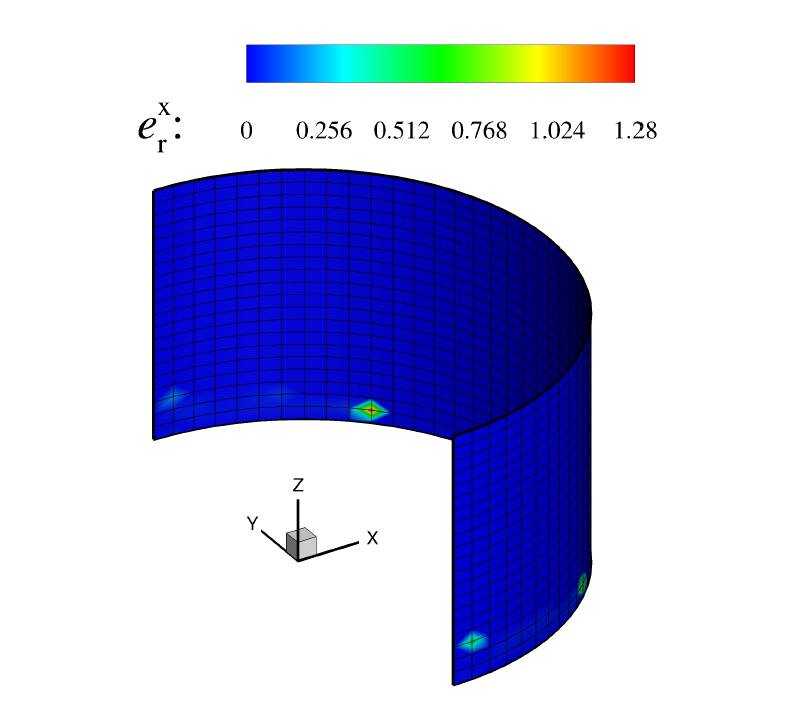}
	\end{subfigure}%
	\caption{Comparison of interpolation via radial basis function mapping for $A^\mathrm{f}/A^\mathrm{s} = 2$ with unstructured triangular mesh for the fluid (left) and structured quadrilateral mesh for the solid (middle) for traction in X direction when $n^\mathrm{s}_z = 20$ and $n^\mathrm{s}_c = 40$. The relative error of the traction values are also shown for the structural mesh (right).}
	\label{RBF_conv5}
\end{figure}

We employ two ratios of mismatch between the fluid and the structural meshes. The discretization in the Z-direction is kept constant at the number of elements of $n^\mathrm{f}_z = n^\mathrm{s}_z = 20$. The number of elements along the circumference of the semi-circle is varied on the fluid ($n^\mathrm{f}_c$) as well as structural ($n^\mathrm{s}_c$) meshes. The element area can thus be written as $A^\mathrm{f} = \pi R z/(n^\mathrm{f}_c n^\mathrm{f}_z)$ and $A^\mathrm{s} = \pi R z/(n^\mathrm{s}_c n^\mathrm{s}_z)$, where $R = 1$ and $z = 1$ are the radius and height of the semi-circular cylinder. The refinement is carried out such that the area mismatch is $A^\mathrm{f}/A^\mathrm{s} \in [0.67, 2]$. The refinement degree of the structural mesh is $n^\mathrm{s}_c \in [16, 32, 64, 128, 256]$ which is kept equivalent for the two cases of mismatch considered. The different meshes on the fluid and structural sides at $\Gamma^\mathrm{fs}$ are shown in Figs. \ref{schematic_RBF_conv1}(b) and (c) for $A^\mathrm{f}/A^\mathrm{s} = 0.67$ and $n^\mathrm{s}_c = 16$.

\begin{figure}[!h]
	\centering
	\begin{subfigure}[b]{0.5\textwidth}
		\centering
		\includegraphics[trim={9cm 0.1cm 8cm 1cm},clip,width=8.5cm]{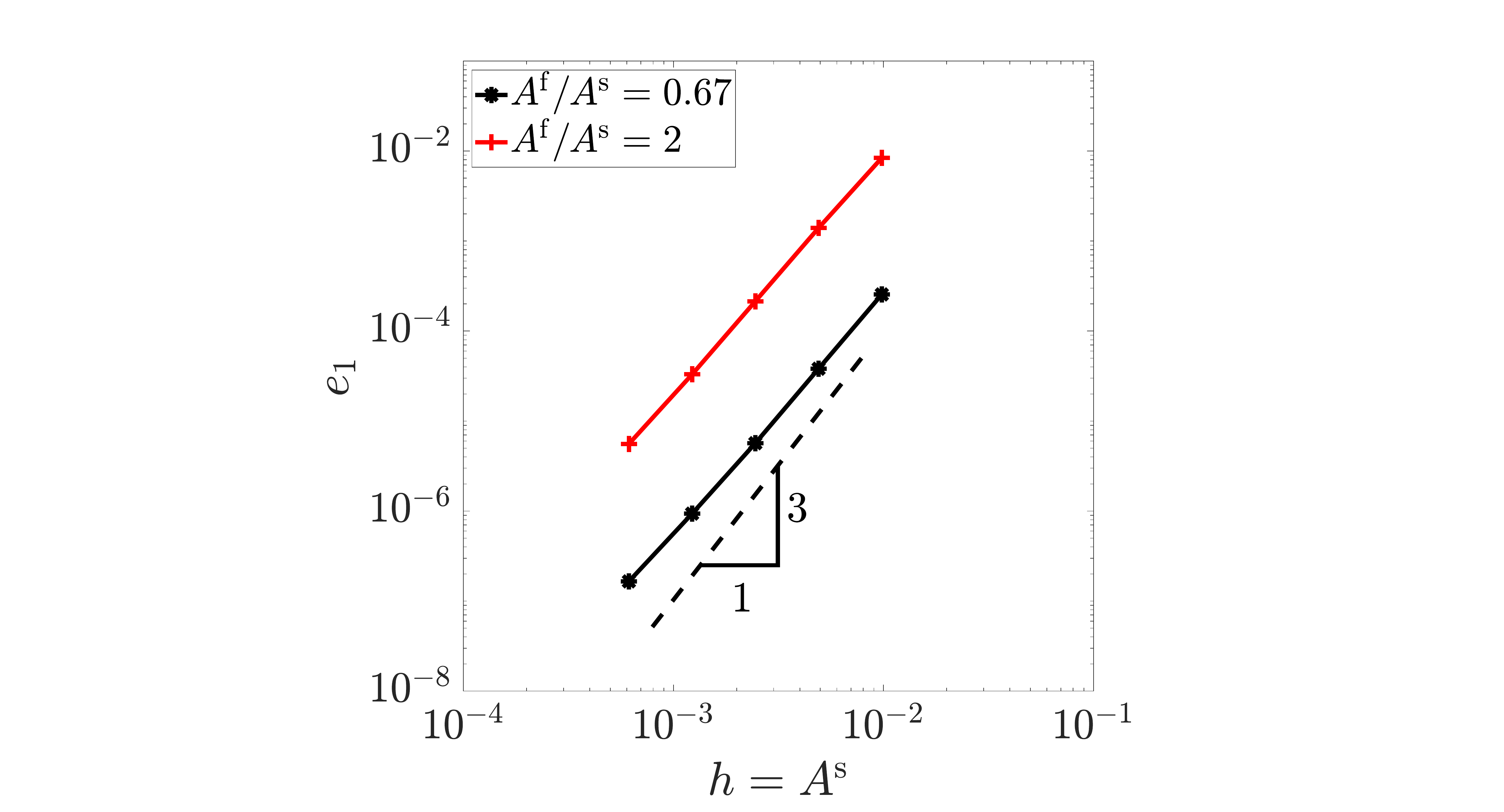}
		\caption{}
	\end{subfigure}%
	\begin{subfigure}[b]{0.5\textwidth}
		\centering
		\includegraphics[trim={9cm 0.1cm 8cm 1cm},clip,width=8.5cm]{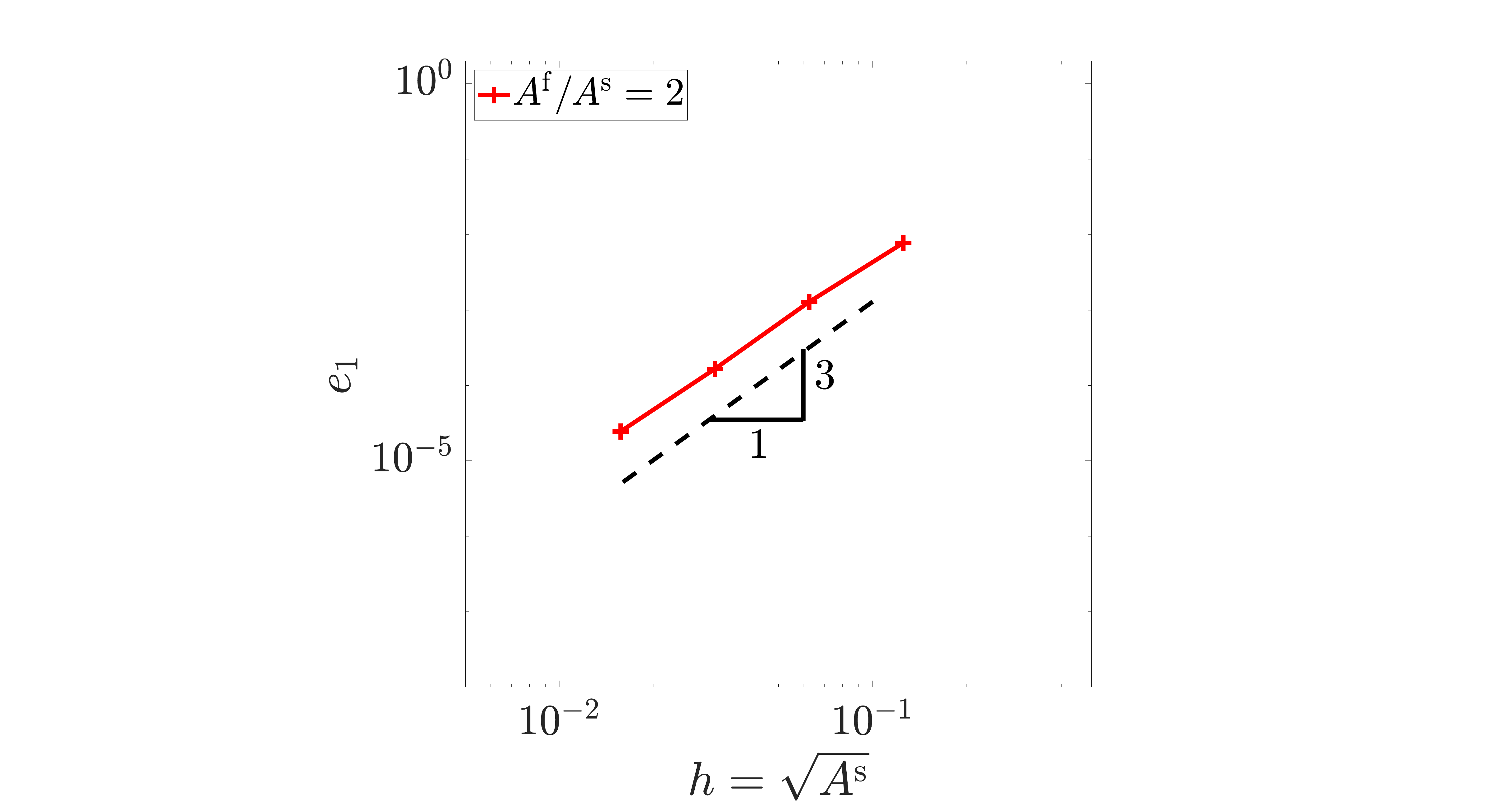}
		\caption{}
	\end{subfigure}%
	\caption{Error convergence for the radial basis function interpolation method for varying mesh mismatch on the fluid and structural sides of the fluid-structural interface for (a) structured grid, and (b) unstructured grid with different shapes.}
	\label{RBF_conv4}
\end{figure}

The interpolated traction values at the structural nodes are shown for the representative case of $A^\mathrm{f}/A^\mathrm{s} = 0.67$ and $n^\mathrm{s}_c = 16$ in Fig. \ref{RBF_conv2}. The relative error in the interpolation is quantified as
\begin{align}
	e_\mathrm{r} = \frac{|\mathbf{t}^\mathrm{s}-\mathbf{t}^\mathrm{s}(\theta,z)|}{|\mathbf{t}^\mathrm{s}(\theta,z)|},
\end{align}
where $\mathbf{t}^\mathrm{s}$ is the interpolated traction values at the structural nodes and $\mathbf{t}^\mathrm{s}(\theta,z)$ is the exact value at the corresponding nodes based on Eq. (\ref{exact_trac}). The contour of the relative error is shown in Fig. \ref{RBF_conv2} (right) for the structural mesh. It is observed that the error is less than 1.5\% for the cases considered. We also quantify the convergence of the RBF interpolation by evaluating the error in the transfer of the traction as
\begin{align}
	e_1 = \frac{||\mathbf{t}^\mathrm{s}-\mathbf{t}^\mathrm{s}(\theta,z)||_2}{||\mathbf{t}^\mathrm{s}(\theta,z)||_2},
\end{align}
The behavior of the error with mesh refinement has been plotted in Fig. \ref{RBF_conv4}(a) where convergence of order close to $3$ is observed which is consistent with the Wendland's $C^2$ function interpolation. Note that in this case, $h = A^\mathrm{s}$ as the number of elements in the Z-direction is constant.

A further analysis is carried out to quantify the error for unstructured non-matching meshes across the fluid and structural domains at the interface consisting of different shapes such as triangular mesh on the fluid and quadrilateral on the structural side. In this case, the number of elements in the structural mesh $n^\mathrm{s}_z \in [10, 20, 40, 80]$ and $n^\mathrm{s}_c \in [20, 40, 80, 160]$ such that $A^\mathrm{f}/A^\mathrm{s} = 2$. As the element size is varying in both the directions, the element size is defined as $h = \sqrt{A^\mathrm{s}}$. The convergence for the unstructured meshes across the interface is shown in Fig. \ref{RBF_conv4}(b) where a third order of convergence is observed, as expected. The interpolation data is shown in Fig. \ref{RBF_conv5} for the representative mesh with $n^\mathrm{s}_z = 20$ and $n^\mathrm{s}_c = 40$.

The error convergence study for the RBF interpolation technique shows a higher order of convergence and independence of the mesh connectivity, emphasizing its generality for scattered data interpolation and efficiency for fluid-structure interaction problems.

\section{Convergence and validation tests: Flow across a pitching plate}
\label{Num_tests}
\begin{figure}[!htbp]
	\centering
	\begin{tikzpicture}[very thick,decoration={markings,mark=at position 0.5 with {\arrow{>}}},scale=1]
	\draw (0,0) node[left]{} -- (0,6) node[right]{} -- (9,6) node[above]{} -- (9,0) node[above]{} -- cycle;
	\draw[black,dotted] (1,1) to (1,7);
	\draw[black] (1,7) to (10,7);
	\draw[black] (10,7) to (10,1);
	\draw[black,dotted] (10,1) to (1,1);
	
	\node (bat) at (4.8,3.25)
	{\includegraphics[trim={0.1cm 0.1cm 0.1cm 0.1cm},clip,width=.3\textwidth]{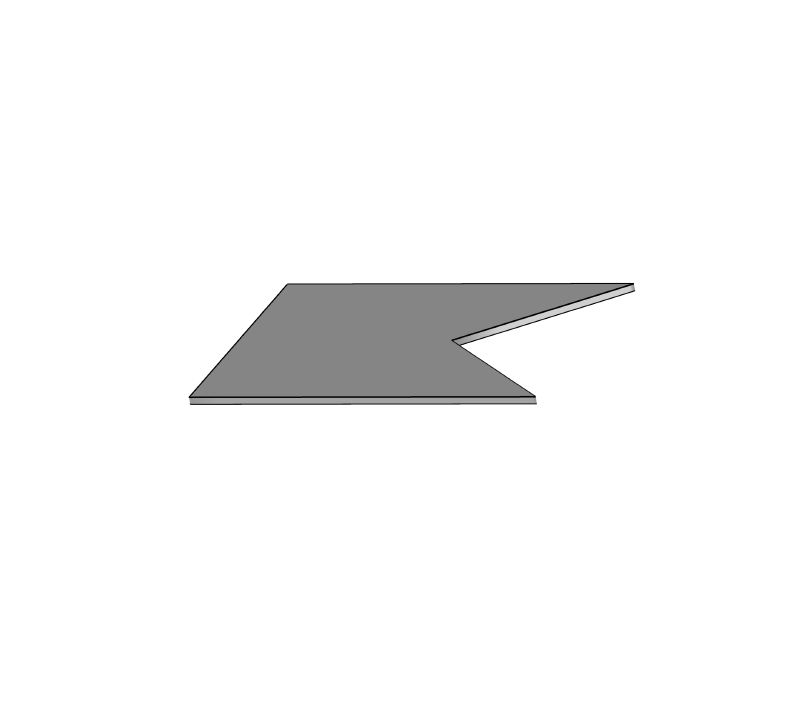}};
	
	\draw[black] (0,6) to (1,7);
	\draw[black] (9,6) to (10,7);
	\draw[black] (9,0) to (10,1);
	\draw[black,dotted] (0,0) to (1,1);

	\draw (5.7,3.05) to (6,3.05);
	\draw (5.7,2.95) to (6,2.95);
	\draw[->] (5.85,3.3) to (5.85,3.05);
	\draw[->] (5.85,2.7) to (5.85,2.95);
	\draw (6,3) node[anchor=west]{$h$};
	
	\draw (3.1,3) to (3.45,3);
	\draw (3.65,3.7) to (4,3.7);
	\draw[<->] (3.275,3) to (3.9,3.7);
	\draw (3.1,3.4) node[anchor=west]{$b$};
	
	\draw[->] (-2.5,3) to (-0.5,3);
	\draw (-1.5,4.5) node[anchor=north]{Inflow};
	\draw (-1.5,4) node[anchor=north]{$|\boldsymbol{u}^\mathrm{f}|=U_\infty$};
	
	\draw (10.2,1) to (10.6,1);
	\draw (10.2,7) to (10.6,7);
	\draw[<->] (10.4,1) to (10.4,7);
	\draw (12,4.5) node[anchor=north]{Outflow};
	\draw (12,3.8) node[anchor=north]{$\boldsymbol{\sigma}^\mathrm{f}\cdot\mathbf{n}^\mathrm{f} = \boldsymbol{0}$};
	\draw (12,3.3) node[anchor=north]{$\nabla\tilde{\nu}\cdot\mathbf{n}^\mathrm{f} = 0$};
	\draw (10.4,4) node[anchor=west]{$H$};
	
	\draw (5,6.6) node[anchor=north]{Slip};
	\draw (5.5,6.75) node{$\boldsymbol{u}^\mathrm{f}\cdot\mathbf{n}^\mathrm{f}=0$, $\boldsymbol{\sigma}^\mathrm{f}\cdot\mathbf{n}^\mathrm{f} = \boldsymbol{0}$};
	
	\draw (5,0.6) node[anchor=north]{Slip};
	\draw (5.5,0.75) node{$\boldsymbol{u}^\mathrm{f}\cdot\mathbf{n}^\mathrm{f}=0$, $\boldsymbol{\sigma}^\mathrm{f}\cdot\mathbf{n}^\mathrm{f} = \boldsymbol{0}$};
	
	\draw[<-] (5.1,3.5) to (5.75,4.75);
	\draw (5.75,4.75) to (6.5,4.75);
	\draw (6.5,4.75) node[anchor=west]{No-slip};
	\draw (6.4,4.4) node[anchor=west] {($\boldsymbol{u}^\mathrm{f}=\boldsymbol{0}$, $\tilde{\nu}=0$)};
	
	\draw (0,-0.2) to (0,-0.6);
	\draw (9,-0.2) to (9,-0.6);
	\draw (4.5,-0.4) node[anchor=north]{$L$};
	\draw[<->] (0,-0.4) to (9,-0.4);
	\draw (9.2,0) to (9.6,0);
	\draw[<->] (9.4,0) to (10.4,1);
	\draw (9.9,0.4) node[anchor=west]{$B$};
	
	\draw[->,red] (2,4.5) to (3,4.5);
	\draw (3,4.5) node[anchor=west,red]{Y};
	\draw[->,red] (2,4.5) to (2,5.5);
	\draw (2,5.5) node[anchor=west,red]{Z};
	\draw[->,red] (2,4.5) to (1.6,4);
	\draw (1.7,4) node[anchor=west,red]{X};
	
	\draw[<->] (3.8,2.6) arc (-30:30:0.75);
	
	\end{tikzpicture}
	\caption{A schematic of the flow past a three-dimensional pitching plate. The computational setup and boundary conditions are shown for the turbulent Navier-Stokes equations. Here, ${\boldsymbol{u}}^\mathrm{f}$ denotes the fluid velocity, $L=80c$, $B=80c$, $H=80c$ are the length, breadth and height of the computational domain respectively, and $b$ and $h$ are the span and thickness of the plate respectively.} 
	\label{schematic_plate}
\end{figure}
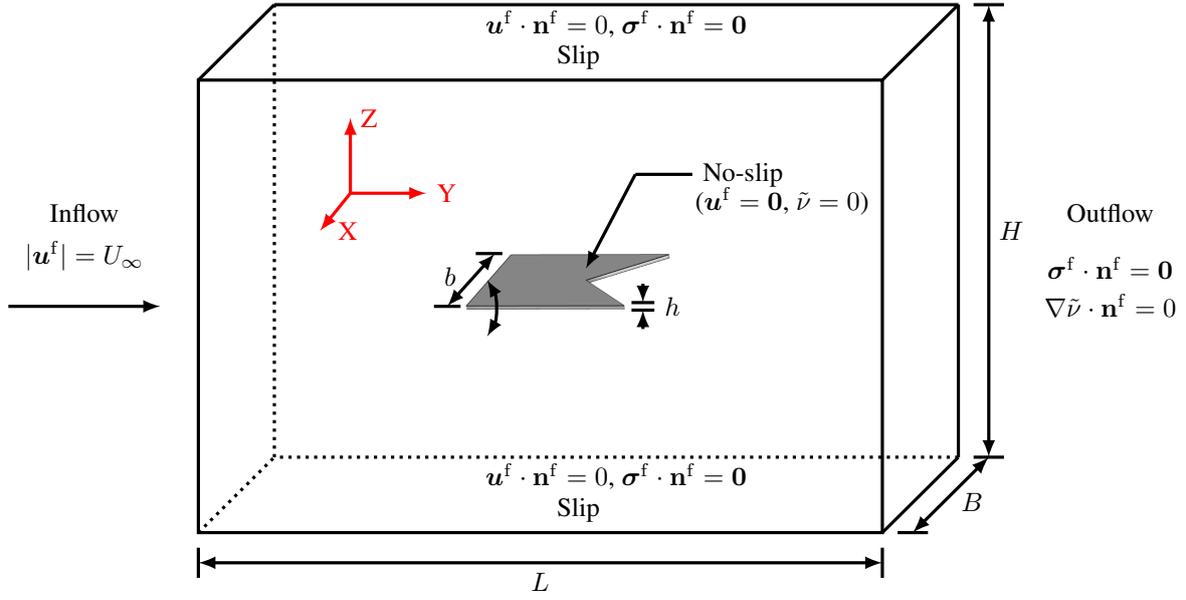

In this section, we carry out numerical tests to verify and validate the coupling between the flexible multibody structural system and the fluid loading via the RBF coupling. To accomplish this, we consider a flow across a pitching plate with a serration of $45^{\circ}$ at the trailing edge as shown in Fig. \ref{schematic_plate}. The width of the plate is $b = 0.1$ m with a thickness of $h = 2.54 \times 10^{-3}$ m and a mean chord of $c = S/b = 0.1$ m. The surface area of the plate is $S = 0.01 $ m$^2$. The computational domain is shown in Fig. \ref{schematic_plate} with an inflow velocity of $U_{\infty} = 0.1$ m/s. The inflow and outflow boundaries are at a distance of $40c$ from the leading edge. The sides (parallel to the $Y-Z$ plane), top and bottom (parallel to $X-Y$ plane) boundaries are around $40c$ from the middle of the plate and satisfy a slip boundary condition. The no-slip boundary condition of $\boldsymbol{u}^\mathrm{f} = \boldsymbol{0}$ is satisfied on all the sides of the plate. A prescribed pitching motion is given at the leading edge with a time varying amplitude as $\theta = \theta_\mathrm{max}\mathrm{sin}(2\pi f_0 t)$, where $\theta_\mathrm{max}$ is the maximum pitching amplitude and $f_0$ is the frequency of pitching. 

\subsection{Computational mesh}
\begin{figure}[!h]
	\centering
	\begin{subfigure}[b]{0.5\textwidth}
		\includegraphics[trim={3cm 4cm 2cm 4cm},clip,width=8cm]{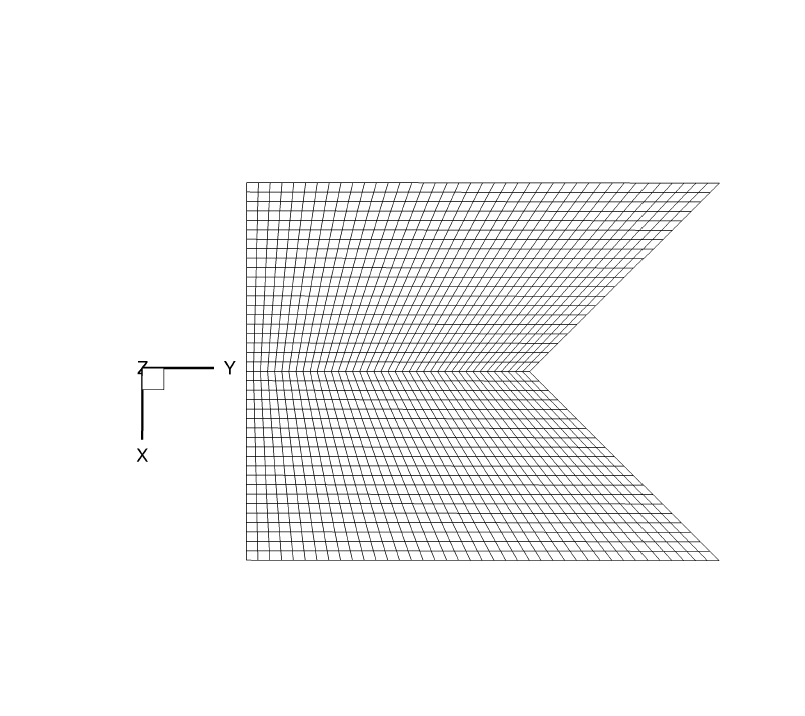}
		\caption{}
	\end{subfigure}%
	\begin{subfigure}[b]{0.5\textwidth}
		\hspace{1cm}
		\includegraphics[trim={12cm 0.01cm 12cm 0.01cm},clip,width=5.7cm, angle=-90]{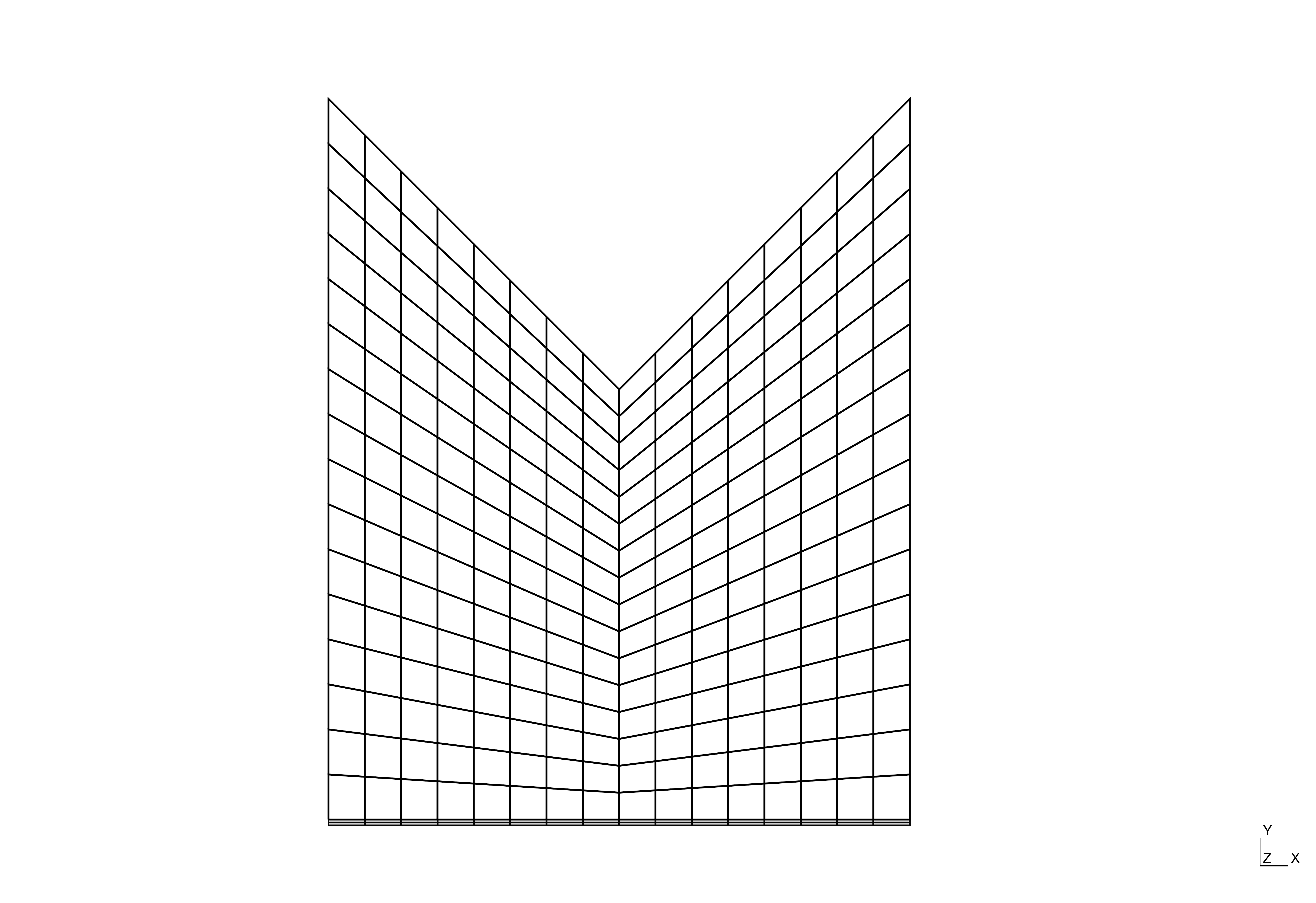}
		\vspace{-0.6cm}
		\caption{}
	\end{subfigure}%
	\caption{Flow across a pitching plate: Representative non-matching mesh at the fluid-structure interface for (a) fluid domain, and (b) structural domain.}
	\label{plate_mesh}
\end{figure}

The computational three-dimensional mesh for the fluid domain is constructed for the pitching plate consisting of eight-node hexahedron elements and the two-dimensional structural mesh is constructed by four-node quadrilaterals. The plate is modeled by two-dimensional flexible thin shell elements in the aeroelastic framework with a prescribed rotation given at the leading edge. The non-matching mesh at the fluid-structure interface for the fluid and structural domains is shown in Fig. \ref{plate_mesh}. A boundary layer mesh around the plate is formed such that the first layer closest to the plate has $y^+ \leq 1$.

\subsection{Non-dimensional parameters}
The key fluid-structure parameters which characterize the problem are the Reynolds number $Re$, Aeroelastic number $Ae$, mass ratio $m^*$, Poisson's ratio $\nu^\mathrm{s}$ and the Strouhal number $St$, defined as follows:
\begin{align}
	Re &= \frac{\rho^\mathrm{f}U_{\infty}c}{\mu^\mathrm{f}} = 10,000, &&Ae = \frac{Eh}{(1/2)\rho^\mathrm{f}U_{\infty}^2 S} = 1.575 \times 10^8,\nonumber \\
	m^* &= \frac{\rho^\mathrm{s}h}{\rho^\mathrm{f} c} = 0.03, 
	&&\nu^\mathrm{s} = 0.37,\qquad St = \frac{f_0 a}{U_{\infty}},
\end{align}
where $E = 3.1 \times 10^9$ N/m$^2$ is the Young's modulus of the structural material of the plate and $a = 2c\mathrm{sin}(\theta)$ is the characteristic width of the wake. In the study, the Strouhal number $St$ is varied while the thrust coefficient $C_T$ and the propulsive efficiency $\eta$ of the pitching motion are quantified. They are defined as
\begin{align}
	C_T = \frac{T}{(1/2)\rho^\mathrm{f}U_{\infty}^2 S},\qquad \eta = \frac{\overline{T}U_{\infty}}{2\tau_\mathrm{max}\theta_\mathrm{max}f_0},
\end{align}
where $T$ is the thrust generated by the pitching plate, $\overline{T}$ is the mean of the thrust and $\tau_\mathrm{max}$ is the maximum value of the spanwise torque.

\renewcommand{\arraystretch}{0.4}
\begin{table}[!htbp]
	\caption{Mesh characteristics and results for interface convergence study.}
	\centering
	\begin{tabular}{ M{1cm} | M{1.5cm} | M{1.5cm} | M{1.5cm} | M{1.5cm} | M{2.5cm} | M{2.5cm} N }
		\hline
		\centering
		\textbf{Mesh} & \multicolumn{2}{c|}{Structure (2D)} & \multicolumn{2}{c|}{Fluid (3D)} & \textbf{$\overline{C}_T$} & $\eta$ &\\[10pt]
		\hline
		& Nodes & Elements & Nodes & Elements & & &\\[10pt]
		\hline
		IC1 &  25 & 16 & 777,508 & 756,147 & $0.5641 (1.79\%)$ & $0.0947 (1.17\%)$ &\\[10pt]
		\centering
		IC2 &  81 & 64 & 777,508 & 756,147 & $0.5558 (0.29\%)$ & $0.0936 (0\%)$ &\\[10pt]
		
		\centering
		\redcolor{IC3} &  289 & 256 & 777,508 & 756,147  & $0.5547 (0.09\%)$ & $0.0936 (0\%)$ &\\[10pt]
		
		\centering
		IC4 &  1089 & 1024 & 777,508 & 756,147 & $0.5543 (0.02\%)$ & $0.0936 (0\%)$ &\\[10pt]
		
		\centering
		IC5 &  4225 & 4096 & 777,508 & 756,147 & $0.5542$ & $0.0936$ &\\[10pt]		
		\hline
	\end{tabular}
	\label{interface_conv1}
\end{table}
\renewcommand{\arraystretch}{0.4}
\begin{table}[!htbp]
	\caption{Mesh characteristics and results for mesh convergence study.}
	\centering
	\begin{tabular}{ M{1cm} | M{1.5cm} | M{1.5cm} | M{1.5cm} | M{1.5cm} | M{2.5cm} | M{2.5cm} N }
		\hline
		\centering
		\textbf{Mesh} & \multicolumn{2}{c|}{Structure (2D)} & \multicolumn{2}{c|}{Fluid (3D)} & \textbf{$\overline{C}_T$} & $\eta$ &\\[10pt]
		\hline
		& Nodes & Elements & Nodes & Elements & & & \\[10pt]
		\hline
		M1 &  25 & 16 & 275,068 & 264,447 & $0.51397 (4.39\%)$ & $0.10416 (0.18\%)$ &\\[10pt]
		\centering
		M2 &  81 & 64 & 415,348 & 401,547 & $0.50744 (3.06\%)$ & $0.10394 (1.59\%)$ &\\[10pt]
		
		\centering
		\redcolor{M3} &  289 & 256 & 777,508 & 756,147  & $0.49740 (1.02\%)$ & $0.10191 (0.39\%)$ &\\[10pt]
		
		\centering
		M4 &  1089 & 1024 & 1,828,228 & 1,786,947 & $0.49236$ & $0.10231$ &\\[10pt]		
		\hline
	\end{tabular}
	\label{mesh_conv1}
\end{table} 

\begin{figure}[!htbp]
	\centering
	\begin{subfigure}[b]{0.5\textwidth}
		\includegraphics[trim={11cm 0.1cm 12cm 0.5cm},clip,width=7.5cm]{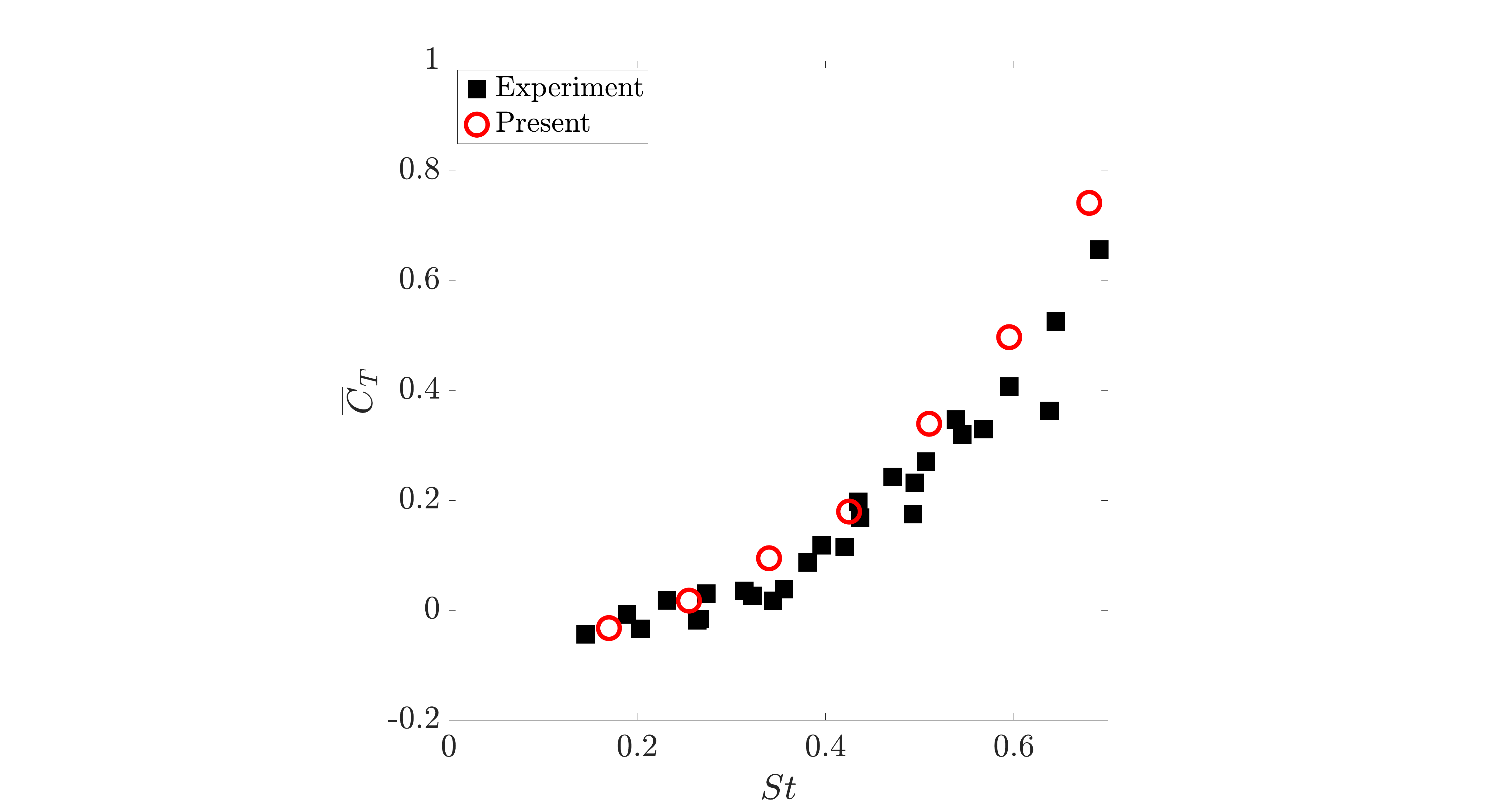}
		\caption{}
	\end{subfigure}%
	\begin{subfigure}[b]{0.5\textwidth}
		\includegraphics[trim={11cm 0.1cm 12cm 0.5cm},clip,width=7.5cm]{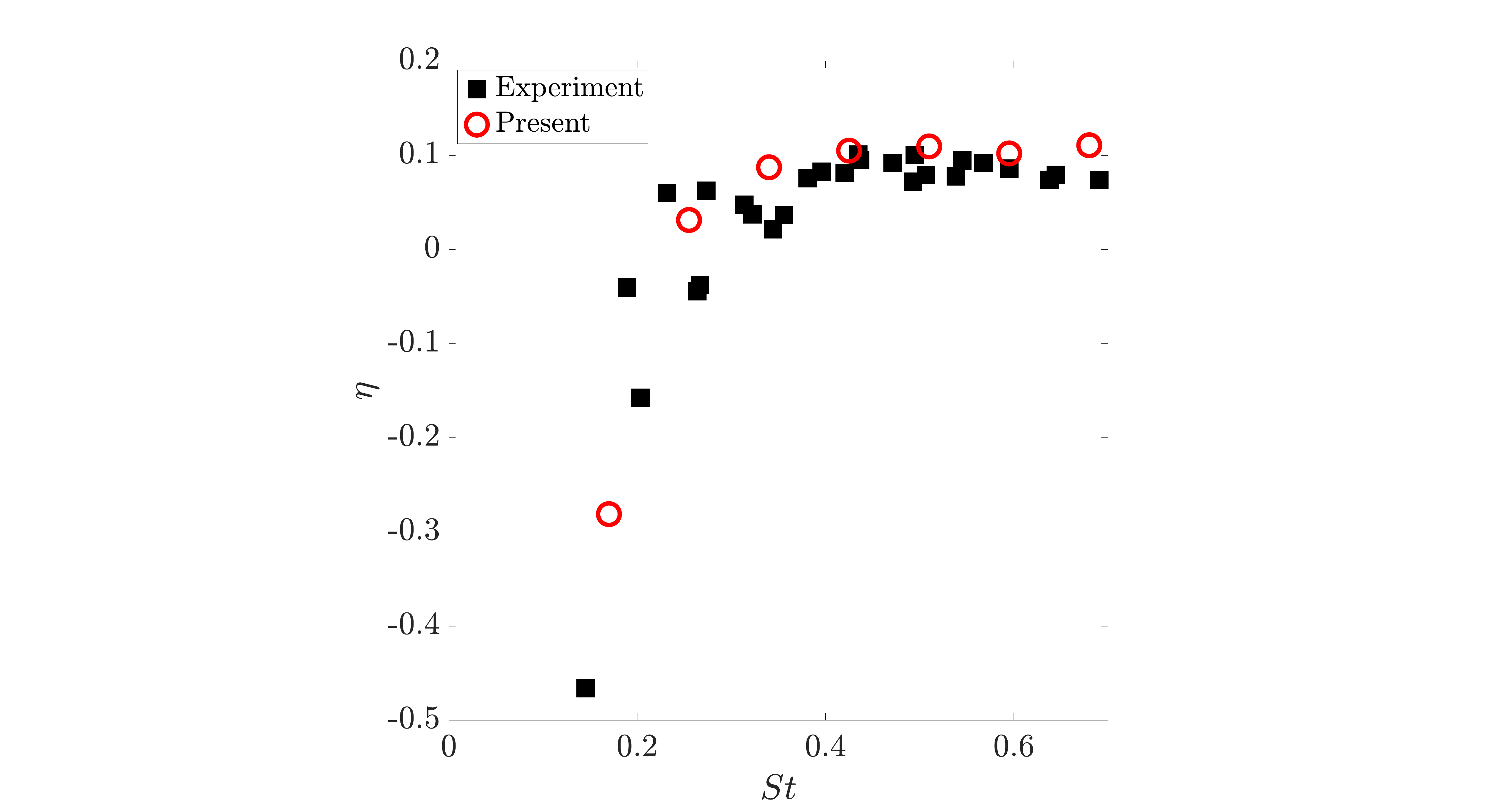}
		\caption{}
	\end{subfigure}%
	\caption{Flow across a pitching plate: Variation with Strouhal number ($St$) of the (a) mean thrust coefficient $\overline{C}_T$, and (b) thrust efficiency $\eta$. The results are compared with that of the experimental study conducted in \cite{VanBuren2017}.}
	\label{plate_validation}
\end{figure}

\begin{figure}[!htbp]
	\centering
	\includegraphics[trim={11cm 0.1cm 12cm 0.5cm},clip,width=6cm]{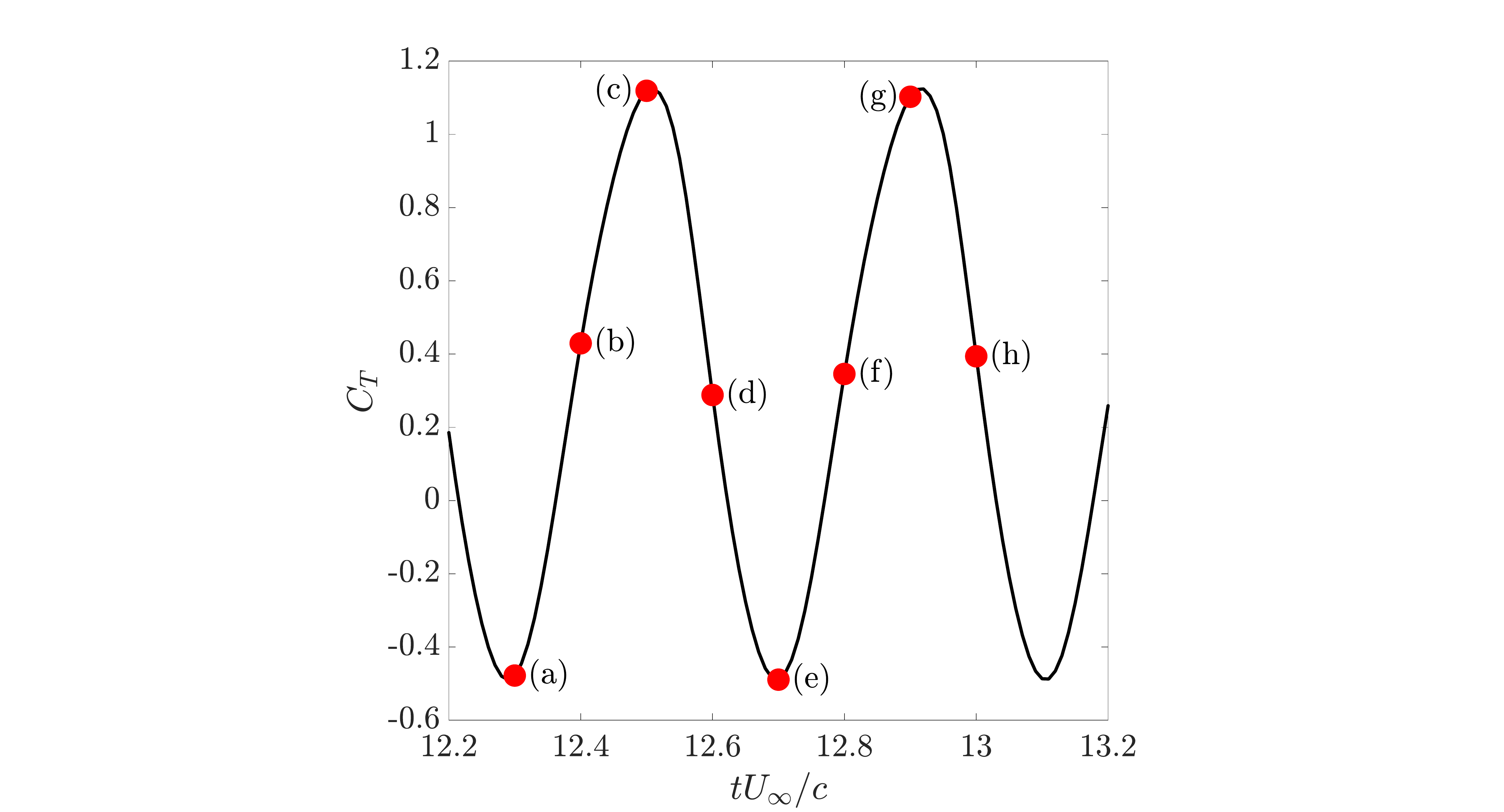}
	\caption{Temporal variation of the thrust coefficient $C_T$ of the pitching plate at $St=0.51$ in a time period of pitching $TU_{\infty}/c = 0.815$. The markers represent the locations on which the three-dimensional $Q$-criterion is plotted in Fig. \ref{plate_Qcriterion_St51}.}
	\label{St51_CT}
\end{figure}
\begin{figure}[!htbp]
	\centering
	\includegraphics[trim={18cm 16cm 0.1cm 0.1cm},clip,width=2cm]{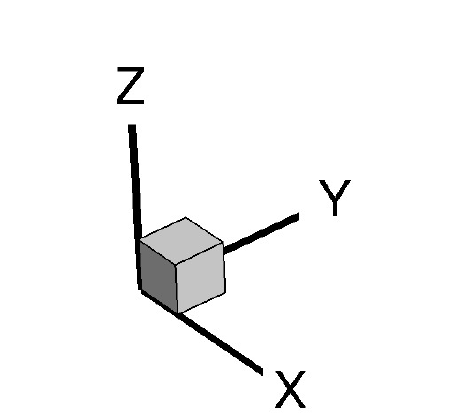}
	\includegraphics[trim={4cm 1cm 4cm 16cm},clip,width=7.5cm]{plate_10_75.pdf}
	
	\begin{subfigure}[b]{0.33\textwidth}
		\includegraphics[trim={2cm 6cm 4cm 2cm},clip,width=5cm]{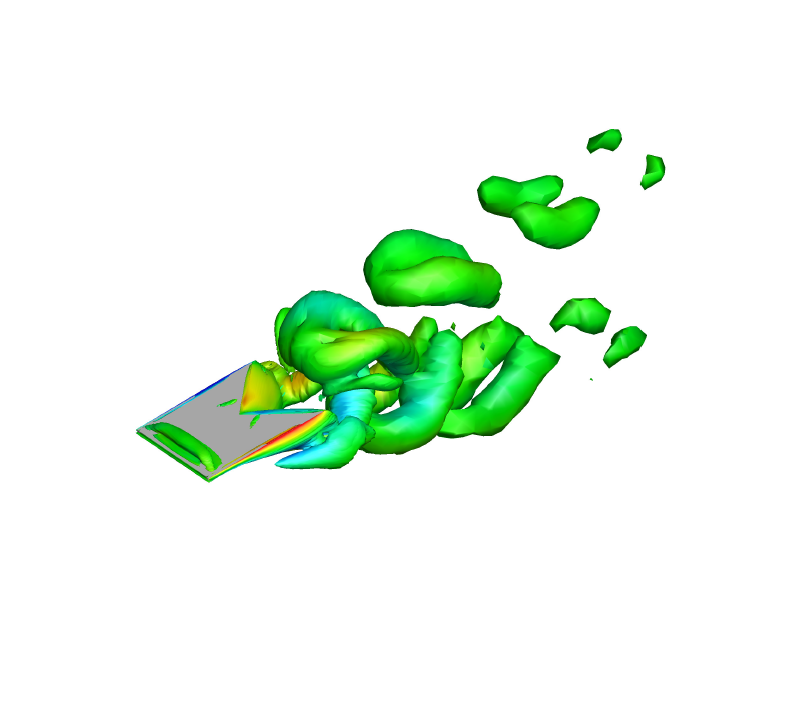}
		\caption{}
	\end{subfigure}%
	\begin{subfigure}[b]{0.33\textwidth}
		\includegraphics[trim={2cm 6cm 4cm 2cm},clip,width=5cm]{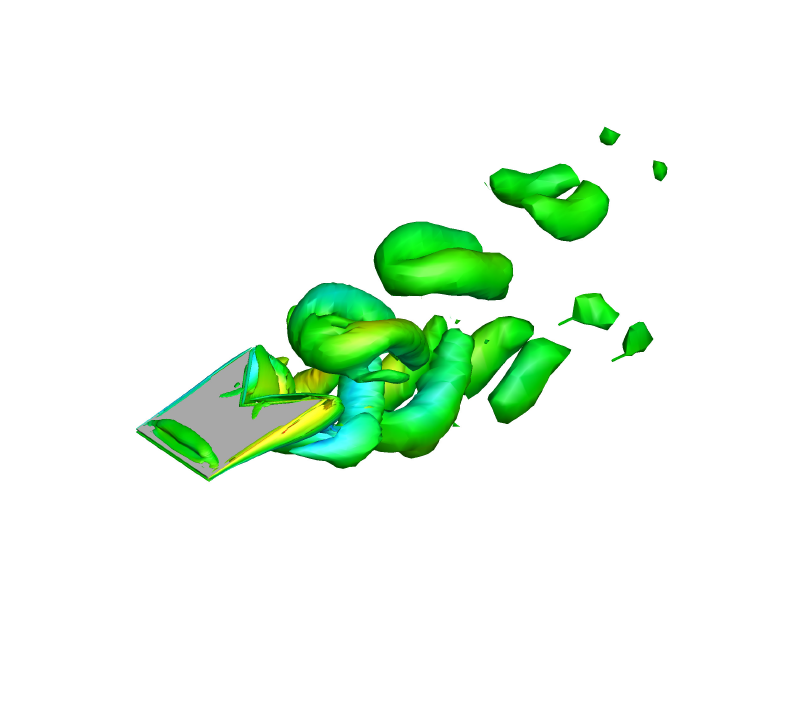}
		\caption{}
	\end{subfigure}%
	\begin{subfigure}[b]{0.33\textwidth}
		\includegraphics[trim={2cm 6cm 4cm 2cm},clip,width=5cm]{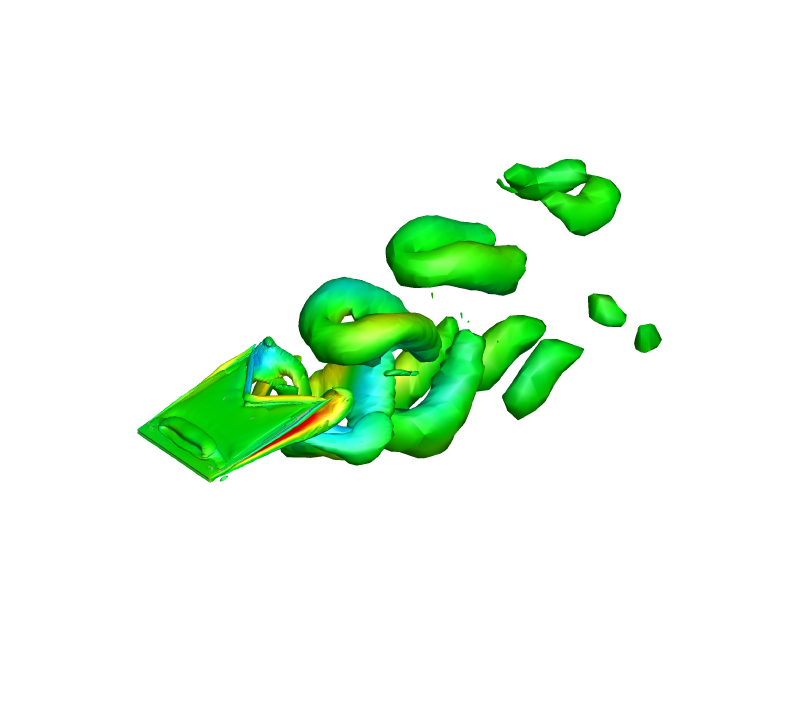}
		\caption{}
	\end{subfigure}%

	\begin{subfigure}[b]{0.33\textwidth}
		\includegraphics[trim={2cm 6cm 4cm 2cm},clip,width=5cm]{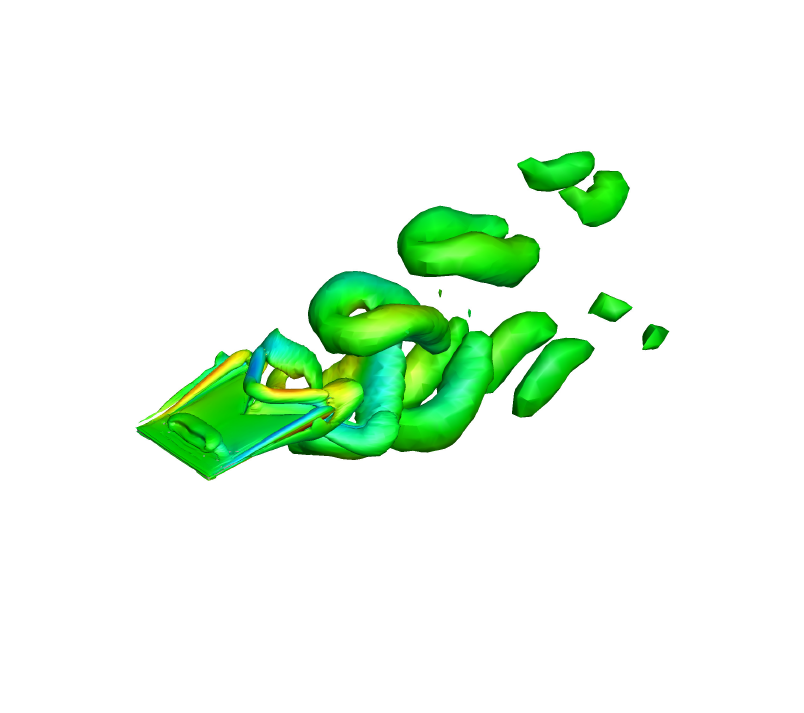}
		\caption{}
	\end{subfigure}%
	\begin{subfigure}[b]{0.33\textwidth}
		\includegraphics[trim={2cm 6cm 4cm 2cm},clip,width=5cm]{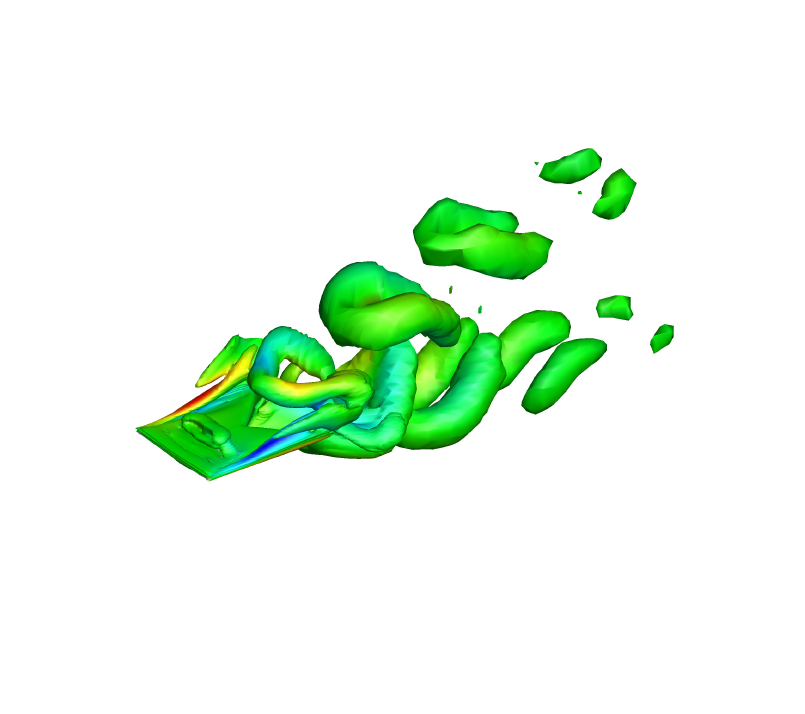}
		\caption{}
	\end{subfigure}%
	\begin{subfigure}[b]{0.33\textwidth}
		\includegraphics[trim={2cm 6cm 4cm 2cm},clip,width=5cm]{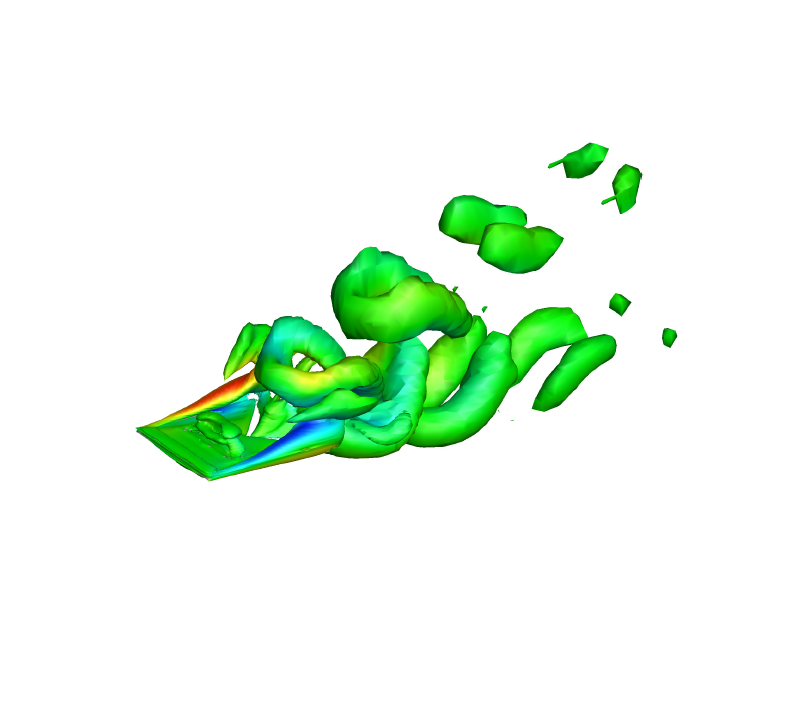}
		\caption{}
	\end{subfigure}%

	\begin{subfigure}[b]{0.33\textwidth}
		\includegraphics[trim={2cm 6cm 4cm 2cm},clip,width=5cm]{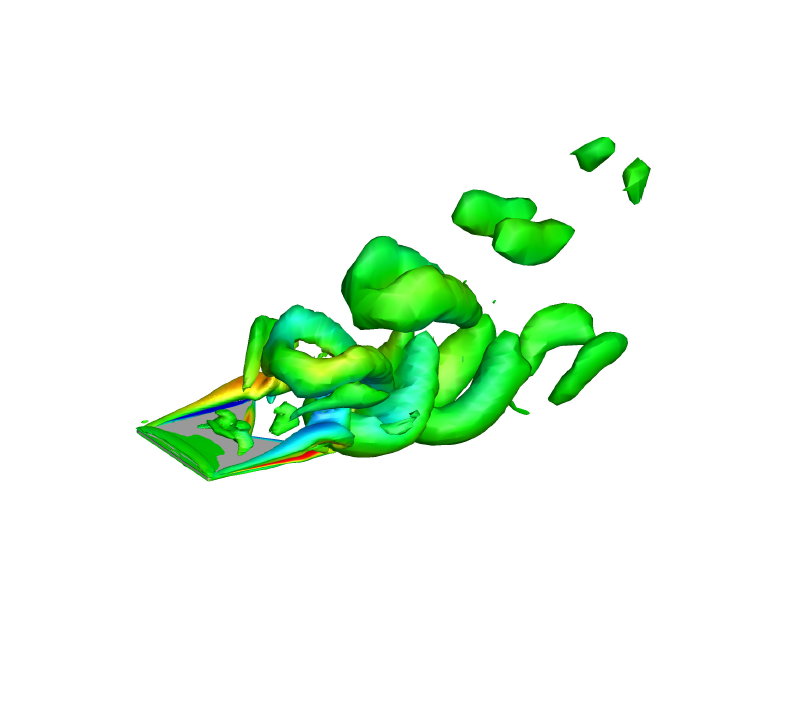}
		\caption{}
	\end{subfigure}%
	\begin{subfigure}[b]{0.33\textwidth}
		\includegraphics[trim={2cm 6cm 4cm 2cm},clip,width=5cm]{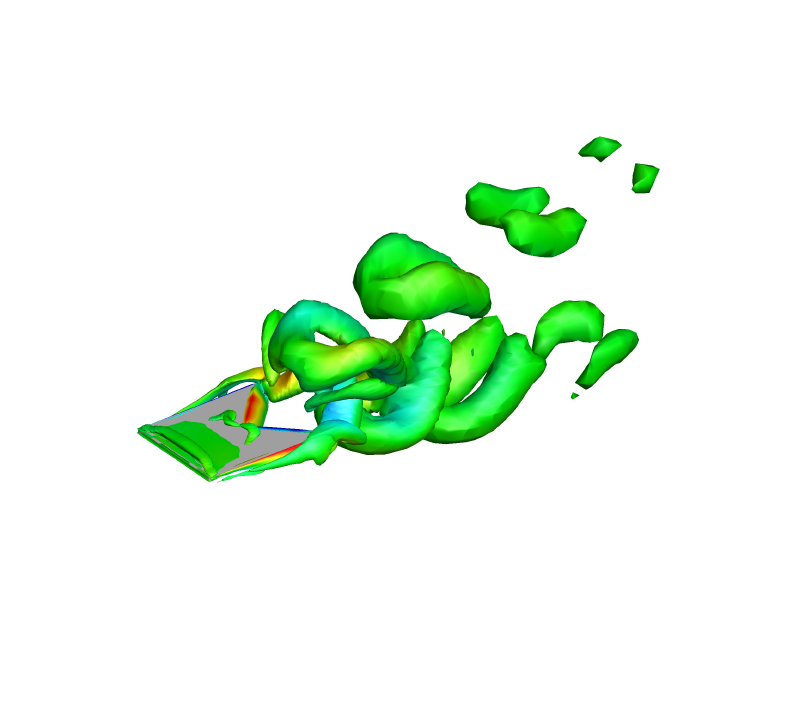}
		\caption{}
	\end{subfigure}%
	\caption{Flow visualization with the help of $Q$-criterion colored by the free-stream velocity at $St = 0.51$ for the pitching plate at $tU_{\infty}/c$: (a) 12.3, (b) 12.4, (c) 12.5, (d) 12.6, (e) 12.7, (f) 12.8, (g) 12.9 and (h) 13.0. These time locations correspond to the markers in Fig. \ref{St51_CT} in one pitching cycle of the plate.}
	\label{plate_Qcriterion_St51}
\end{figure}

\subsection{Interface convergence}
With the problem setup and the computational parameters defined, we next carry out an interface convergence study where the three-dimensional fluid mesh is chosen with a systematic refinement of the structural mesh to eliminate the effect of the interface discretization errors. The fluid mesh contains 777,508 eight-node hexahedron elements and 756,147 nodes. The refinement details of the structural mesh are given in Table \ref{interface_conv1}. The interface convergence study is conducted by fixing the Strouhal number at $St = 0.595$ and the results are also summarized in Table \ref{interface_conv1} where the percentage error compared to the finest mesh (IC5) is quantified. It can be concluded that the mesh IC3 is sufficiently converged with error $< 0.1\%$ with respect to interface mesh.

\subsection{Mesh convergence}
After the interface convergence criteria have been established, we fixed the ratio of the size of the structural to the fluid elements at the interface, and simultaneously refined the structural as well as fluid meshes for the mesh convergence study. The details about the different meshes employed in the study are shown in Table \ref{mesh_conv1} along with the quantities of interest. It can be seen that the mesh M3 is the optimal mesh with respect to errors being $< 1\%$ compared to the finest mesh M4. Therefore, M3 is chosen for the further validation study. It consists of 289 structural nodes with 256 four-node quadrilaterals in the two-dimensional mesh.

\subsection{Validation}
With the help of the interface and mesh convergence studies, we selected the optimal mesh to perform validation of the flexible multibody aeroelastic framework at various Strouhal numbers. We conduct the numerical experiment of flow across a pitching foil for the Strouhal numbers in the range $St \in [0.17, 0.68]$ and quantify the average thrust coefficient and the propulsive efficiency which are plotted in Fig. \ref{plate_validation}. The results show a good agreement with the experimental studies in \cite{VanBuren2017}. The temporal evolution of the thrust coefficient with time for a cycle of pitching is shown for $St = 0.51$ in Fig. \ref{St51_CT}. It can be observed that the frequency of the thrust oscillation is twice that of the pitching frequency. The position of the pitching plate in a pitching time period along with the three-dimensional vortical structures generated in the wake are shown in Fig. \ref{plate_Qcriterion_St51}. The vortical structures are visualized by the iso-contours of $Q$-criterion colored by the streamwise velocity. The markers in Fig. \ref{St51_CT} represent the temporal locations shown in Fig. \ref{plate_Qcriterion_St51}. One can observe the horseshoe-like vortices created due to the $45^{\circ}$ serration at the trailing edge. The tip vortices merge with these horseshoe vortices as they convect downstream.

\section{Three-dimensional flapping dynamics of a bat at $Re=12,000$}
\label{demons}
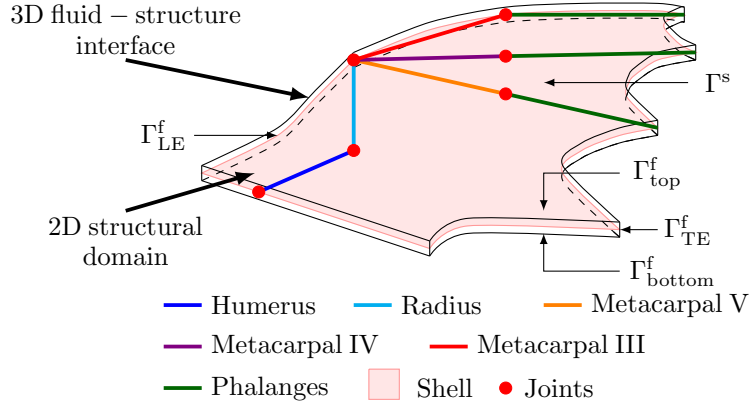
\begin{figure}[!htbp]
\centering
\def\th{0.1}
\def\ty{0.5}
\begin{tikzpicture}[decoration={markings,mark=at position 0.5 with {\arrow[scale=2]{>}}},scale=1]
\draw plot [smooth] coordinates  {(5.5,-0.25-\th) (4.75,0.5-\th) (5,0.8-\th) (5.5,1.0-\th) (6,1.1-\th)};
		\draw plot [smooth] coordinates  { (6,1.1-\th) (5.6,1.5-\th) (5.62,1.7-\th) (5.7,1.8-\th) (6,2-\th) (6.5,2.1-\th)};
		\draw plot [smooth] coordinates  { (6.5,2.1-\th) (6.3,2.2-\th) (6.1,2.4-\th) (6.1,2.5-\th) (6.3,2.6-\th) (6.35,2.6-\th)};
		
		\draw[fill,color=white!90!red,fill opacity=1] plot [smooth] coordinates  { (3,-0.5) (3.5,-0.2) (5.5,-0.25)}-- plot [smooth] coordinates{(5.5,-0.25) (4.75,0.5) (5,0.8) (5.5,1.0) (6,1.1)} -- plot [smooth] coordinates  { (6,1.1) (5.6,1.5) (5.62,1.7) (5.7,1.8) (6,2) (6.5,2.1)} -- plot [smooth] coordinates  { (6.5,2.1) (6.3,2.2) (6.1,2.4) (6.1,2.5) (6.3,2.6) (6.35,2.6)} -- plot [smooth] coordinates  { (6.35,2.6) (6,2.65) (5,2.65) (4,2.6) (3,2.4)  (2,2)} -- plot [smooth] coordinates  {(2,2) (1.5,1.5) (1,1) (0,0.5)} -- (0,0.5)-- (3,-0.5);
		\draw[color=white!60!red] plot [smooth] coordinates  { (3,-0.5) (3.5,-0.2) (5.5,-0.25)}-- plot [smooth] coordinates{(5.5,-0.25) (4.75,0.5) (5,0.8) (5.5,1.0) (6,1.1)} -- plot [smooth] coordinates  { (6,1.1) (5.6,1.5) (5.62,1.7) (5.7,1.8) (6,2) (6.5,2.1)} -- plot [smooth] coordinates  { (6.5,2.1) (6.3,2.2) (6.1,2.4) (6.1,2.5) (6.3,2.6) (6.35,2.6)} -- plot [smooth] coordinates  { (6.35,2.6) (6,2.65) (5,2.65) (4,2.6) (3,2.4)  (2,2)} -- plot [smooth] coordinates  {(2,2) (1.5,1.5) (1,1) (0,0.5)} -- (0,0.5)-- (3,-0.5);
		
		\draw (0,0.5+\th) to (3,-0.5+\th);
		\draw plot [smooth] coordinates  { (3,-0.5+\th) (3.5,-0.2+\th) (5.5,-0.25+\th)};
		\draw plot [smooth] coordinates  { (5.5,-0.25+\th) (4.75,0.5+\th) (5,0.8+\th) (5.5,1.0+\th) (6,1.1+\th)};
		\draw plot [smooth] coordinates  { (6,1.1+\th) (5.6,1.5+\th) (5.62,1.7+\th) (5.7,1.8+\th) (6,2+\th) (6.5,2.1+\th)};
		\draw plot [smooth] coordinates  { (6.5,2.1+\th) (6.3,2.2+\th) (6.1,2.4+\th) (6.1,2.5+\th) (6.3,2.6+\th) (6.35,2.6+\th)};
		\draw plot [smooth] coordinates  { (0,0.5+\th) (1,1+\th) (1.5,1.5+\th) (2,2+\th)};
		\draw plot [smooth] coordinates  { (2,2+\th) (3,2.4+\th) (4,2.6+\th) (5,2.65+\th) (6,2.65+\th) (6.35,2.6+\th)};
		
		\draw (0,0.5-\th) to (3,-0.5-\th);
		\draw plot [smooth] coordinates  { (3,-0.5-\th) (3.5,-0.2-\th) (5.5,-0.25-\th)};
		\draw[dashed] plot [smooth] coordinates  { (5.5,-0.25-\th) (4.75,0.5-\th) (5,0.8-\th) (5.5,1.0-\th) (6,1.1-\th)};
		\draw[dashed] plot [smooth] coordinates  { (6,1.1-\th) (5.6,1.5-\th) (5.62,1.7-\th) (5.7,1.8-\th) (6,2-\th) (6.5,2.1-\th)};
		\draw[dashed] plot [smooth] coordinates  { (6.5,2.1-\th) (6.3,2.2-\th) (6.1,2.4-\th) (6.1,2.5-\th) (6.3,2.6-\th) (6.35,2.6-\th)};
		\draw[dashed] plot [smooth] coordinates  { (0,0.5-\th) (1,1-\th) (1.5,1.5-\th) (2,2-\th)};
		\draw[dashed] plot [smooth] coordinates  { (2,2-\th) (3,2.4-\th) (4,2.6-\th) (5,2.65-\th) (6,2.65-\th) (6.35,2.6-\th)};
		
		\draw[-,black] (0,0.5+\th) to (0,0.5-\th);
		\draw[-,black] (3,-0.5+\th) to (3,-0.5-\th);
		\draw[-,black] (5.5,-0.25+\th) to (5.5,-0.25-\th);
		\draw[-,black] (6,1.1+\th) to (6,1.1-\th);
		\draw[-,black] (6.5,2.1+\th) to (6.5,2.1-\th);
		\draw[-,black] (6.35,2.6+\th) to (6.35,2.6-\th);
		
		\draw[blue,line width=0.5mm] (0.75,0.25) to (2.0,0.8);
		\draw[cyan,line width=0.5mm] (2.0,0.8) to (2,2);
		\draw[red!50!yellow,line width=0.5mm] (2,2) to (4,1.55);
		\draw[black!60!green,line width=0.5mm] (4,1.55) to (6,1.1);
		\draw[red!50!blue,line width=0.5mm] (2,2) to (4,2.05);
		\draw[black!60!green,line width=0.5mm] (4,2.05) to (6.5,2.1);
		\draw[red,line width=0.5mm] (2,2) to (4,2.6);
		\draw[black!60!green,line width=0.5mm] (4,2.6) to (6.35,2.6);
		\draw[red,fill=red] (0.75,0.25) circle (0.5ex);
		\draw[red,fill=red] (2.0,0.8) circle (0.5ex);
		\draw[red,fill=red] (2,2) circle (0.5ex);
		\draw[red,fill=red] (4,2.6) circle (0.5ex);
		\draw[red,fill=red] (4,1.55) circle (0.5ex);
		\draw[red,fill=red] (4,2.05) circle (0.5ex);
		
		\draw[<-,line width=0.5mm] (1.42,1.5) -- (-1,2);
		\draw (-1,2.6) node{$\mathrm{3D\ fluid-structure}$};
		\draw (-1,2.2) node {$\mathrm{interface}$};
		
		\draw[<-,line width=0.5mm] (0.7,0.5) -- (-1,0);
		\draw (-1,-0.2) node{$\mathrm{2D\ structural}$};
		\draw (-1,-0.6) node {$\mathrm{domain}$};
	
	\draw[<-] (4.5,0) -- (4.5,0.5) -- (5.5,0.5);
	\draw (6,0.5) node {$\Gamma^\mathrm{f}_\mathrm{top}$};
	\draw[<-] (4.5,-0.2-\th) -- (4.5,-0.8) -- (5.5,-0.8);
	\draw (6.2,-0.8) node {$\Gamma^\mathrm{f}_\mathrm{bottom}$};
	\draw[<-] (5.5,-0.25) -- (6,-0.25);
	\draw (6.4,-0.25) node {$\Gamma^\mathrm{f}_\mathrm{TE}$};
	\draw[<-] (1,1) -- (-0.1,1);
	\draw (-0.5,1) node {$\Gamma^\mathrm{f}_\mathrm{LE}$};
	\draw[<-] (4.5,1.7) -- (6.4,1.7);
	\draw (6.8,1.7) node {$\Gamma^\mathrm{s}$};
	
		\draw[blue,line width=0.5mm] (-0.5,-1.25) to (0,-1.25);
		\draw (0,-1.25) [anchor=west] node {$\mathrm{Humerus}$};
		\draw[cyan,line width=0.5mm] (2,-1.25) to (2.5,-1.25);
		\draw (2.5,-1.25) [anchor=west] node {$\mathrm{Radius}$};
		\draw[red!50!yellow,line width=0.5mm] (4.5,-1.25) to (5,-1.25);
		\draw (5,-1.25) [anchor=west] node {$\mathrm{Metacarpal\ V}$};
		\draw[red!50!blue,line width=0.5mm] (-0.5,-1.8) to (0,-1.8);
		\draw (0,-1.8) [anchor=west] node {$\mathrm{Metacarpal\ IV}$};
		\draw[red,line width=0.5mm] (3,-1.8) to (3.5,-1.8);
		\draw (3.5,-1.8) [anchor=west] node {$\mathrm{Metacarpal\ III}$};
		\draw[black!60!green,line width=0.5mm] (-0.5,-2.35) to (0,-2.35);
		\draw (0,-2.35) [anchor=west] node {$\mathrm{Phalanges}$};	
		\draw[fill,color=white!90!red,fill opacity=1.0] (2.2,-1.6-\ty) -- (2.2,-2-\ty) -- (2.6,-2-\ty) -- (2.6,-1.6-\ty) -- (2.2,-1.6-\ty);	
		\draw[color=white!60!red] (2.2,-1.6-\ty) -- (2.2,-2-\ty) -- (2.6,-2-\ty) -- (2.6,-1.6-\ty) -- (2.2,-1.6-\ty);	
		\draw (3.2,-2.35) node {$\mathrm{Shell}$};
		\draw[red,fill=red] (4,-1.85-\ty) circle (0.5ex);
		\draw (4.7,-2.35) node {$\mathrm{Joints}$};
\end{tikzpicture}
\caption{The structural model of the bat wing where the bone fingers and membranes are represented by multibody components such as beams (lines) and thin shells (surfaces) connected by joints (dots). The structural boundary of the interface $\Gamma^\mathrm{s} = \Gamma^\mathrm{s}_\mathrm{Beam} \cup \Gamma^\mathrm{s}_\mathrm{Shell} \cup \Gamma^\mathrm{s}_\mathrm{Joints}$ and the fluid boundary is $\Gamma^\mathrm{f} = \Gamma^\mathrm{f}_\mathrm{top} \cup \Gamma^\mathrm{f}_\mathrm{bottom} \cup \Gamma^\mathrm{f}_\mathrm{LE} \cup \Gamma^\mathrm{f}_\mathrm{TE}$.}
\label{multibody_bat}
\end{figure}

We next demonstrate the developed flexible multibody aeroelastic framework to the application of flapping dynamics of a bat. As discussed in the introduction, the wing of a bat is a prime example of a multibody system consisting of bone fingers connected via joints of varying degrees of freedom. These joints with bones form a skeleton for the flexible membrane of the wing as shown in Fig. \ref{multibody_bat}. The bones consist of humerus, radius, metacarpals and phalanges. In a bat flight, the joints (shown as dots in the figure) can have multiple degrees of freedom and allows for active and passive movements. This complex kinematics of the bat wing makes its computational modeling a challenge. Moreover, changing the structural properties of the bones as well as thin membranes in the spanwise and chordwise directions of the wing adds to the modeling complexity. 

In this work, we presented the developed flexible multibody aeroelastic formulation for such complex problems. With the help of the framework, we now demonstrate the flapping flight of a bat with passive movement at the joints considering the multiple components by modeling the bones as Euler-Bernoulli beams and the membranes as thin shell structures. There are many joints in the actual bat wing, but we have simplified the model by considering only the major 6 joints, as shown in Fig. \ref{multibody_bat}. All the joints are modeled as revolute joints. The geometric model of the bat is based on the Pallas' long tongued bat species \textit{Glossophaga soricina}. The physical parameters of the geometry of the bat are given in Table \ref{geometry_bat}.
\renewcommand{\arraystretch}{0.5}
\begin{table}[!h]
	\caption{Physical parameters of the geometry of the Pallas' long tongued bat \textit{Glossophaga soricina}.}
	\centering
	\begin{tabular}{  M{5cm} | M{3cm} N }
		\hline
		\centering
		\textbf{Parameter} & \textbf{Value} &\\[10pt]
		\hline
		Wing span ($b$) & $0.2369$ $\mathrm{m}$ &\\[10pt]
		\centering
		Chord length at root ($c_\mathrm{root}$) & $0.0384$ $\mathrm{m}$ &\\[10pt]
		
		\centering
		Wing surface area ($S$) & $8.3217\times 10^{-3}$ $\mathrm{m}^2$ &\\[10pt]
		
		\centering
		Mean chord length ($c=S/b$) & $0.03512$ $\mathrm{m}$ &\\[10pt]		
		\hline
	\end{tabular}
	\label{geometry_bat}
\end{table}

\begin{figure}[!htbp]
	\centering
	\begin{tikzpicture}[very thick,decoration={markings,mark=at position 0.5 with {\arrow{>}}},scale=1]
	\draw (0,0) node[left]{} -- (0,6) node[right]{} -- (9,6) node[above]{} -- (9,0) node[above]{} -- cycle;
	\draw[black,dotted] (1,1) to (1,7);
	\draw[black] (1,7) to (10,7);
	\draw[black] (10,7) to (10,1);
	\draw[black,dotted] (10,1) to (1,1);
	
	\draw[black] (0,6) to (1,7);
	\draw[black] (9,6) to (10,7);
	\draw[black] (9,0) to (10,1);
	\draw[black,dotted] (0,0) to (1,1);
	
	\node (bat) at (4.8,3.25)
	{\includegraphics[trim={1cm 1cm 1cm 1cm},clip,width=.3\textwidth]{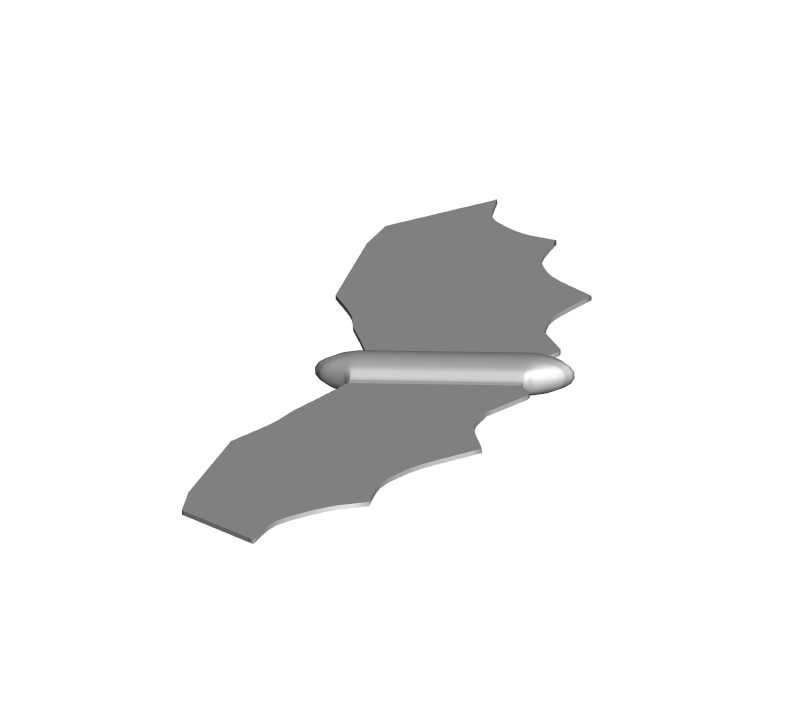}};

	\draw (4.45,3.15) to (4.45,2);
	\draw (5.6,3.15) to (5.6,2);
	\draw[<->] (4.45,2.1) to (5.6,2.1);
	\draw (4.95,2.05) node[anchor=north]{$c_\mathrm{root}$};
	
	\draw[dotted] (-2.5,2.5) to (-0.5,2.5);
	\draw[->] (-2.5,2.5) to (-0.5,3.5);
	\draw (-1.5,2.5) arc (0:25:1cm);
	\draw (-1.5,4.5) node[anchor=north]{Inflow};
	\draw (-1.5,4) node[anchor=north]{$|\boldsymbol{u}^\mathrm{f}|=U_\infty$};
	\draw (-1.05,3.1) node[anchor=north]{$AOA$};
	
	\draw (10.2,1) to (10.6,1);
	\draw (10.2,7) to (10.6,7);
	\draw[<->] (10.4,1) to (10.4,7);
	\draw (12,4.5) node[anchor=north]{Outflow};
	\draw (12,3.8) node[anchor=north]{$\boldsymbol{\sigma}^\mathrm{f}\cdot\mathbf{n}^\mathrm{f} = \boldsymbol{0}$};
	\draw (12,3.3) node[anchor=north]{$\nabla\tilde{\nu}\cdot\mathbf{n}^\mathrm{f} = 0$};
	\draw (10.4,4) node[anchor=west]{$H$};
	
	\draw (5,6.75) node[anchor=north]{Periodic};
	
	\draw (5,0.75) node[anchor=north]{Periodic};
	
	\draw[<-] (5.25,4) to (5.75,4.75);
	\draw (5.75,4.75) to (6.5,4.75);
	\draw (6.5,4.75) node[anchor=west]{No-slip};
	\draw (6.4,4.4) node[anchor=west] {($\boldsymbol{u}^\mathrm{f}=\boldsymbol{0}$, $\tilde{\nu}=0$)};
	
	\draw (0,-0.2) to (0,-0.6);
	\draw (9,-0.2) to (9,-0.6);
	\draw (4.5,-0.4) node[anchor=north]{$L$};
	\draw[<->] (0,-0.4) to (9,-0.4);
	\draw (9.2,0) to (9.6,0);
	\draw[<->] (9.4,0) to (10.4,1);
	\draw (9.9,0.4) node[anchor=west]{$B$};
	
	\draw[->,red] (2,4.5) to (3,4.5);
	\draw (3,4.5) node[anchor=west,red]{X};
	\draw[->,red] (2,4.5) to (2,5.5);
	\draw (2,5.5) node[anchor=west,red]{Z};
	\draw[->,red] (2,4.5) to (2.75,5);
	\draw (2.6,5) node[anchor=west,red]{Y};
	
	\end{tikzpicture}
	\caption{A schematic of  the flow past a bat. The computational setup and boundary conditions are shown for the turbulent Navier-Stokes equations. Here, ${\boldsymbol{u}}^\mathrm{f}$ denotes the fluid velocity, $\theta$ is the angle of attack at the inflow boundary, $L$, $B$, $H$ are the length, breadth and height of the computational domain respectively, and $c_\mathrm{root}$ is the chord at the root of the bat wing.} 
	\label{fig:schematics_bat}
\end{figure}
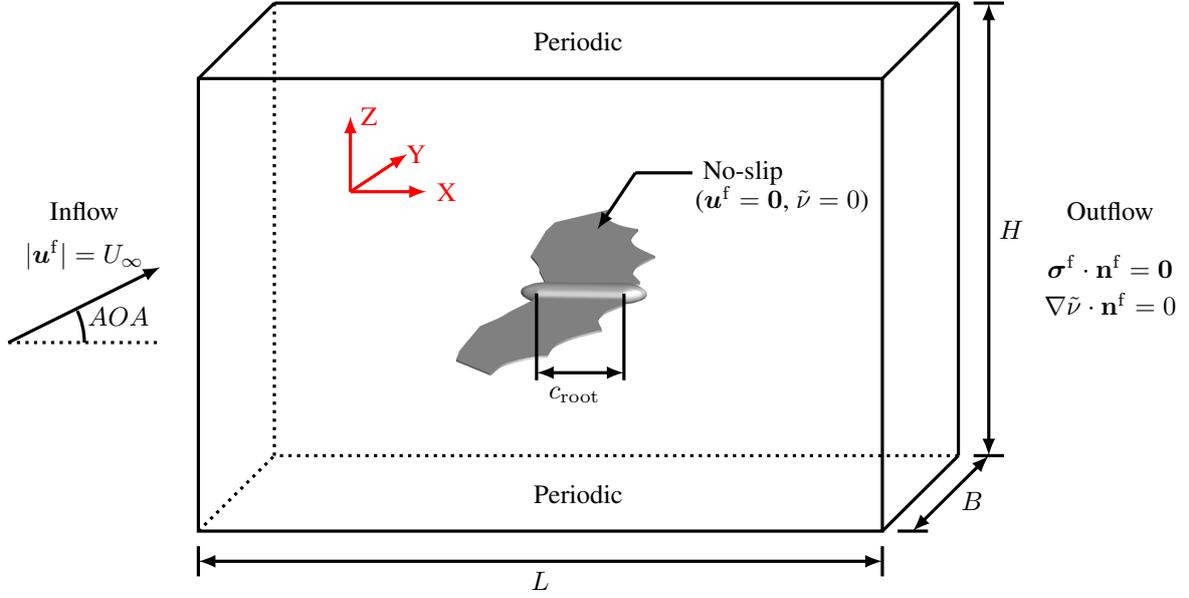

\subsection{Computational domain and key fluid-structure parameters}
The computational domain for the numerical experiment is shown in Fig. \ref{fig:schematics_bat}. It consists of a cuboid domain with dimensions $L\times B\times H$, where $L\approx 65c$, $B\approx 68c$ and $H\approx 68c$, with $c$ being the mean chord of the bat wing. A periodic boundary condition is imposed on the top and bottom boundaries while a symmetric slip condition is satisfied on the side boundaries parallel to the $X$-$Z$ plane. The inflow and outflow boundaries are at a distance of about $30c$ and $35c$ from the center of the bat body respectively. The no-slip boundary condition is satisfied on the body of the bat and both wings. A Dirichlet boundary condition is imposed on the inflow boundary with a velocity of $\boldsymbol{u}^\mathrm{f} = (U_\infty \mathrm{cos}(AOA), 0, U_\infty \mathrm{sin}(AOA))$, where $AOA = 10^{\circ}$ is the angle of attack for the flow across the bat wings. A stress-free boundary condition is satisfied at the outflow boundary for the flow variables and $\nabla\tilde{\nu}\cdot\mathbf{n}^\mathrm{f} = 0$ is satisfied for the turbulence eddy viscosity $\tilde{\nu}$.

The flapping motion to the wings is prescribed at the revolute joint where the Humerus bone is connected with the body. The prescribed motion is a sinusoidal rotational motion along the $X$-axis with an amplitude of $\theta_\mathrm{max} = 25^{\circ}$ and frequency of $1$ Hz. The reference velocity $U_\mathrm{ref}$ is selected as the magnitude of the inflow velocity, $U_\infty$. Thus, the non-dimensional frequency of rotation, $fc/U_\mathrm{ref} = 0.03512$, which gives a non-dimensional time period of $T_{w}U_\mathrm{ref}/c=28.5$, where $T_w$ is the flapping time period.

\renewcommand{\arraystretch}{0.5}
\begin{table}[!h]
	\caption{Anisotropic elastic properties for the bone fingers (modeled as beam) in the flexible wing.}
	\centering
	\begin{tabular}{  M{3cm} | M{3.5cm} | M{3cm} N }
		\hline
		\centering
		\textbf{Component} & \textbf{Flexural rigidity ($EI$)} & \textbf{Length ($l$)} &\\[6pt]
		& (Nm$^2$) & (cm) &\\[10pt]
		\hline
		 Humerus & $1.56\times 10^{-3}$ & $1.55$  &\\[10pt]
		\centering
		 Radius & $1.31\times 10^{-3}$ & $3.00$ &\\[10pt]	
		 \centering
		 Metacarpal V & $0.12\times 10^{-3}$ & $2.05$  &\\[10pt]
		 \centering
		 Metacarpal IV & $0.98\times 10^{-3}$ & $2.84$ &\\[10pt]
		 \centering
		 Metacarpal III & $0.23\times 10^{-3}$ & $4.62$  &\\[10pt]
		 \centering
		 Phalanx (digit V) & $0.04\times 10^{-3}$ & $3.29$  &\\[10pt]
		 \centering
		 Phalanx (digit IV) & $0.04\times 10^{-3}$ & $2.74$ &\\[10pt]
		 \centering
		 Phalanx (digit III) & $0.04\times 10^{-3}$ & $2.67$  &\\[10pt]
		\hline
	\end{tabular}
	\label{par_diff_beam}
\end{table} 
The Reynolds number for the study is defined as $Re = \rho^\mathrm{f}U_\mathrm{ref}c/\mu^\mathrm{f} = 12,000$. The bone fingers, modeled as Euler-Bernoulli beams, have varying flexural rigidity $EI$ and length $l$, where $E$ and $I$ are the Young's modulus and second moment of area of the cross-section of the beam. The cross-section of the beam is assumed to be circular with a diameter of $0.5$ cm and density of the beam is $\rho^\mathrm{s} = 2200$ kg/m$^3$ while evaluating the parameters. The properties of the different bone fingers are given in Table \ref{par_diff_beam}, which are based on the experimental estimates in \cite{Swartz2008}. The structural properties of the flexible membranes enveloping the skeleton of bones and joints are listed in Table \ref{par_diff_membrane}. 
\renewcommand{\arraystretch}{0.5}
\begin{table}[!h]
	\caption{Anisotropic elastic properties for the flexible membrane (modeled as thin shells) in the flexible wing.}
	\centering
	\begin{tabular}{  M{2cm} | M{3.5cm} | M{2.5cm}| M{2cm}| M{3cm} N }
		\hline
		\centering
		\textbf{Component} & \textbf{Young's modulus ($E$)} & \textbf{Thickness ($h$)} & \textbf{Density ($\rho^\mathrm{s}$)} & \textbf{Poisson's ratio ($\nu^\mathrm{s}$)} &\\[10pt]
		& (N/m$^2$) & (cm) & (kg/m$^3$) & &\\[10pt]
		\hline
		Membrane & $7\times 10^{5}$ & $0.046$ & $373.05$ & $0.33$  &\\[10pt]
		\hline
	\end{tabular}
	\label{par_diff_membrane}
\end{table} 

All the dimensional variables presented in the subsequent sections are non-dimensionalized by the reference velocity $U_\mathrm{ref}$, mean chord length $c$, fluid density $\rho^\mathrm{f}$ and the flapping time period $T_w$ as required. The aerodynamic coefficients are quantified as
\begin{align}
C_L &= \frac{1}{\frac{1}{2}\rho^\mathrm{f}U_\mathrm{ref}^2 S} \int_\Gamma (\bar{\boldsymbol{\sigma}}^\mathrm{f} \cdot \mathbf{n})\cdot\mathbf{n}_z \mathrm{d}\Gamma, \label{Cl}\\
C_D &= \frac{1}{\frac{1}{2}\rho^\mathrm{f}U_\mathrm{ref}^2 S} \int_\Gamma (\bar{\boldsymbol{\sigma}}^\mathrm{f} \cdot \mathbf{n})\cdot\mathbf{n}_x \mathrm{d}\Gamma, \label{Cd}
\end{align}
where $\mathbf{n}$ is the normal to the surface of the bat. $\overline{X}$, $\overline{X}_\mathrm{,max}$ and $\overline{X}_\mathrm{,rms}$ denote the mean, maximum and the root mean square from the mean of the coefficient $X$.

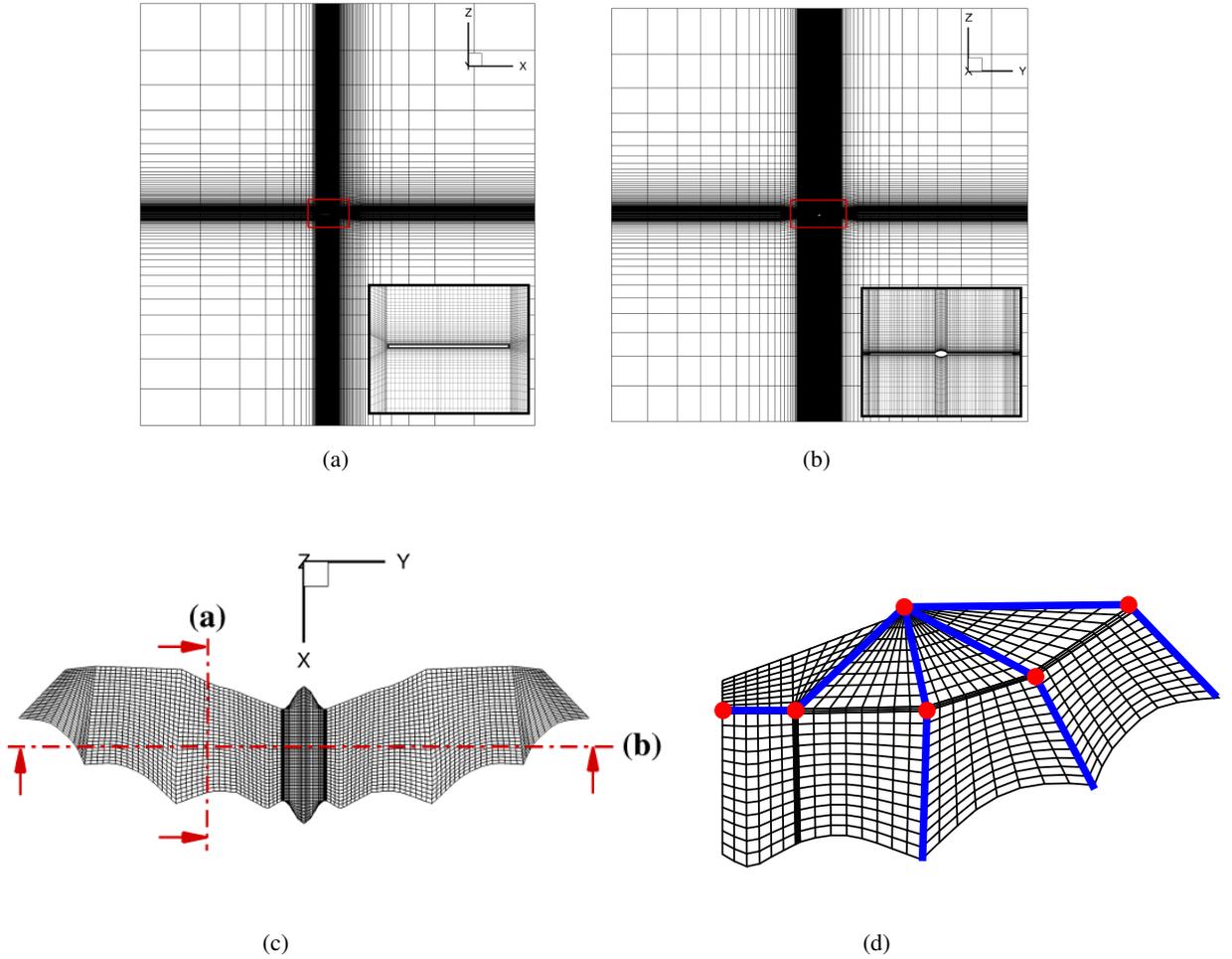
\begin{figure}[!h]
	\centering
	\begin{subfigure}[b]{0.4\textwidth}
		\includegraphics[trim={0.1cm 0.2cm 0.1cm 0.1cm},clip,width=6.5cm]{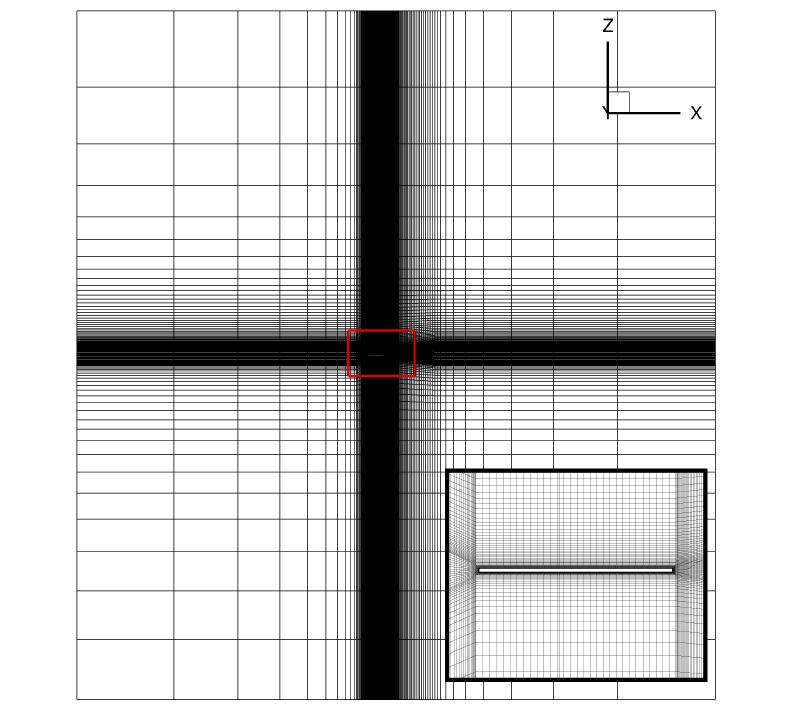}
		\caption{}
	\end{subfigure}%
	\begin{subfigure}[b]{0.4\textwidth}
		\includegraphics[trim={0.1cm 0.2cm 0.1cm 0.1cm},clip,width=6.5cm]{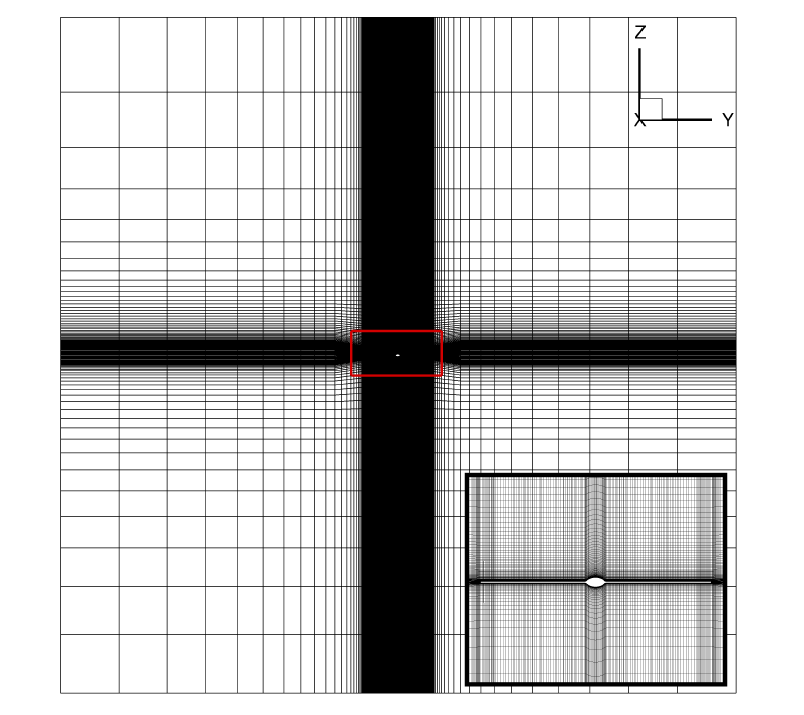}
		\caption{}
	\end{subfigure}%
	
	\begin{subfigure}[b]{0.5\textwidth}
		\centering
		\includegraphics[trim={3cm 5cm 3cm 6.5cm},clip,width=9.5cm]{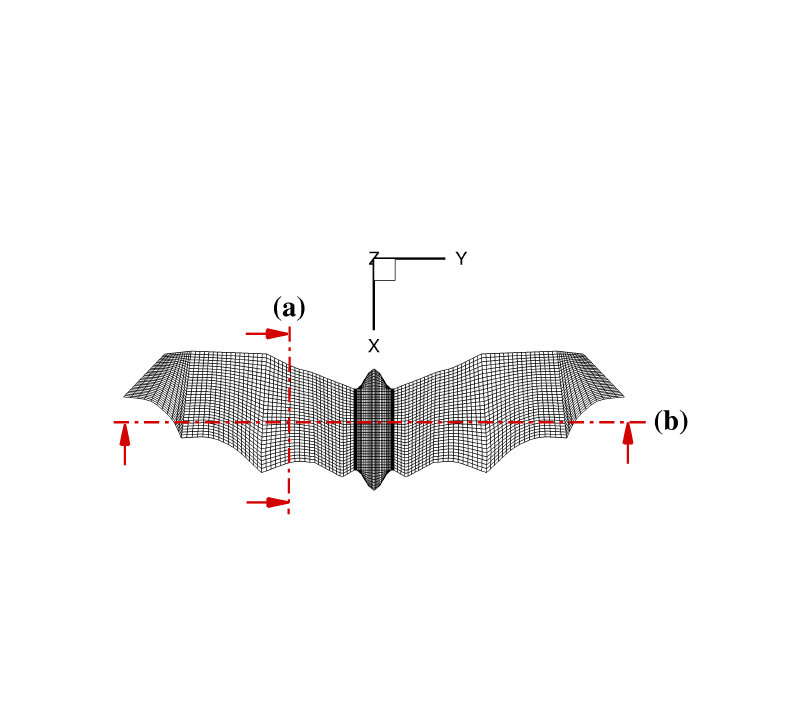}
		\caption{}
	\end{subfigure}%
	\begin{subfigure}[b]{0.5\textwidth}
		\hspace{1cm}
		\begin{tikzpicture}[very thick,decoration={markings,mark=at position 0.5 with {\arrow{>}}},scale=0.6]
		\node (bat) at (0,0)
		{\includegraphics[trim={15cm 0.1cm 20cm 0.1cm},clip,width=0.6\textwidth,angle=-90]{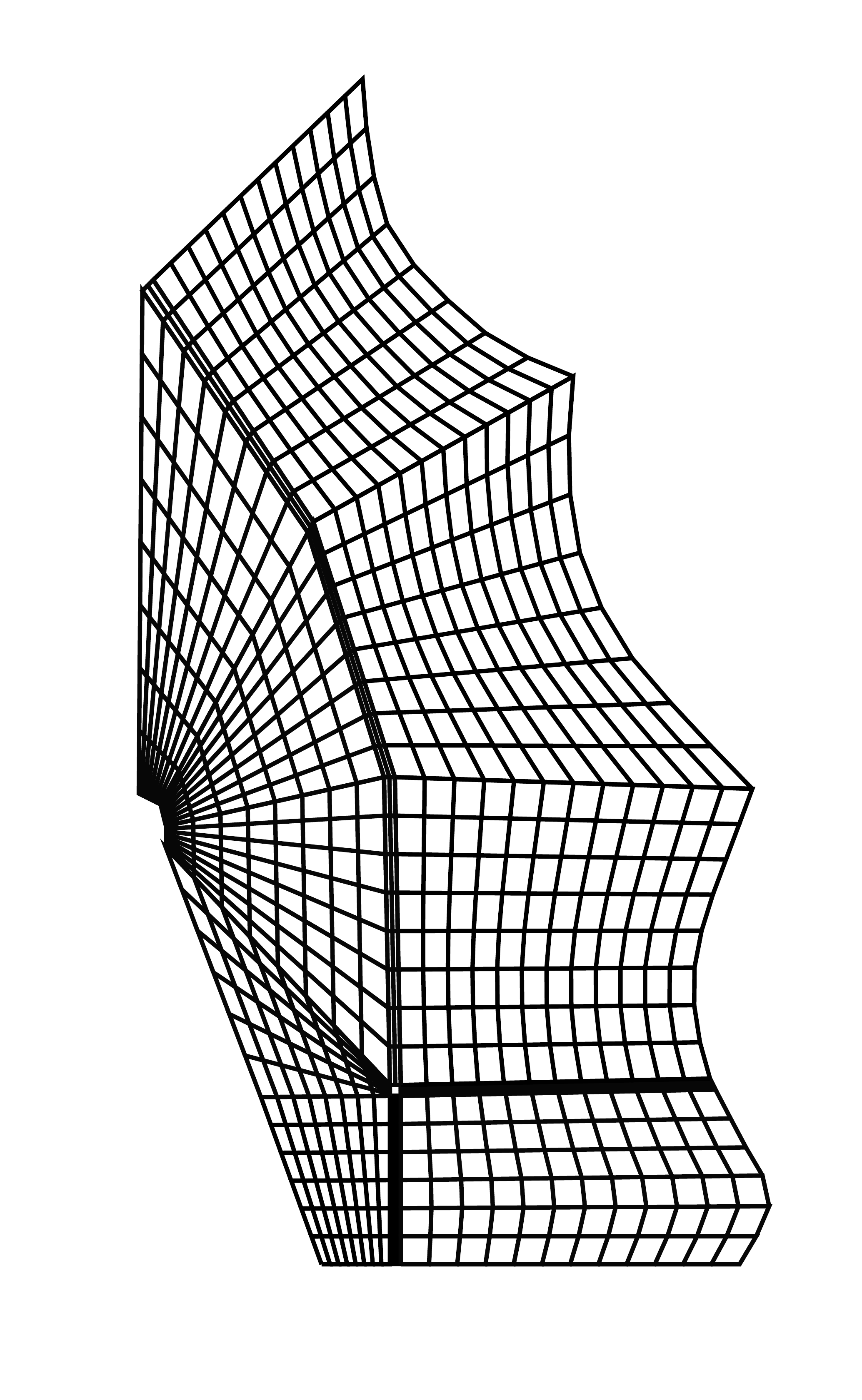}};
		
		\draw[blue,line width=1mm] (-5.2,0.35) to (-3.6,0.35);
		\draw[blue,line width=1mm] (-3.6,0.35) to (-1.2,2.65);
		\draw[blue,line width=1mm] (-1.2,2.65) to (-0.7,0.35);
		\draw[blue,line width=1mm] (-1.2,2.65) to (1.7,1.1);
		\draw[blue,line width=1mm] (-1.2,2.65) to (3.75,2.7);
		\draw[blue,line width=1mm] (1.7,1.1) to (3,-1.4);
		\draw[blue,line width=1mm] (-0.7,0.35) to (-0.8,-3);
		\draw[blue,line width=1mm] (3.75,2.7) to (5.7,0.65);
		
		\draw[red,fill=red] (-3.6,0.35) circle (1ex);
		\draw[red,fill=red] (-1.2,2.65) circle (1ex);
		\draw[red,fill=red] (1.7,1.1) circle (1ex);
		\draw[red,fill=red] (-0.7,0.35) circle (1ex);
		\draw[red,fill=red] (3.75,2.7) circle (1ex);
		\draw[red,fill=red] (-5.2,0.35) circle (1ex);
		\end{tikzpicture}
		\caption{}
	\end{subfigure}%
	\caption{Forward flight of a bat: Three-dimensional fluid mesh in the (a) $X-Z$ plane, (b) $Y-Z$ plane, (c) bat body; (d) Two-dimensional structural mesh consisting of multibody components as beams, shells and revolute joints for the right wing.}
	\label{bat_mesh}
\end{figure}
\subsection{Discretization characteristics}
\begin{figure}[!htbp]
\centering
\def\th2{4.5}
\def\tha{0.1}
\def\thh{4.5}
\def\tx{6}
\def\ty{4}
\begin{tikzpicture}[decoration={markings,mark=at position 0.5 with {\arrow[scale=2]{>}}},scale=0.65]
	\draw[fill,color=white!90!red,fill opacity=1] plot [smooth] coordinates  { (3,-0.5) (3.5,-0.2) (5.5,-0.25)}-- plot [smooth] coordinates{(5.5,-0.25) (4.75,0.5) (5,0.8) (5.5,1.0) (6,1.1)} -- plot [smooth] coordinates  { (6,1.1) (5.6,1.5) (5.62,1.7) (5.7,1.8) (6,2) (6.5,2.1)} -- plot [smooth] coordinates  { (6.5,2.1) (6.3,2.2) (6.1,2.4) (6.1,2.5) (6.3,2.6) (6.35,2.6)} -- plot [smooth] coordinates  { (6.35,2.6) (6,2.65) (5,2.65) (4,2.6) (3,2.4)  (2,2)} -- plot [smooth] coordinates  {(2,2) (1.5,1.5) (1,1) (0,0.5)} -- (0,0.5)-- (3,-0.5);
	\draw[color=white!60!red] plot [smooth] coordinates  { (3,-0.5) (3.5,-0.2) (5.5,-0.25)}-- plot [smooth] coordinates{(5.5,-0.25) (4.75,0.5) (5,0.8) (5.5,1.0) (6,1.1)} -- plot [smooth] coordinates  { (6,1.1) (5.6,1.5) (5.62,1.7) (5.7,1.8) (6,2) (6.5,2.1)} -- plot [smooth] coordinates  { (6.5,2.1) (6.3,2.2) (6.1,2.4) (6.1,2.5) (6.3,2.6) (6.35,2.6)} -- plot [smooth] coordinates  { (6.35,2.6) (6,2.65) (5,2.65) (4,2.6) (3,2.4)  (2,2)} -- plot [smooth] coordinates  {(2,2) (1.5,1.5) (1,1) (0,0.5)} -- (0,0.5)-- (3,-0.5);
	\draw[fill,pattern=hatch, hatch angle=117, hatch size=10pt] plot [smooth] coordinates  { (3,-0.5) (3.5,-0.2) (5.5,-0.25)}-- plot [smooth] coordinates{(5.5,-0.25) (4.75,0.5) (5,0.8) (5.5,1.0) (6,1.1)} -- plot [smooth] coordinates  { (6,1.1) (5.6,1.5) (5.62,1.7) (5.7,1.8) (6,2) (6.5,2.1)} -- plot [smooth] coordinates  { (6.5,2.1) (6.3,2.2) (6.1,2.4) (6.1,2.5) (6.3,2.6) (6.35,2.6)} -- plot [smooth] coordinates  { (6.35,2.6) (6,2.65) (5,2.65) (4,2.6) (3,2.4)  (2,2)} -- plot [smooth] coordinates  {(2,2) (1.5,1.5) (1,1) (0,0.5)} -- (0,0.5)-- (3,-0.5);

	\draw plot [smooth] coordinates  { (3,-0.5+\th2) (3.5,-0.2+\th2) (5.5,-0.25+\th2)}-- plot [smooth] coordinates{(5.5,-0.25+\th2) (4.75,0.5+\th2) (5,0.8+\th2) (5.5,1.0+\th2) (6,1.1+\th2)} -- plot [smooth] coordinates  { (6,1.1+\th2) (5.6,1.5+\th2) (5.62,1.7+\th2) (5.7,1.8+\th2) (6,2+\th2) (6.5,2.1+\th2)} -- plot [smooth] coordinates  { (6.5,2.1+\th2) (6.3,2.2+\th2) (6.1,2.4+\th2) (6.1,2.5+\th2) (6.3,2.6+\th2) (6.35,2.6+\th2)} -- plot [smooth] coordinates  { (6.35,2.6+\th2) (6,2.65+\th2) (5,2.65+\th2) (4,2.6+\th2) (3,2.4+\th2)  (2,2+\th2)} -- plot [smooth] coordinates  {(2,2+\th2) (1.5,1.5+\th2) (1,1+\th2) (0,0.5+\th2)} -- (0,0.5+\th2)-- (3,-0.5+\th2);
	\draw[fill,pattern=hatch, hatch angle=117, hatch size=5pt] plot [smooth] coordinates  { (3,-0.5+\th2) (3.5,-0.2+\th2) (5.5,-0.25+\th2)}-- plot [smooth] coordinates{(5.5,-0.25+\th2) (4.75,0.5+\th2) (5,0.8+\th2) (5.5,1.0+\th2) (6,1.1+\th2)} -- plot [smooth] coordinates  { (6,1.1+\th2) (5.6,1.5+\th2) (5.62,1.7+\th2) (5.7,1.8+\th2) (6,2+\th2) (6.5,2.1+\th2)} -- plot [smooth] coordinates  { (6.5,2.1+\th2) (6.3,2.2+\th2) (6.1,2.4+\th2) (6.1,2.5+\th2) (6.3,2.6+\th2) (6.35,2.6+\th2)} -- plot [smooth] coordinates  { (6.35,2.6+\th2) (6,2.65+\th2) (5,2.65+\th2) (4,2.6+\th2) (3,2.4+\th2)  (2,2+\th2)} -- plot [smooth] coordinates  {(2,2+\th2) (1.5,1.5+\th2) (1,1+\th2) (0,0.5+\th2)} -- (0,0.5+\th2)-- (3,-0.5+\th2);
	\draw plot [smooth] coordinates  { (3,-0.5-\thh) (3.5,-0.2-\thh) (5.5,-0.25-\thh)}-- plot [smooth] coordinates{(5.5,-0.25-\thh) (4.75,0.5-\thh) (5,0.8-\thh) (5.5,1.0-\thh) (6,1.1-\thh)} -- plot [smooth] coordinates  { (6,1.1-\thh) (5.6,1.5-\thh) (5.62,1.7-\thh) (5.7,1.8-\thh) (6,2-\thh) (6.5,2.1-\thh)} -- plot [smooth] coordinates  { (6.5,2.1-\thh) (6.3,2.2-\thh) (6.1,2.4-\thh) (6.1,2.5-\thh) (6.3,2.6-\thh) (6.35,2.6-\thh)} -- plot [smooth] coordinates  { (6.35,2.6-\thh) (6,2.65-\thh) (5,2.65-\thh) (4,2.6-\thh) (3,2.4-\thh)  (2,2-\thh)} -- plot [smooth] coordinates  {(2,2-\thh) (1.5,1.5-\thh) (1,1-\thh) (0,0.5-\thh)} -- (0,0.5-\thh)-- (3,-0.5-\thh);
	\draw[fill,pattern=hatch, hatch angle=117, hatch size=5pt] plot [smooth] coordinates  { (3,-0.5-\thh) (3.5,-0.2-\thh) (5.5,-0.25-\thh)}-- plot [smooth] coordinates{(5.5,-0.25-\thh) (4.75,0.5-\thh) (5,0.8-\thh) (5.5,1.0-\thh) (6,1.1-\thh)} -- plot [smooth] coordinates  { (6,1.1-\thh) (5.6,1.5-\thh) (5.62,1.7-\thh) (5.7,1.8-\thh) (6,2-\thh) (6.5,2.1-\thh)} -- plot [smooth] coordinates  { (6.5,2.1-\thh) (6.3,2.2-\thh) (6.1,2.4-\thh) (6.1,2.5-\thh) (6.3,2.6-\thh) (6.35,2.6-\thh)} -- plot [smooth] coordinates  { (6.35,2.6-\thh) (6,2.65-\thh) (5,2.65-\thh) (4,2.6-\thh) (3,2.4-\thh)  (2,2-\thh)} -- plot [smooth] coordinates  {(2,2-\thh) (1.5,1.5-\thh) (1,1-\thh) (0,0.5-\thh)} -- (0,0.5-\thh)-- (3,-0.5-\thh);

	\draw plot [smooth] coordinates  { (6.35-\tx,2.6+\tha) (6-\tx,2.65+\tha) (5-\tx,2.65+\tha) (4-\tx,2.6+\tha) (3-\tx,2.4+\tha)  (2-\tx,2+\tha)} -- plot [smooth] coordinates  {(2-\tx,2+\tha) (1.5-\tx,1.5+\tha) (1-\tx,1+\tha) (0-\tx,0.5+\tha)} -- (0-\tx,0.5+\tha)-- (0-\tx,0.5-\tha) -- plot [smooth] coordinates  {(0-\tx,0.5-\tha) (1-\tx,1-\tha) (1.5-\tx,1.5-\tha) (2-\tx,2-\tha)} -- plot [smooth] coordinates  {(2-\tx,2-\tha) (3-\tx,2.4-\tha) (4-\tx,2.6-\tha) (5-\tx,2.65-\tha) (6-\tx,2.65-\tha) (6.35-\tx,2.6-\tha)} -- (6.35-\tx,2.6-\tha) -- (6.35-\tx,2.6+\tha) ;
	\draw[fill,pattern=hatch, hatch angle=45, hatch size=5pt] plot [smooth] coordinates  { (6.35-\tx,2.6+\tha) (6-\tx,2.65+\tha) (5-\tx,2.65+\tha) (4-\tx,2.6+\tha) (3-\tx,2.4+\tha)  (2-\tx,2+\tha)} -- plot [smooth] coordinates  {(2-\tx,2+\tha) (1.5-\tx,1.5+\tha) (1-\tx,1+\tha) (0-\tx,0.5+\tha)} -- (0-\tx,0.5+\tha)-- (0-\tx,0.5-\tha) -- plot [smooth] coordinates  {(0-\tx,0.5-\tha) (1-\tx,1-\tha) (1.5-\tx,1.5-\tha) (2-\tx,2-\tha)} -- plot [smooth] coordinates  {(2-\tx,2-\tha) (3-\tx,2.4-\tha) (4-\tx,2.6-\tha) (5-\tx,2.65-\tha) (6-\tx,2.65-\tha) (6.35-\tx,2.6-\tha)} -- (6.35-\tx,2.6-\tha) -- (6.35-\tx,2.6+\tha) ;

	\draw plot [smooth] coordinates  { (3+\ty,-0.5+\tha) (3.5+\ty,-0.2+\tha) (5.5+\ty,-0.25+\tha)}-- plot [smooth] coordinates{(5.5+\ty,-0.25+\tha) (4.75+\ty,0.5+\tha) (5+\ty,0.8+\tha) (5.5+\ty,1.0+\tha) (6+\ty,1.1+\tha)} -- plot [smooth] coordinates  { (6+\ty,1.1+\tha) (5.6+\ty,1.5+\tha) (5.62+\ty,1.7+\tha) (5.7+\ty,1.8+\tha) (6+\ty,2+\tha) (6.5+\ty,2.1+\tha)} -- plot [smooth] coordinates  { (6.5+\ty,2.1+\tha) (6.3+\ty,2.2+\tha) (6.1+\ty,2.4+\tha) (6.1+\ty,2.5+\tha) (6.3+\ty,2.6+\tha) (6.35+\ty,2.6+\tha)} -- (6.35+\ty,2.6+\tha) -- (6.35+\ty,2.6-\tha) -- plot [smooth] coordinates  { (6.35+\ty,2.6-\tha) (6.3+\ty,2.6-\tha) (6.1+\ty,2.5-\tha) (6.1+\ty,2.4-\tha) (6.3+\ty,2.2-\tha) (6.5+\ty,2.1-\tha)}-- plot [smooth] coordinates  { (6.5+\ty,2.1-\tha) (6+\ty,2-\tha) (5.7+\ty,1.8-\tha) (5.62+\ty,1.7-\tha) (5.6+\ty,1.5-\tha) (6+\ty,1.1-\tha)} -- plot [smooth] coordinates{(6+\ty,1.1-\tha) (5.5+\ty,1.0-\tha) (5+\ty,0.8-\tha) (4.75+\ty,0.5-\tha) (5.5+\ty,-0.25-\tha)} -- plot [smooth] coordinates  { (5.5+\ty,-0.25-\tha) (3.5+\ty,-0.2-\tha) (3+\ty,-0.5-\tha)} -- (3+\ty,-0.5-\tha) -- (3+\ty,-0.5+\tha)   ;
	\draw[fill,pattern=hatch, hatch angle=45, hatch size=5pt] plot [smooth] coordinates  { (3+\ty,-0.5+\tha) (3.5+\ty,-0.2+\tha) (5.5+\ty,-0.25+\tha)}-- plot [smooth] coordinates{(5.5+\ty,-0.25+\tha) (4.75+\ty,0.5+\tha) (5+\ty,0.8+\tha) (5.5+\ty,1.0+\tha) (6+\ty,1.1+\tha)} -- plot [smooth] coordinates  { (6+\ty,1.1+\tha) (5.6+\ty,1.5+\tha) (5.62+\ty,1.7+\tha) (5.7+\ty,1.8+\tha) (6+\ty,2+\tha) (6.5+\ty,2.1+\tha)} -- plot [smooth] coordinates  { (6.5+\ty,2.1+\tha) (6.3+\ty,2.2+\tha) (6.1+\ty,2.4+\tha) (6.1+\ty,2.5+\tha) (6.3+\ty,2.6+\tha) (6.35+\ty,2.6+\tha)} -- (6.35+\ty,2.6+\tha) -- (6.35+\ty,2.6-\tha) -- plot [smooth] coordinates  { (6.35+\ty,2.6-\tha) (6.3+\ty,2.6-\tha) (6.1+\ty,2.5-\tha) (6.1+\ty,2.4-\tha) (6.3+\ty,2.2-\tha) (6.5+\ty,2.1-\tha)}-- plot [smooth] coordinates  { (6.5+\ty,2.1-\tha) (6+\ty,2-\tha) (5.7+\ty,1.8-\tha) (5.62+\ty,1.7-\tha) (5.6+\ty,1.5-\tha) (6+\ty,1.1-\tha)} -- plot [smooth] coordinates{(6+\ty,1.1-\tha) (5.5+\ty,1.0-\tha) (5+\ty,0.8-\tha) (4.75+\ty,0.5-\tha) (5.5+\ty,-0.25-\tha)} -- plot [smooth] coordinates  { (5.5+\ty,-0.25-\tha) (3.5+\ty,-0.2-\tha) (3+\ty,-0.5-\tha)} -- (3+\ty,-0.5-\tha) -- (3+\ty,-0.5+\tha);
	\draw[-,black] (5.5+\ty,-0.25+\tha) to (5.5+\ty,-0.25-\tha);
	\draw[-,black] (6+\ty,1.1+\tha) to (6+\ty,1.1-\tha);
	\draw[-,black] (6.5+\ty,2.1+\tha) to (6.5+\ty,2.1-\tha);
	\draw[-,black] (6.35+\ty,2.6+\tha) to (6.35+\ty,2.6-\tha);

	\draw[blue,line width=0.5mm] (0.75,0.25) to (2.0,0.8);
	\draw[blue,line width=0.5mm] (2.0,0.8) to (2,2);
	\draw[blue,line width=0.5mm] (2,2) to (6,1.1);
	\draw[blue,line width=0.5mm] (2,2) to (6.5,2.1);
	\draw[blue,line width=0.5mm] (2,2) to (4,2.6) to (6.35,2.6);	
	\draw[red,fill=red] (0.75,0.25) circle (0.5ex);
	\draw[red,fill=red] (2.0,0.8) circle (0.5ex);
	\draw[red,fill=red] (2,2) circle (0.5ex);
	\draw[red,fill=red] (4,2.6) circle (0.5ex);
	\draw[red,fill=red] (4,1.55) circle (0.5ex);
	\draw[red,fill=red] (4,2.05) circle (0.5ex);

	\draw[red,->,line width=0.5mm] (-1,0) arc (-90:-160:1.5);
	\draw[blue,dashed,<-,line width=0.5mm] (0,1) arc (30:100:1.5);

	\draw[red,->,line width=0.5mm] (2.7,2.8) arc (-110:-180:1.5);
	\draw[blue,dashed,<-,line width=0.5mm] (6.5,3) arc (-10:60:1.5);

	\draw[red,->,line width=0.5mm] (7,2.2) arc (110:40:1.5);
	\draw[blue,dashed,->,line width=0.5mm] (8,1) arc (-70:-140:1.5);

	\draw[blue,dashed,->,line width=0.5mm] (2.7,-2) arc (-110:-180:1.5);
	\draw[red,<-,line width=0.5mm] (6.5,-2) arc (-10:60:1.5);
	\draw (6.5,6) node {$\Gamma^\mathrm{f}_\mathrm{top}$};
	\draw (10.5,1.5) node {$\Gamma^\mathrm{f}_\mathrm{TE}$};
	\draw (-3.5,1.5) node {$\Gamma^\mathrm{f}_\mathrm{LE}$};
	\draw (6.5,-4) node {$\Gamma^\mathrm{f}_\mathrm{bottom}$};
	\draw (3.4,-0.7) node {$\Gamma^\mathrm{s}$};
	\draw (-1.2,1) node {$\mathbf{RBF}$};
	\draw (4.6,3.5) node {$\mathbf{RBF}$};
	\draw (7.5,1.6) node {$\mathbf{RBF}$};
	\draw (4.2,-1.25) node {$\mathbf{RBF}$};
	\draw[red,->,line width=0.5mm] (-7,-2) -- (-6,-2);
	\draw (-5.8,-2) node [anchor=west]{$\mathrm{Displacement\ transfer}$};
	\draw[blue,dashed,->,line width=0.5mm] (-7,-3) -- (-6,-3);
	\draw (-5.8,-3) node [anchor=west]{$\mathrm{Force\ transfer}$};
	\draw (-7.5,-1.5) -- (0,-1.5) -- (0,-3.5) -- (-7.5,-3.5) -- (-7.5,-1.5);

\end{tikzpicture}
\caption{The structural displacements of the discretized structural nodes is mapped onto the fluid mesh (which may be non-matching) by the radial basis function (RBF) mapping and the fluid forces from the fluid nodes to the structural nodes are also transferred in the same manner.}
\label{RBF_bat}
\end{figure}
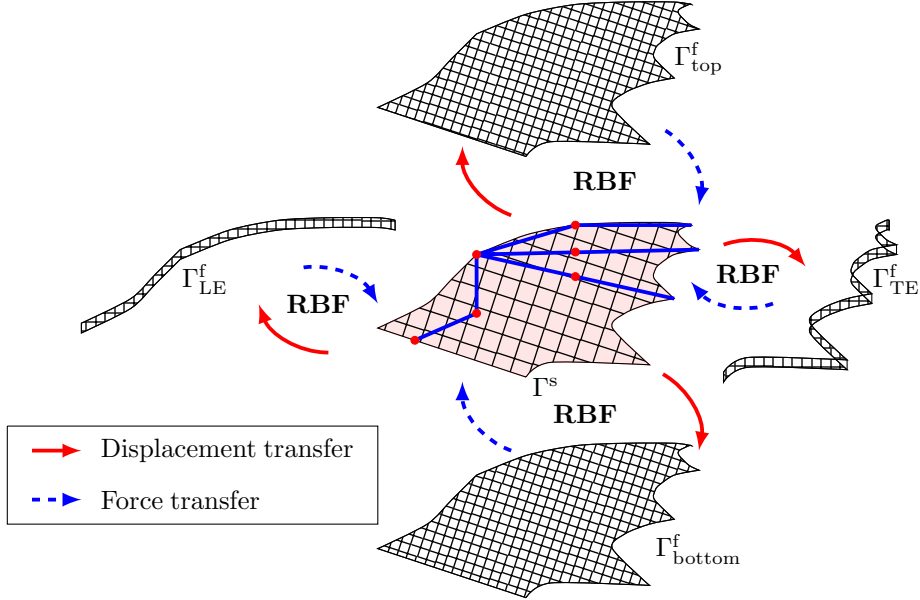

A typical mesh for the bat model considered in the present study is shown in Fig. \ref{bat_mesh}. A boundary layer mesh is constructed around the two wings of the bat consisting of 4.04 million grid points and 3.96 million eight-node hexahedron elements. The size of the elements is progressively increased from the bat to the outer computational domain. The two-dimensional structural mesh, shown in Fig. \ref{bat_mesh}(d) consists of 792 four-node quadrilateral elements for the shell along with 74 two-node beam elements for each wing. The non-dimensional time step size employed in the simulation is $\Delta t U_\mathrm{ref}/c = 0.071$ or $\Delta t/T_w = 2.5 \times 10^{-3}$. The non-matching characteristic of the discretization along the fluid-structure interface is a challenge to be addressed for such aeroelastic problems. As mentioned in previous sections, we utilize the radial basis function (RBF) interpolation to transfer the displacements as well as the tractions across the interface. A schematic of this transfer for the case of a discretized bat wing is shown in Fig. \ref{RBF_bat}. 

\subsection{Numerical predictions}
\begin{figure}[]
	\centering
	\begin{subfigure}[b]{0.5\textwidth}
		\includegraphics[trim={10.5cm 0 12cm 0},clip,width=8cm]{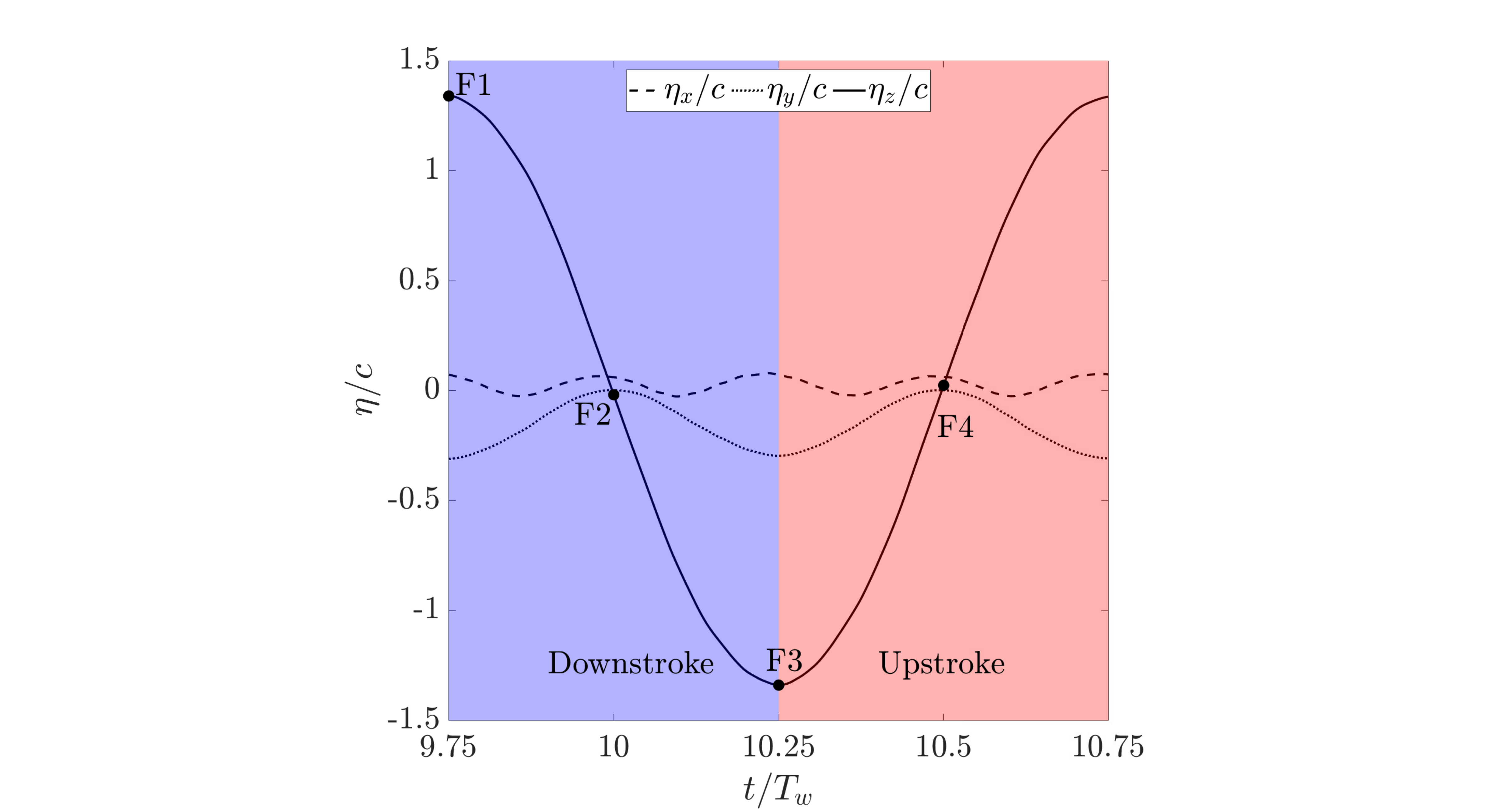}
		\caption{}
	\end{subfigure}%
	\begin{subfigure}[b]{0.5\textwidth}
		\includegraphics[trim={10.5cm 0 12cm 0},clip,width=8cm]{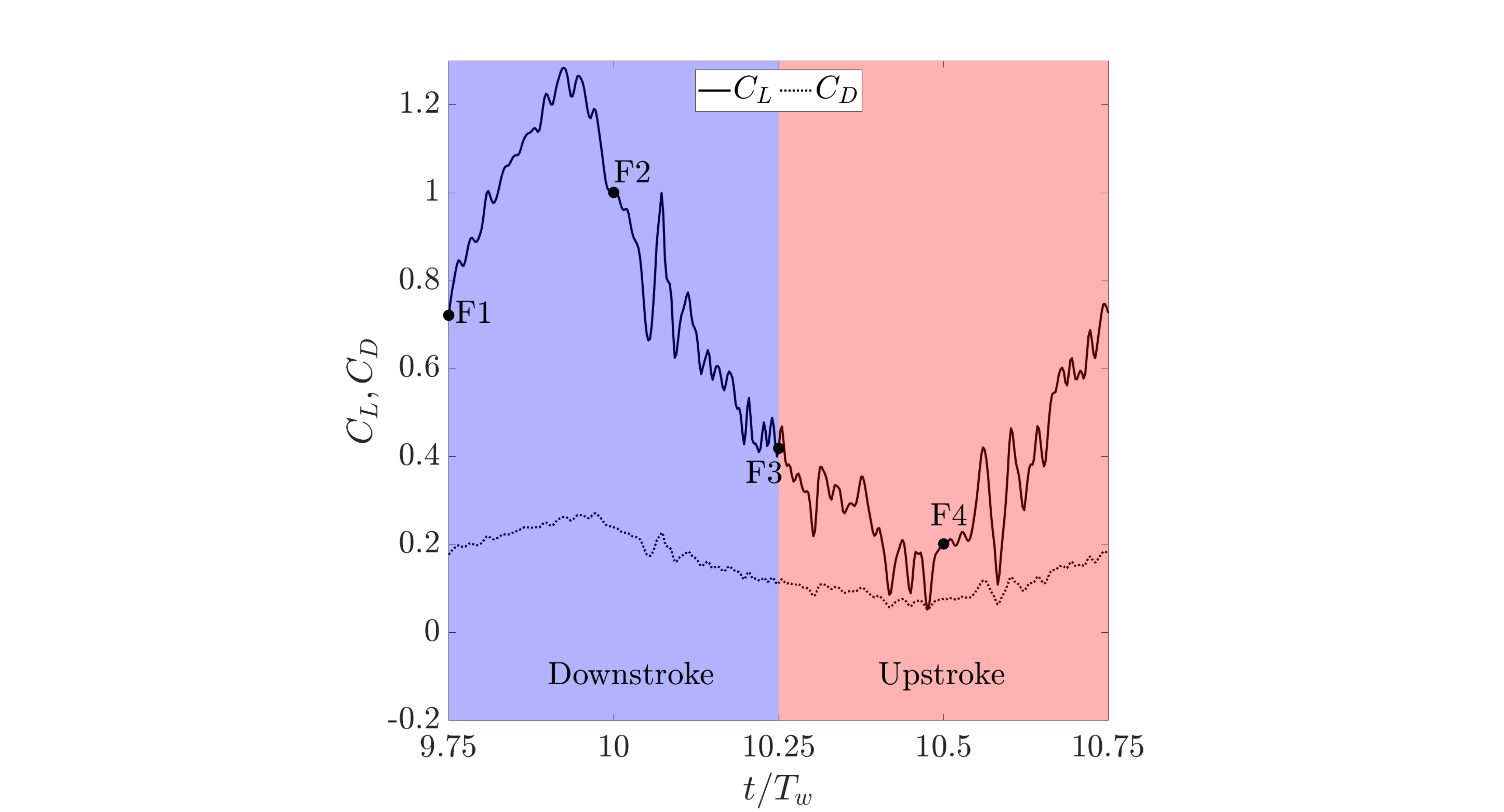}
		\caption{}
	\end{subfigure}%
	\caption{Time histories of (a) normalized displacement at the right wing tip, and (b) lift and drag coefficients for a time period of oscillation at angle of attack of $10^{\circ}$. F1, F2, F3 and F4 indicate the points at one-quarter times of a period of flapping which have been utilized for further analysis.}
	\label{disp_force_AOA10}
\end{figure}

\begin{figure}[!htbp]
	\centering
	\begin{subfigure}[b]{0.25\textwidth}
		\centering
		\includegraphics[trim={2cm 4cm 2cm 4cm},clip,width=4.5cm]{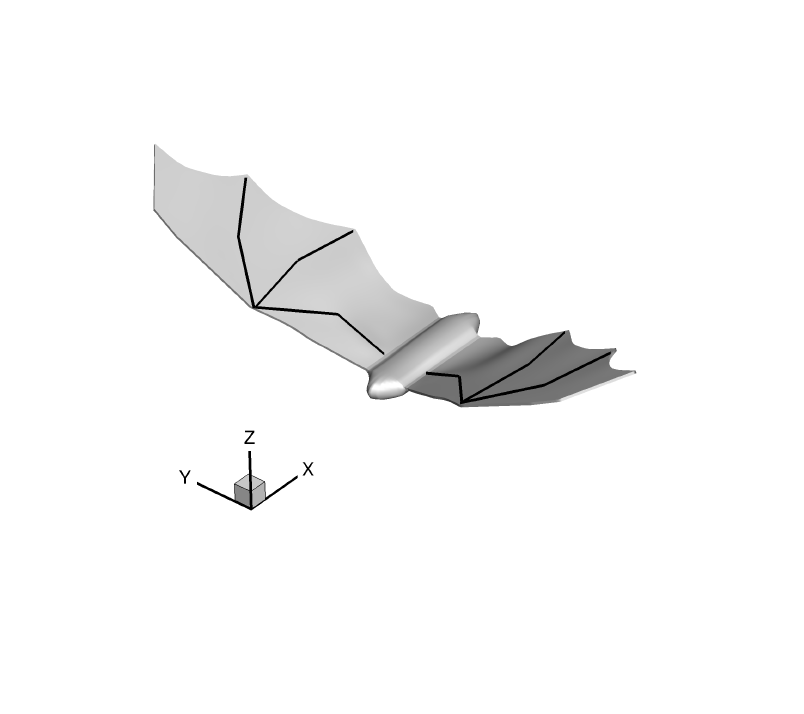}
		\caption{}
	\end{subfigure}%
	\begin{subfigure}[b]{0.25\textwidth}
		\centering
		\includegraphics[trim={2cm 4cm 2cm 4cm},clip,width=4.5cm]{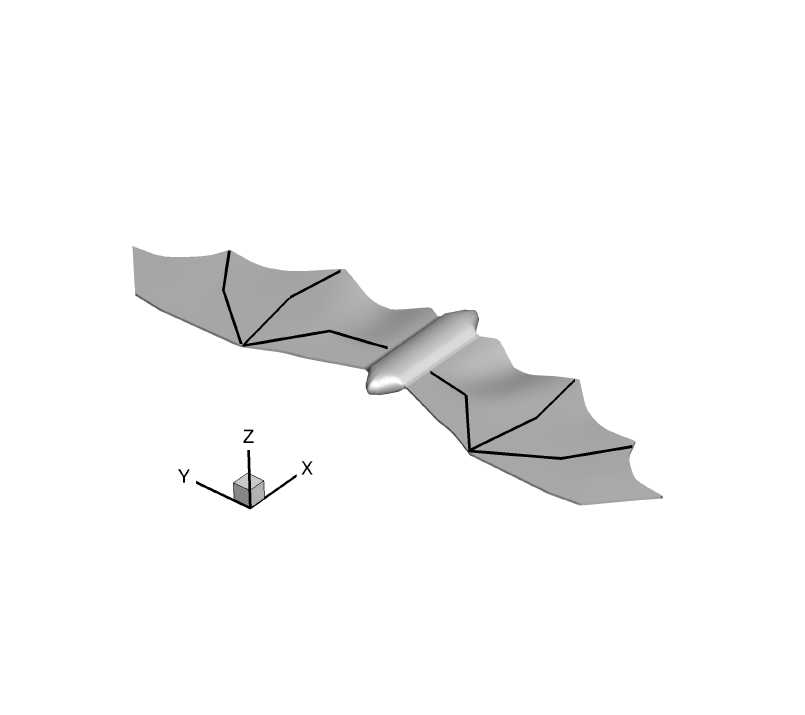}
		\caption{}
	\end{subfigure}%
	\begin{subfigure}[b]{0.25\textwidth}
		\centering
		\includegraphics[trim={2cm 4cm 2cm 4cm},clip,width=4.5cm]{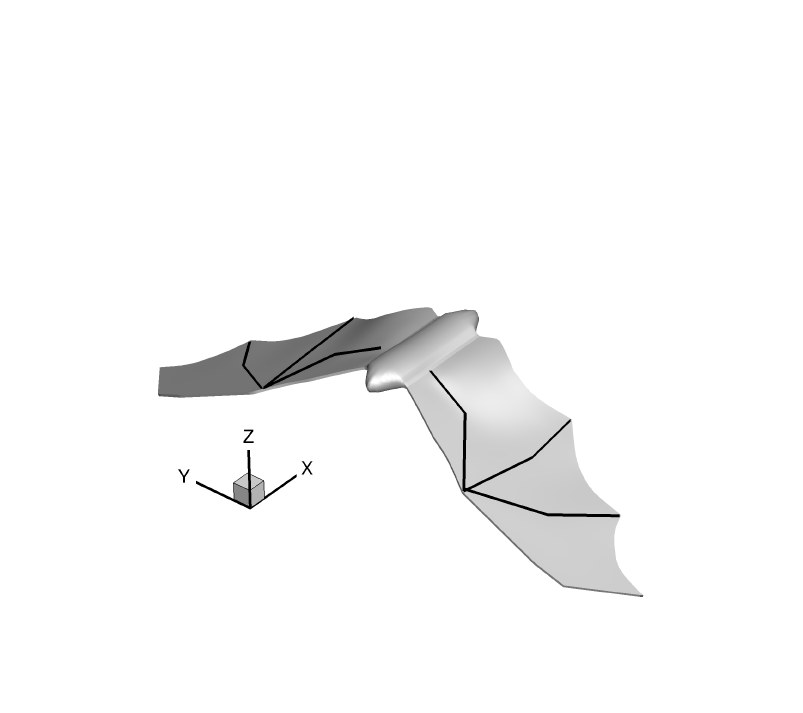}
		\caption{}
	\end{subfigure}%
	\begin{subfigure}[b]{0.25\textwidth}
		\centering
		\includegraphics[trim={2cm 4cm 2cm 4cm},clip,width=4.5cm]{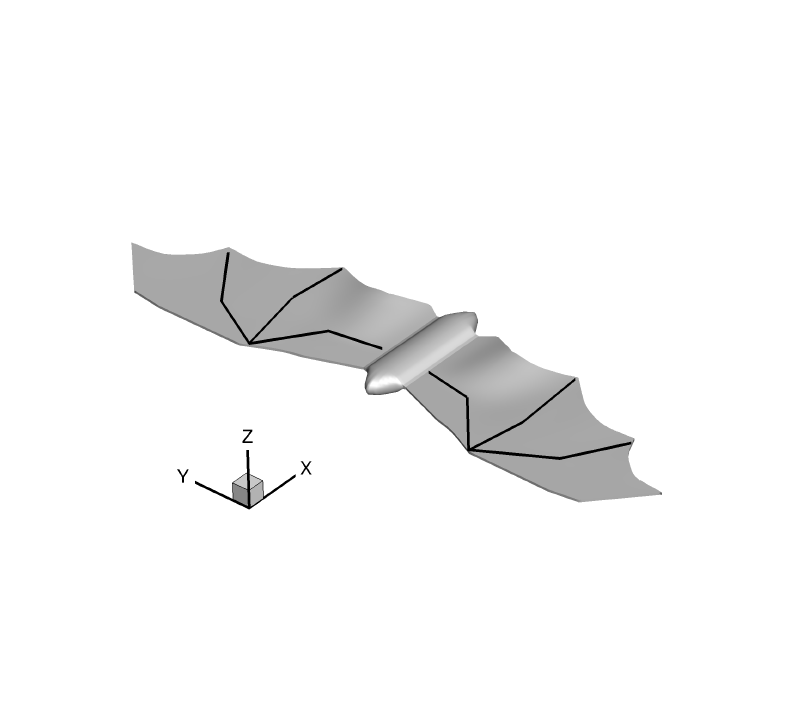}
		\caption{}
	\end{subfigure}%
	\caption{The deformation of the wing with approximate locations of the joints and bone fingers for the flexible flapping wings at: (a) $t/T_w = 9.75$ (F1), (b) $t/T_w = 10$ (F2), (c) $t/T_w=10.25$ (F3) and (d) $t/T_w=10.5$ (F4).}
	\label{Deformation_AOA10}
\end{figure}

In this section, we investigate the amplitude response, aerodynamic coefficients and wake dynamics of the flapping flight of a bat at $10^{\circ}$ angle of attack. The response of the amplitude of the tip of the right wing for the flexible flapping wing in a flapping cycle is shown in Fig. \ref{disp_force_AOA10}(a). In the figure, F1, F2, F3 and F4 denote the four temporal locations in a flapping cycle chosen for further analysis and post-processing purposes. The time history of the integrated lift and drag coefficients on the surface of the wings of the bat are shown in Fig. \ref{disp_force_AOA10}(b). We observe that most of the lift is generated during the downstroke, as observed in the literature. A larger effective angle of attack during the downstroke is also responsible for the high lift. High frequency oscillations are also observed which may be due to the different modes of vibrations across the flexible membrane. The deformation along the flexible wings of the bat are shown in Fig. \ref{Deformation_AOA10} with approximate locations of the bone fingers and joints.

\renewcommand{\arraystretch}{0.5}
\begin{table}[!htbp]
	\caption{Aerodynamic coefficients for the different wing configurations at $AOA = 10^{\circ}$.}
	\centering
	\begin{tabular}{  M{1.5cm} | M{5cm} |  M{5cm} N }
		\hline
		\centering
		\textbf{Parameter} & \textbf{Rigid non-flapping wing} & \textbf{Flexible flapping wing} &\\[10pt]
		\hline
		\centering
		$\overline{C_L}$ & 0.5623 & 0.6059 &\\[10pt]
		
		\centering
		$\overline{C_D}$ &  0.1462 & 0.1522 &\\[10pt]
		
		\centering
		$\overline{C_L}_\mathrm{,rms}$ &  0.0069 & 0.3446 &\\[10pt]
		
		\centering
		$\overline{C_D}_\mathrm{,rms}$ &   0 & 0.0623 &\\[10pt]
		
		\centering
		${C_L}_\mathrm{,max}$  &  0.5822 & 1.2838 &\\[10pt]
		
		\centering
		$\overline{C_L}/\overline{C_D}$ &  3.8452 & 3.9805 &\\[10pt]
		\hline
	\end{tabular}
	\label{aerody_par}
\end{table}
The statistical data for the aerodynamic coefficients are provided in Table \ref{aerody_par}. For comparison, we also considered a rigid non-flapping wing case, which is similar to the case of a fixed-wing flying vehicle.
We observe improvements in both the mean lift coefficient (8 \%) as well as the mean lift-to-drag coefficient ratio (3.5 \%). In particular, the higher value of the unsteady lift coefficient (120 \%) for the flexible flapping wing is observed which is the basis mechanism for the generation of lift for the natural flyers.

\begin{figure}[]
	\centering
	\begin{subfigure}[b]{0.5\textwidth}
		\centering
		\includegraphics[trim={0.5cm 2cm 0.5cm 0.5cm},clip,width=7cm]{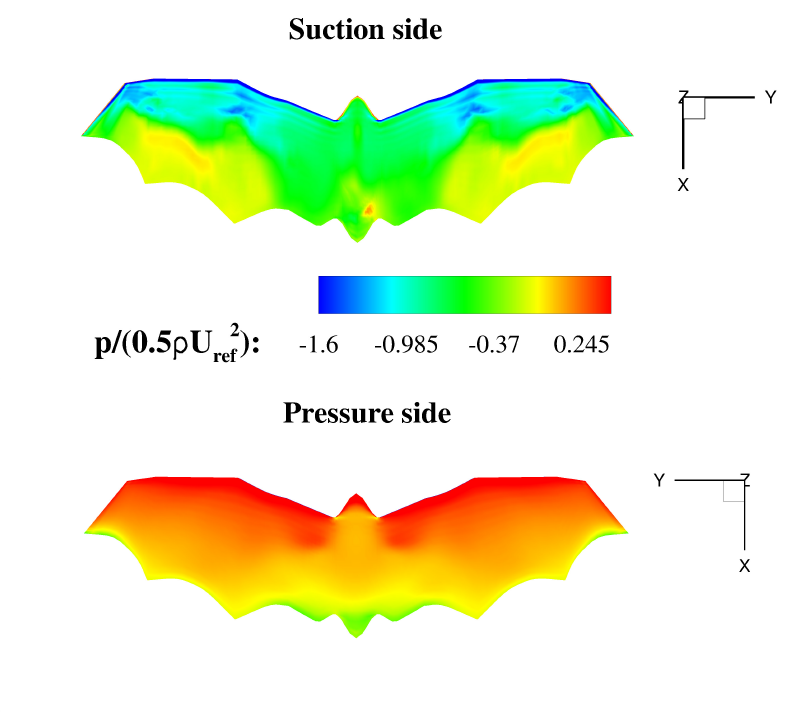}
		\caption{$t/T_w= 9.75$ (F1)}
	\end{subfigure}%
	\begin{subfigure}[b]{0.5\textwidth}
		\centering
		\includegraphics[trim={0.5cm 2cm 0.5cm 0.5cm},clip,width=7cm]{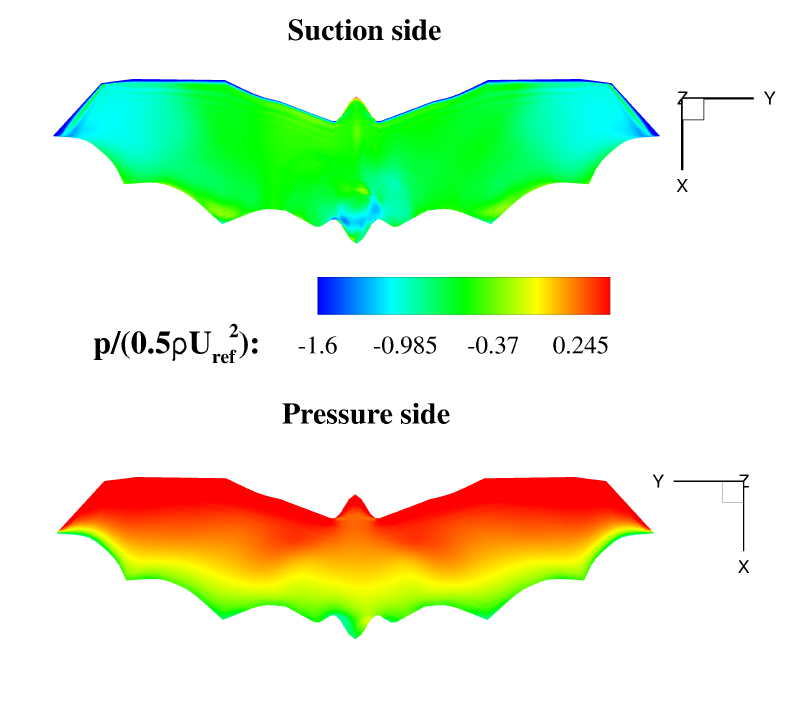}
		\caption{$t/T_w = 10$ (F2)}
	\end{subfigure}%
	\vspace{0.5cm}
	
	\begin{subfigure}[b]{0.5\textwidth}
		\centering
		\includegraphics[trim={0.5cm 2cm 0.5cm 0.5cm},clip,width=7cm]{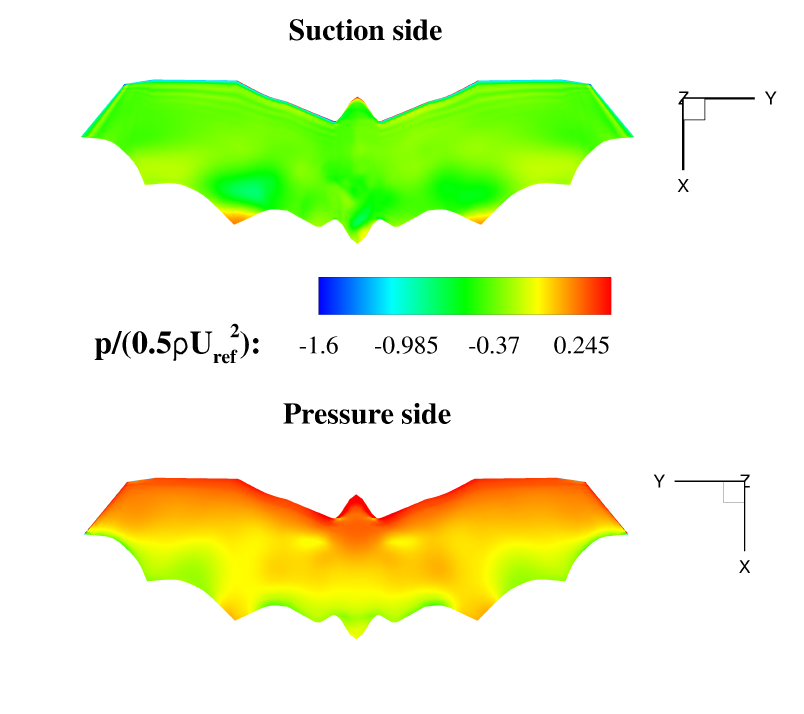}
		\caption{$t/T_w=10.25$ (F3)}
	\end{subfigure}%
	\begin{subfigure}[b]{0.5\textwidth}
		\centering
		\includegraphics[trim={0.5cm 2cm 0.5cm 0.5cm},clip,width=7cm]{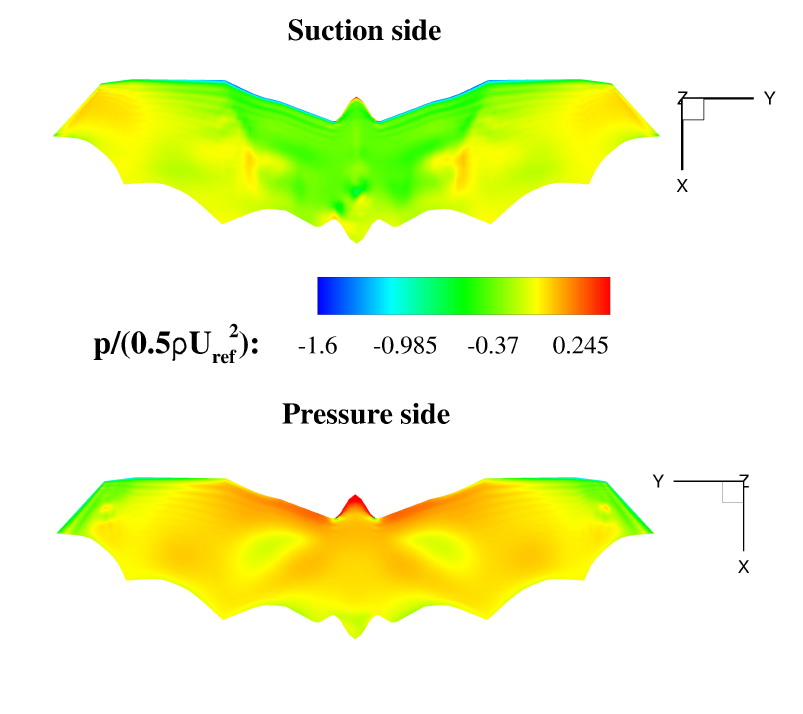}
		\caption{$t/T_w=10.5$ (F4)}
	\end{subfigure}%
	\caption{Pressure distribution on the pressure and suction sides of the bat wing at $AOA=10^{\circ}$.}
	\label{pres_AOA10}
\end{figure}

To explore the reason behind the higher value of the unsteady lift, we analyze the pressure distribution on the suction and the pressure sides of the wings for the temporal locations F1, F2, F3 and F4 in Fig. \ref{pres_AOA10}. It can be observed that the positive pressure regions on the pressure side and negative regions on the suction side creates a larger pressure differential across the wings during the downstroke (F1 and F2) compared to the upstroke (F3 and F4). 

\begin{figure}[]
	\centering
	\begin{subfigure}[b]{0.5\textwidth}
		\centering
		\includegraphics[trim={0.5cm 0.1cm 0.1cm 0.1cm},clip,width=7cm]{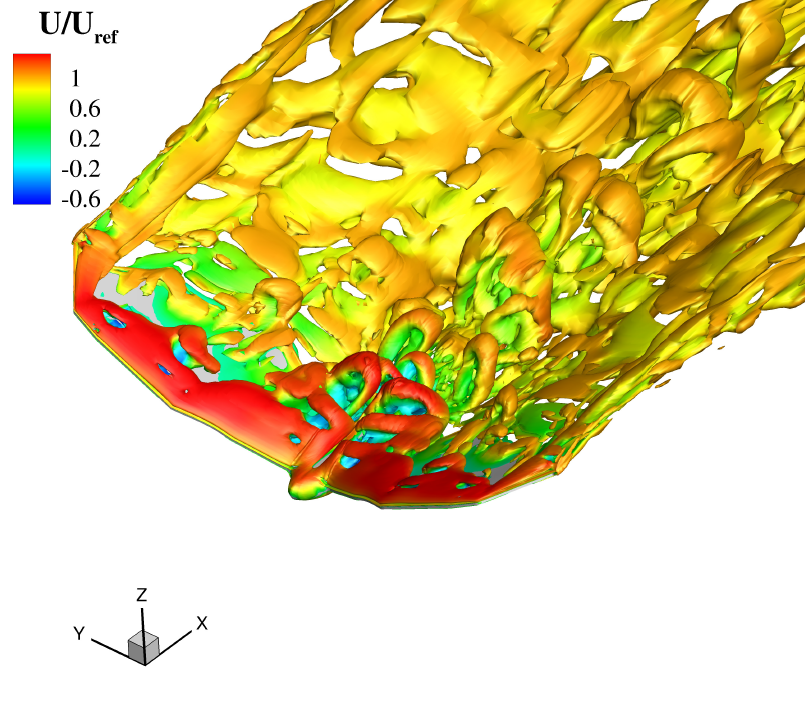}
		\caption{$t/T_w = 9.75$ (F1)}
	\end{subfigure}%
	\begin{subfigure}[b]{0.5\textwidth}
		\centering
		\includegraphics[trim={0.5cm 0.1cm 0.1cm 0.1cm},clip,width=7cm]{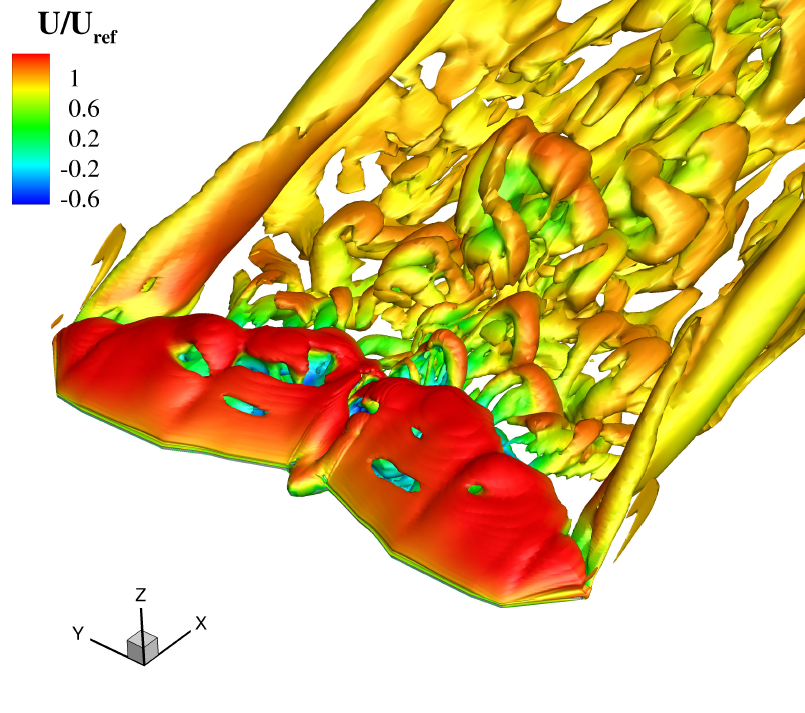}
		\caption{$t/T_w = 10$ (F2)}
	\end{subfigure}%
	\vspace{0.5cm}
	
	\begin{subfigure}[b]{0.5\textwidth}
		\centering
		\includegraphics[trim={0.5cm 0.1cm 0.1cm 0.1cm},clip,width=7cm]{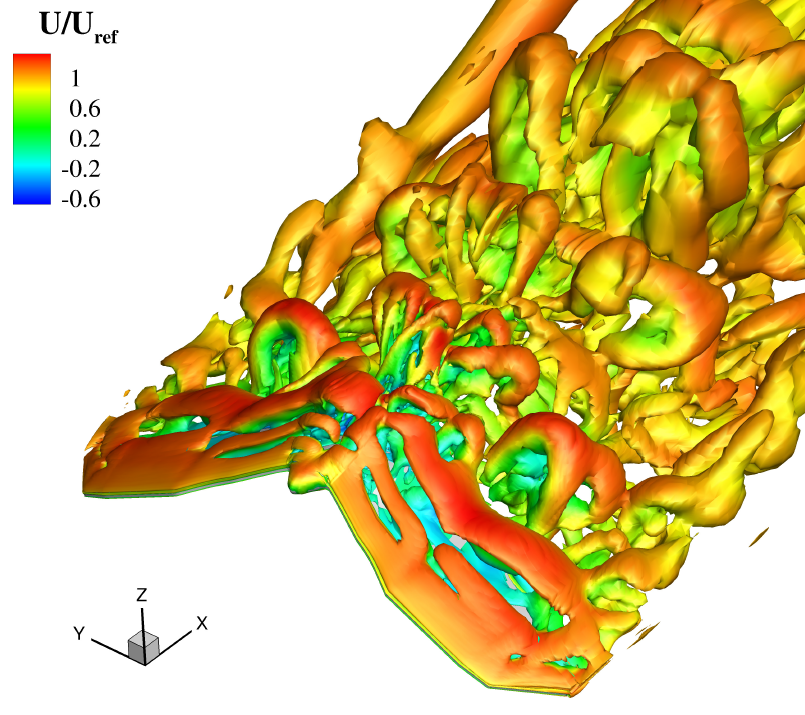}
		\caption{$t/T_w=10.25$ (F3)}
	\end{subfigure}%
	\begin{subfigure}[b]{0.5\textwidth}
		\centering
		\includegraphics[trim={0.5cm 0.1cm 0.1cm 0.1cm},clip,width=7cm]{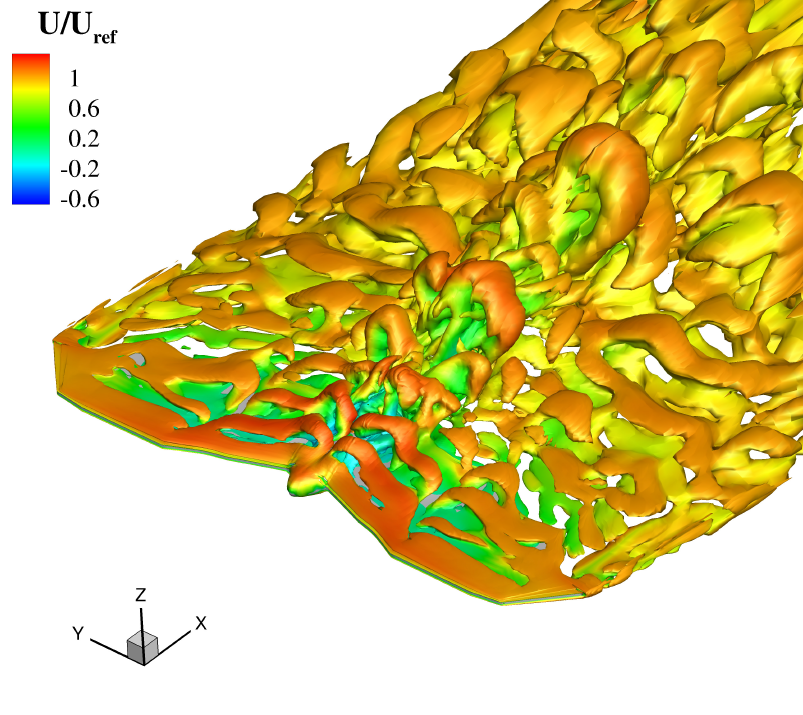}
		\caption{$t/T_w=10.5$ (F4)}
	\end{subfigure}%
	\caption{The three-dimensional vorticity patterns visualized by the iso-surface of Q-criterion at $Q^+ = 100$ colored by the normalized velocity magnitude for the flexible flapping wings.}
	\label{Qcrit_AOA10}
\end{figure}

\begin{figure}[]
	\centering
	\begin{subfigure}[b]{\textwidth}
		\centering
		\includegraphics[trim={0.5cm 0.1cm 0.1cm 0.1cm},clip,width=5cm]{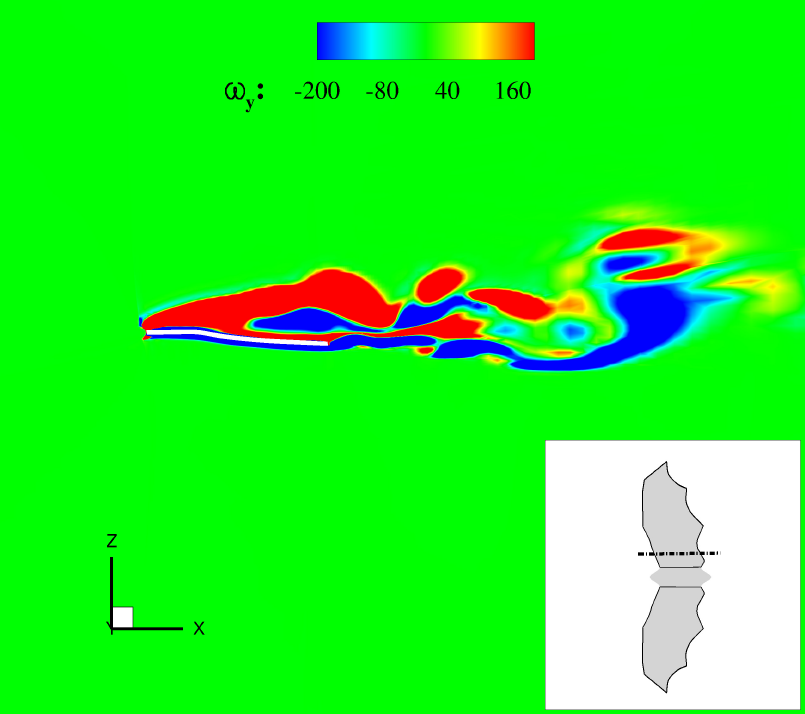}
	\hspace{0.2cm}
		\includegraphics[trim={0.5cm 0.1cm 0.1cm 0.1cm},clip,width=5cm]{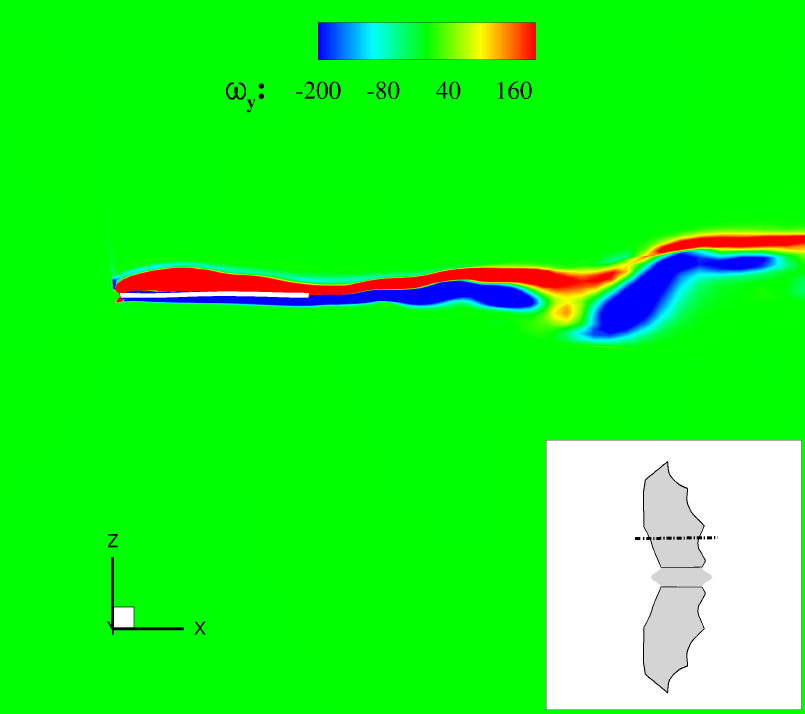}
	\hspace{0.2cm}
		\includegraphics[trim={0.5cm 0.1cm 0.1cm 0.1cm},clip,width=5cm]{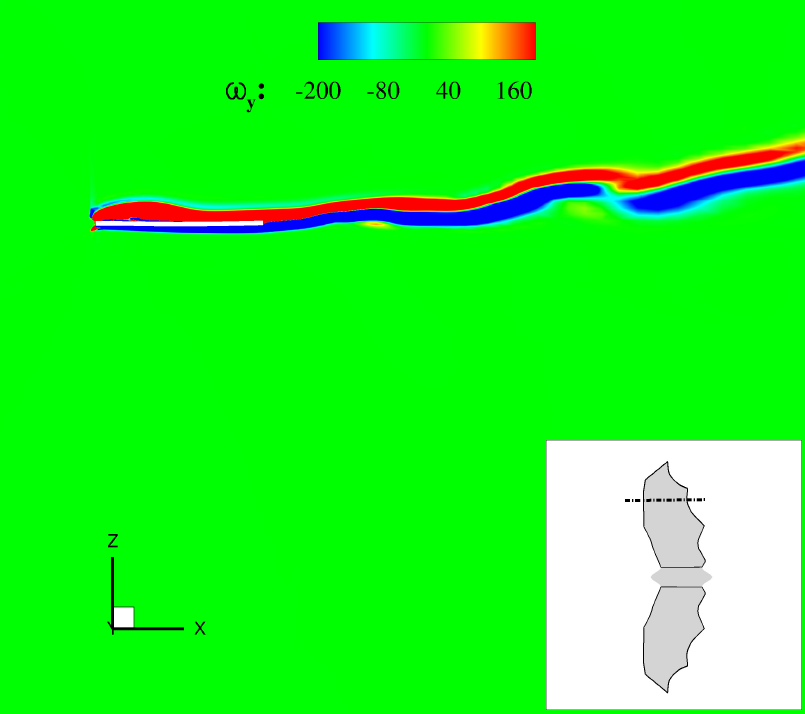}
		\caption{$t/T_w = 9.75$ (F1)}
	\end{subfigure}%
	\vspace{0.5cm}
	
	\begin{subfigure}[b]{\textwidth}
		\centering
		\includegraphics[trim={0.5cm 0.1cm 0.1cm 0.1cm},clip,width=5cm]{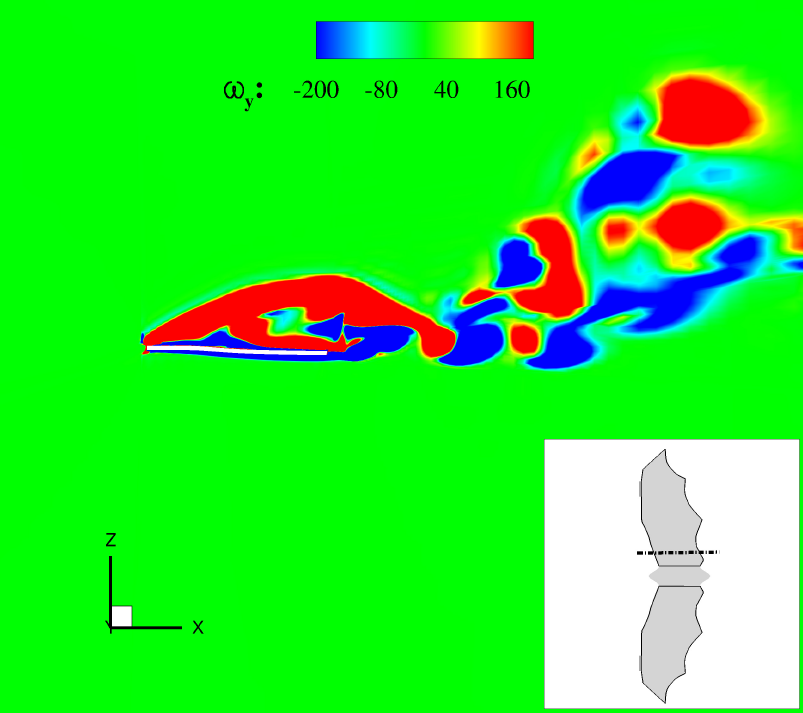}
	\hspace{0.2cm}
		\includegraphics[trim={0.5cm 0.1cm 0.1cm 0.1cm},clip,width=5cm]{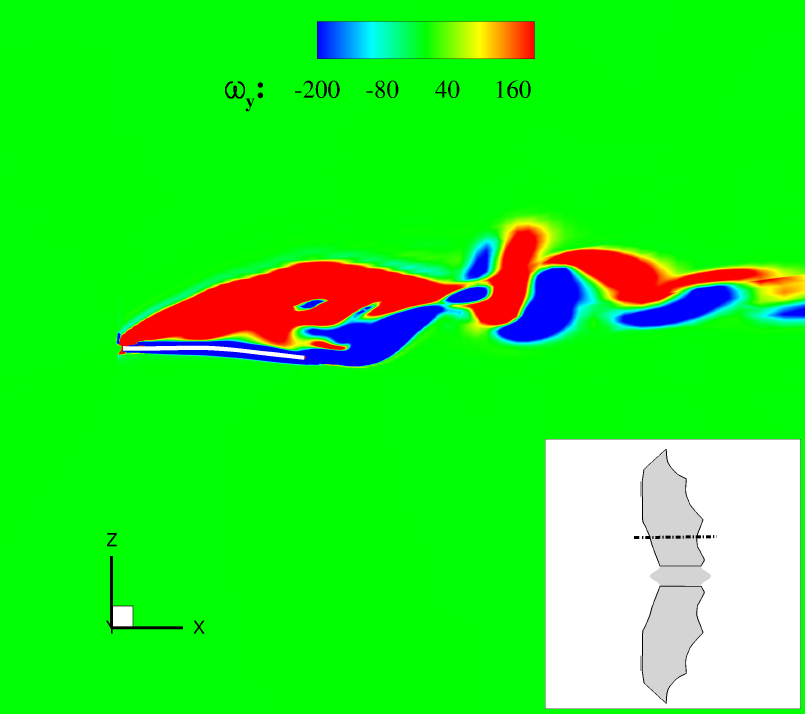}
	\hspace{0.2cm}
		\includegraphics[trim={0.5cm 0.1cm 0.1cm 0.1cm},clip,width=5cm]{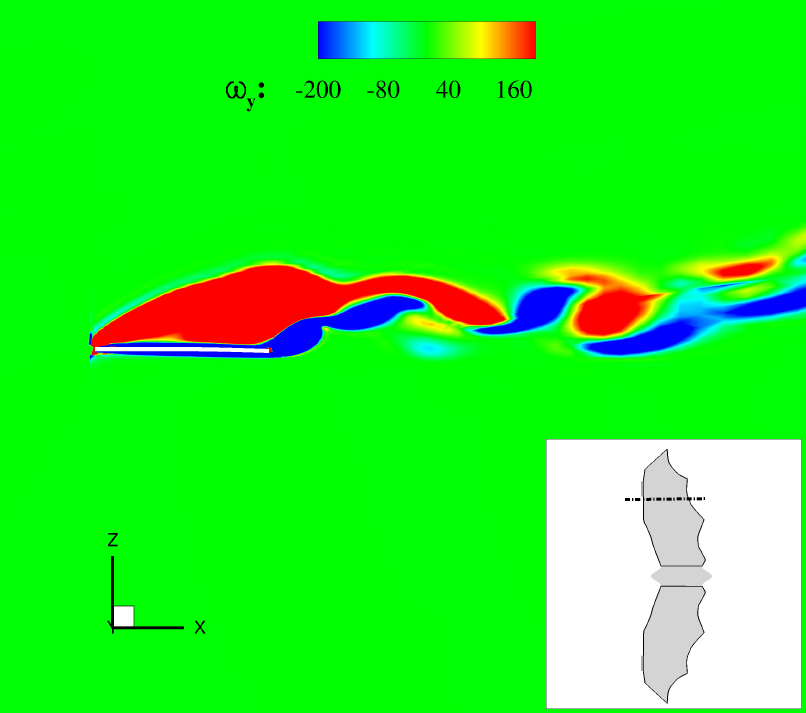}
		\caption{$t/T_w = 10$ (F2)}
	\end{subfigure}%
	\vspace{0.5cm}
	
	\begin{subfigure}[b]{\textwidth}
		\centering
		\includegraphics[trim={0.5cm 0.1cm 0.1cm 0.1cm},clip,width=5cm]{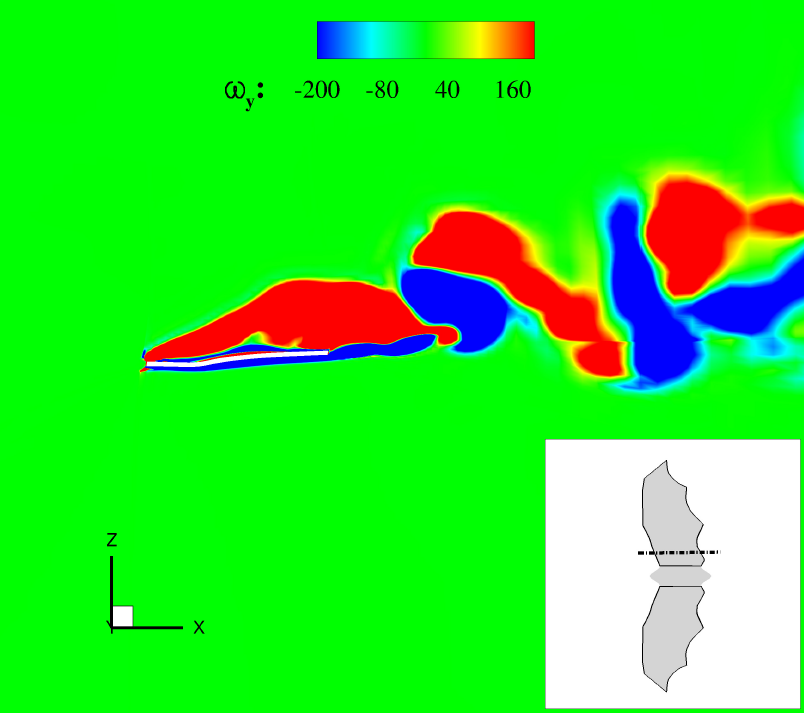}
	\hspace{0.2cm}
		\includegraphics[trim={0.5cm 0.1cm 0.1cm 0.1cm},clip,width=5cm]{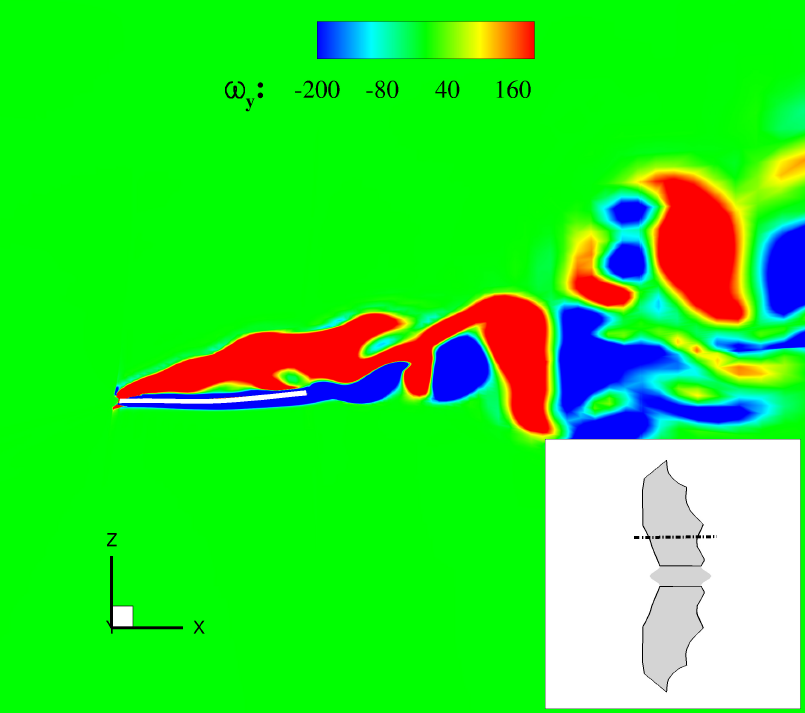}
	\hspace{0.2cm}
		\includegraphics[trim={0.5cm 0.1cm 0.1cm 0.1cm},clip,width=5cm]{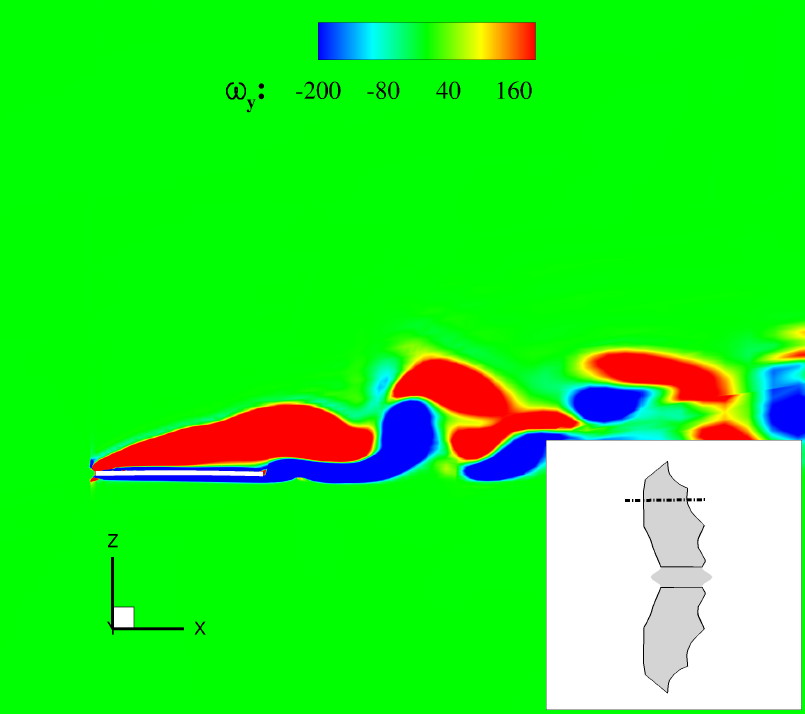}
		\caption{$t/T_w=10.25$ (F3)}
	\end{subfigure}%
	\vspace{0.5cm}
	
	\begin{subfigure}[b]{\textwidth}
		\centering
		\includegraphics[trim={0.5cm 0.1cm 0.1cm 0.1cm},clip,width=5cm]{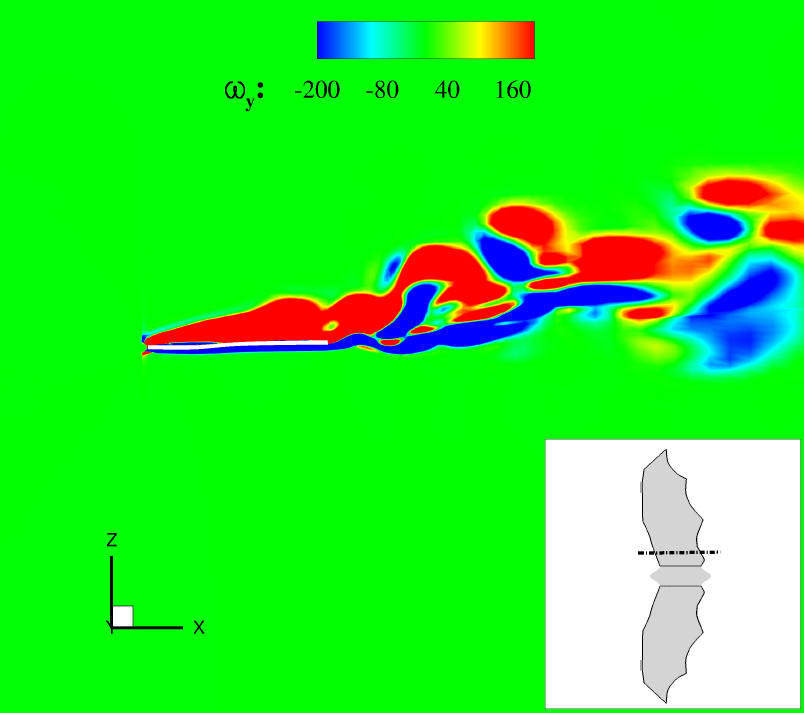}
	\hspace{0.2cm}
		\includegraphics[trim={0.5cm 0.1cm 0.1cm 0.1cm},clip,width=5cm]{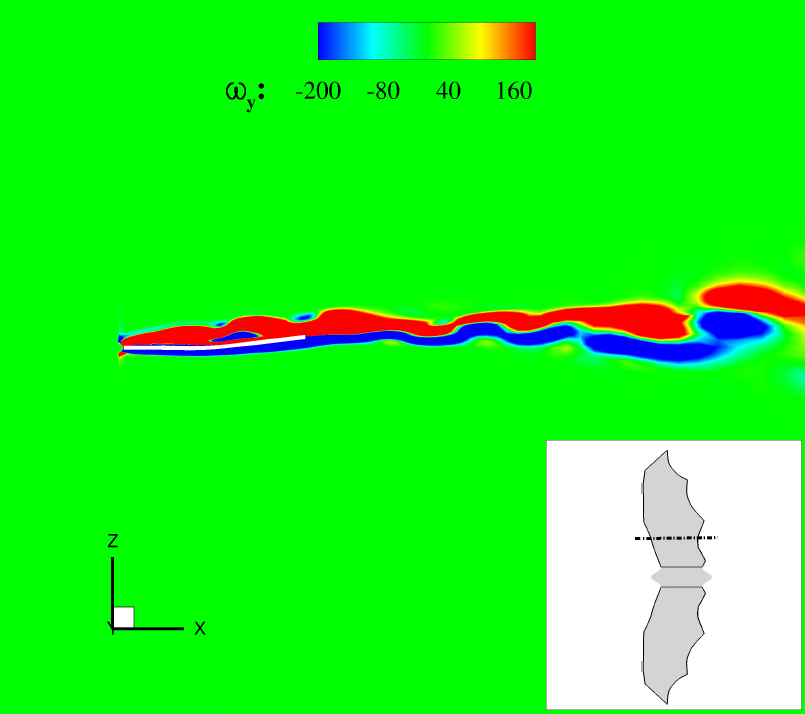}
	\hspace{0.2cm}
		\includegraphics[trim={0.5cm 0.1cm 0.1cm 0.1cm},clip,width=5cm]{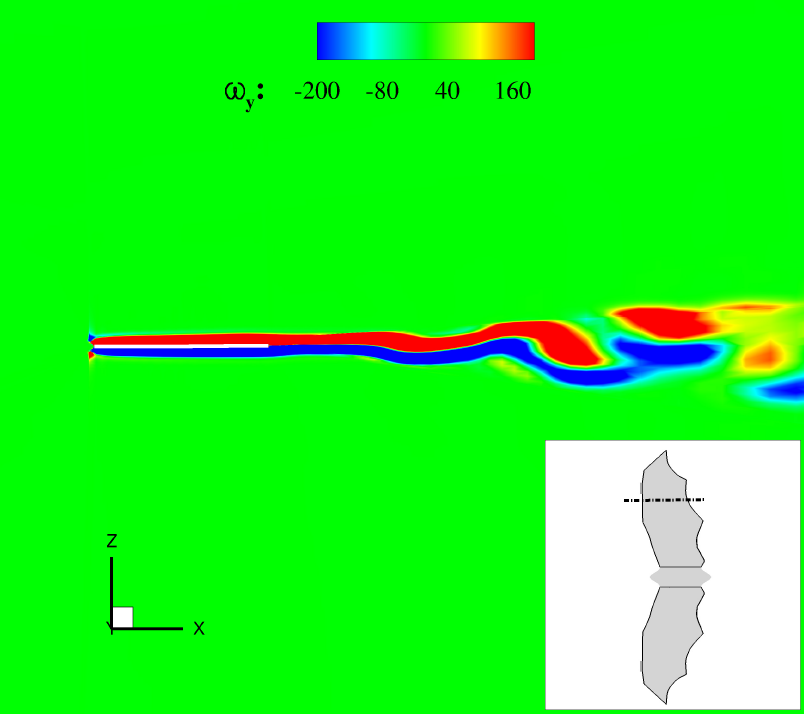}
		\caption{$t/T_w=10.5$ (F4)}
	\end{subfigure}%
	\caption{The $Y$-vorticity shown at different slices of the right wing (0.51$c$, 1.02$c$ and 2.05$c$ from the body center) for the flexible flapping wings.}
	\label{vor_AOA10}
\end{figure}

The wake of the flapping multibody wing is visualized by the iso-contours of $Q$-criterion colored by the streamwise velocity in Fig. \ref{Qcrit_AOA10}. Horseshoe-like vortices are observed in the near as well as far wake which are generated due to the inward serration of the trailing edge, similar to the pattern seen in the wake of the pitching plate in Fig. \ref{plate_Qcriterion_St51}. A complex interaction between the tip vortices and leading-edge vortices is observed. Furthermore, the $Y$-vorticity contours at different cross-sections of along the span of the right wing are shown in Fig. \ref{vor_AOA10}. As the structural properties are varying along the cross-section (depending upon the presence of bone fingers and membrane), the deformation of the wing is affected passively. In reality, due to the active kinematics of the bone fingers and joints, the prime bone fingers (humerus, radius and metacarpals) are controlled explicitly by the bat during flight which leads to further deformation in the membrane wings and rich vortex dynamics. This complex investigation is a topic of future study.

\section{Conclusions}
\label{conclusion}

In this paper, a variational flexible multibody aeroelastic framework has been presented for small strain problems. The connections between the multiple components of the structural system are imposed as constraints by the Lagrange multiplier technique. The fluid loads are solved by the incompressible Navier-Stokes equations and are transferred to the structural interface by the radial basis function interpolation with compact support. This interpolation technique is found to have around third-order of convergence based on the error analysis conducted. The framework provides us with a partitioned staggered scheme with a nonlinear correction of fluid forces to impart stability and accuracy for small structure-to-fluid mass ratios. Flow across a pitching plate with serration is considered for mesh convergence and validation of the developed framework. It is found that the numerical results are in very close agreement with the experimental observations. As an application to the passive flapping of flexible wings of a bat, the framework is then demonstrated with multiple components in the bat wing. The bone fingers are modeled as Euler-Bernoulli beams, the flexible membrane between the bone fingers is assumed to be a thin shell and the joints connecting the bone fingers are modeled as revolute joints. The results show an 8\% increase in the mean lift coefficient compared to the fixed wing counterpart. A higher value of maximum unsteady lift of about 120\% is also observed as a result of flexible flapping phenomenon. Horseshoe-like vortices form a complex pattern in the wake of the flapping bat wing. The presented high-fidelity variational formulation is generic and can model fluid-structure interaction with flexible multibody applications in biological, marine/offshore, automobile and aerospace industries. 

\section*{Acknowledgements}
This research was supported in part through computational resources and services provided by Advanced Research Computing at the University of British Columbia.

\section*{References}
\bibliographystyle{unsrt}
\bibliography{biblio}

\end{document}